\RequirePackage{fix-cm}
\pdfoutput=1
\documentclass[twoside]{svjour3} 
\usepackage{aas_macros}
\smartqed
\usepackage{natbib}
\usepackage{graphicx}
\usepackage{hyperref}
\usepackage{amsmath}
\usepackage{amssymb}

\usepackage{todonotes}
\usepackage[draft,commentmarkup=todo]{changes}

\definechangesauthor{SA}
\definechangesauthor{PA}
\definechangesauthor{AI}
\definechangesauthor{EK}
\definechangesauthor{JM}
\definechangesauthor{DK}
\definechangesauthor{GN}
\definechangesauthor{DP}
\definechangesauthor{KP}
\definechangesauthor{DY}

\usepackage{hyperref}

\usepackage{geometry}
\def\ion#1#2{#1$\;${\small\rm\@{#2}}\relax}

\newcommand{\arcsec}{^{\prime\prime}}

\title{Novel data analysis techniques in coronal seismology}
\author{Sergey A. Anfinogentov  \and Patrick Antolin   \and  Andrew R. Inglis \and 	Dmitrii Kolotkov \and Elena G. Kupriyanova \and 	James A. McLaughlin \and	Giuseppe Nistic\`o  \and David J. Pascoe \and S. Krishna Prasad \and  Ding Yuan }

\institute{S. A. Anfinogentov \at
              Institute of solar-terrestrial physics, \\
              Lermontov st., 126a Irkutsk 664033, Russia
              Tel.: +7-3952-428265 \\
              Fax: +7-3952-425557\\
              \email{anfinogentov@iszf.irk.ru}           
        \and
           James A. McLaughlin [0000-0002-7863-624X] \at
            Northumbria University, Newcastle upon Tyne, NE1 8ST, UK
           \and
            Patrick Antolin [0000-0003-1529-4681] \at
            Northumbria University, Newcastle upon Tyne, NE1 8ST, UK
        \and
            David J. Pascoe [0000-0002-0338-3962] \at
            Centre for mathematical Plasma Astrophysics, Mathematics Department, KU Leuven, Celestijnenlaan 200B bus 2400, B-3001 Leuven, Belgium
            \email{david.pascoe@kuleuven.be}
        \and S. Krishna Prasad [0000-0002-0735-4501] \at
        Centre for mathematical Plasma Astrophysics, KU Leuven, Celestijnenlaan 200B, 3001 Leuven, Belgium
        \and
          D.Y. Kolotkov [0000-0002-0687-6172] \at
          Centre for Fusion, Space and Astrophysics, Physics Department, University of Warwick,
          Coventry CV4 7AL, United Kingdom,\\ 
          Institute of Solar-Terrestrial Physics SB RAS, Irkutsk 664033, Russia\\
          \email{d.kolotkov.1@warwick.ac.uk}
}

\date{Received: date / Accepted: date}

\begin{document}

\maketitle

\begin{abstract}
We review novel data analysis techniques developed or adapted for the field of coronal seismology. We focus on methods from the last ten years that were developed for extreme ultraviolet (EUV) imaging observations of the solar corona, as well as for light curves from radio and X-ray. The review covers methods for the analysis of transverse and longitudinal waves; spectral analysis of oscillatory signals in time series; automated detection and processing of large data sets; empirical mode decomposition; motion magnification; and reliable detection, including the most common pitfalls causing artefacts and false detections. We also consider techniques for the detailed investigation of MHD waves and seismological inference of physical parameters of the coronal plasma, including restoration of the three-dimensional geometry of oscillating coronal loops, forward modelling and Bayesian parameter inference.

\end{abstract}

\section{Introduction}


During the recent decade, a number of new instruments for observing the solar corona were commissioned including space-borne instruments such as the Atmospheric Imaging Assembly on-board the Solar Dynamics Observatory \citep[SDO/AIA, ][]{2012SoPh..275...17L}, the TESIS experiment on the CORONAS-PHOTON spacecraft \citep{2011SoSyR..45..162K}, and the Interface Region Imaging Spectrograph \citep[IRIS, ][]{2014SoPh..289.2733D}, as well as ground-based optical instruments such as the Coronal Multi-Channel Polarimeter \citep[CoMP, ][]{2008SoPh..247..411T}. 
These additional  observational capabilities triggered a new wave of research in the field of coronal seismology, the results of which are discussed in detail in other reviews of this series.

Among the current observational instruments, the most important is SDO/AIA which is an imaging instrument continuously observing the Sun and providing full-disk images in 7 channels in the EUV band with a cadence of 12 seconds, as well as in 2 channels in the UV range with a cadence of 24 seconds. The spatial resolution is around one arcsecond, with a pixel size of 0.6 arcseconds for all channels.
In contrast to the previous-generation instrument TRACE, SDO/AIA has higher sensitivity and a wider field-of-view and observes the Sun simultaneously in several filters that are sensitive to different temperatures.
Higher sensitivity allows for precise measurements of the brightness variations associated with MHD waves. Thanks to the field-of-view covering the whole solar disk and almost-continuous observational setup, every event is recorded and can be investigated afterwards.
The simultaneous high cadence observations at multiple wavelengths allows the differential emission measure (DEM) to be computed, and hence   the density and temperature in coronal structures to be estimated. 

Since seismological information comes from spatially- and temporally-resolved observations of the solar corona, a typical workflow of data analysis consists of three steps: 
\begin{enumerate}
	\item Dimension reduction in order to extract a time-series corresponding to some physical quantity (e.g. displacement or brightness of a coronal structure);
	\item Spectral analysis of the time-series obtained in the previous step;
	\item Inferring physical properties of the coronal plasma from parameters of the oscillatory signals obtained with spectral analysis. 
\end{enumerate}

Although imaging data obtained by SDO/AIA are essentially 3D, with two spatial and one temporal dimension, traditional analysis techniques such as Fast Fourier Transform (FFT) and Continuous Wavelet Transform (CWT), as well as a number of novel techniques reviewed here (see Section \ref{sec:detection_in_time_series}) are designed to process 1D time-series.
Therefore, before applying these techniques, one needs to reduce the number of dimensions.
A typical approach of dimension reduction is to put an artificial slit across or along the oscillating structure (depending upon the polarisation of the analysed oscillation) and produce a 2D distribution of the emission intensity in the time-distance space.
Then, the dimensions are reduced further by extracting an oscillation profile (e.g. by measuring the structure's position or its brightness at every instant of time).
This approach is very common and is also used by some of the advanced data analysis techniques reviewed here (see Sections \ref{sec:NUWT} and \ref{sec:longitudinal_waves}).
However, for some techniques the spatial information in imaging data is essential and used without dimension reductions (see Sections \ref{sec:motion_magnification}, \ref{sec:3D_shape}, and \ref{sec:forward_modelling}).

The rapid development of coronal MHD seismology and the new observational capabilities have caused new challenges for the research field, which are listed below:
\begin{enumerate}
    \item Oscillatory signals in observational time series need to be discriminated from the noise and background on which they are superimposed;
	\item Oscillatory signals associated with MHD waves and oscillations are often non-stationary and  only a few oscillation cycles are observed.
	\item Modern instruments such as SDO/AIA produce a large amount of observational data that needs to be stored and processed;
	\item Spatial and temporal resolution of the available data is limited;
	\item MHD seismology problems are usually ill-posed and the available observational information needed to infer model parameters is often limited and incomplete;

\end{enumerate}


Time-series analysis is one of the basic elements in a typical data processing pipelines in the field of coronal seismology.
For instance, a time-series can represent the coronal loop displacement or its brightness variation associated with an MHD wave.
Such an oscillatory signal is always superimposed with noise and non-periodic background processes.
Therefore, it should be discriminated from the noise and a background trend. The methods of detecting oscillatory signals in time-series are discussed in Section \ref{sec:detection_in_time_series}.

In contrast to classical helioseismology, coronal seismology often deals with the rapidly-evolving medium of the solar corona.
The characteristic evolution time of coronal structures (e.g. coronal loops) is comparable to the typical periods of MHD oscillations.
Thus, the key parameters of an oscillating structure, such as density and temperature,  can change significantly during an oscillation event causing pronounced amplitude and frequency modulation, resulting in the  non-stationarity of the observed oscillatory signals.
The analysis of such signals with traditional spectral analysis, based on decomposition of the signal into fixed basis functions such as in Fourier or Wavelet transforms, is less reliable and can lead to both false positive and false negative detections of oscillatory phenomena.
Thus, there is a need for new data processing techniques targeted at the analysis of non-stationary signals modulated in both frequency and amplitude, including the adoption of approaches intensively used in other research fields.
One such method, empirical mode decomposition (EMD), discussed in Section \ref{sec:EMD}.

The majority of coronal MHD seismology research is concentrated on case studies where one or several individual events are analysed in detail. However, there is a need for much broader statistical studies covering a large number of events and covering several years of observations.
The gathering of such a statistical sample requires new techniques with automated processing of a large amount of data.
Thus, we need  data processing techniques with the capabilities of fast automated detection and analysis for imaging observations of MHD waves in the coronal plasma.
One of the responses to this challenge is the Northumbria University Wave Tracking (NUWT) code aimed at automatic detection of transverse oscillations in time-distance plots.
This technique is  described in Section~\ref{sec:NUWT}.

Despite there being a large volume of observational data from instruments observing the solar corona in different bands, including EUV and radio, the available observables are incomplete in terms of seismological inference of the physical parameters of coronal plasma.
For instance, the determination of the coronal magnetic field from kink oscillations of coronal loops requires measurements of: the oscillation period of the fundamental kink mode, density, internal-to-external density ratio (density contrast),  and the length of the oscillating coronal loop.
From these four required parameters only the oscillation period can be  measured directly, whereas the measurement of the other three parameters has some ambiguities  because the observed EUV emission is optically thin. 
For instance, the estimation of the density contrast requires measurement of the background plasma density, which is complicated by the line-of-sight integration effects of the optically-thin EUV emission of coronal plasma.
Such observational limitations are a source of uncertainty that should be quantified and propagated to the inference results.
The solution of this problem is addressed by the Bayesian analysis discussed in Section \ref{sec:bayesian_analysis}. 
Likewise, the reconstruction of the 3D geometry of the oscillating coronal structure is crucial for the estimation of loop length and is a non-trivial problem. Solutions to this are discussed in Section \ref{sec:3D_shape}.

As mentioned, the analysis of observations of the coronal plasma in EUV and radio is complicated by line-of-sight effects.
Thus, the interpretation of such observations requires accounting for these  effects by modelling of the radiation transfer in the coronal plasma.
This problem is addressed in Section \ref{sec:forward_modelling} where forward modelling of EUV emission is discussed.

Another problem of available observations in the EUV band is the limited spatial resolution.
This is especially important for low-amplitude decayless kink oscillation of coronal loops recently discovered in SDO/AIA data (See Section 12 in Nakariakov et al., 2021).
These oscillations are characterised by very low spatial displacements of oscillating loops which are usually lower than the pixel size of the SDO/AIA instrument and, therefore are hard to detect and analyse.
This challenge motivated the development of the motion magnification algorithm which is discussed in Section \ref{sec:motion_magnification}.

This review is organised as follows:
Firstly, we discuss  methods for detection and preliminary analysis of waves and oscillations in different kinds of data. 
Section \ref{sec:detection_in_time_series} is devoted to the processing  of time-series data, while the analysis of transverse and longitudinal motions in imaging data is reviewed in Sections \ref{sec:transverse_motions}  and \ref{sec:longitudinal_waves}, respectively.
We then consider advanced data analysis techniques used for the detailed investigation of MHD waves and seismological inference of physical parameters of coronal plasma.
Methods for restoration of the 3D geometry of oscillating coronal loops are reviewed in Section \ref{sec:3D_shape}, forward modelling of the instrumental response upon waves and oscillations is considered in Section \ref{sec:forward_modelling},
and inference of the physical parameters of coronal plasma within the Bayesian paradigm is discussed in Section \ref{sec:bayesian_analysis}.

This review does not cover all existing data processing methods used in solar physics.
Instead, we focused only on the novel techniques developed in the last ten years that are connected with the subject of MHD seismology via EUV observations.
Therefore, many methods beyond the scope of our review are not considered here.
These are mainly methods developed more than ten years ago that have already become standard tools such as the wavelet analysis \citep{1998BAMS...79...61T} and tools based on it \citep{2008SoPh..248..395S}.
Since our focus is on EUV data, we do not cover new advances in forward modelling and analysis of radio observations \citep{2011SPD....42.1811N,2015SoPh..290.1173K}.
Also, we do not cover new promising techniques based on machine-learning, which can be used for forecasting solar flares and space weather events \citep[see e.g.][]{2015ApJ...798..135B, 2019SpWea..17.1166C, 2020ApJ...904L...7B} since these have not been applied in coronal seismology yet.

\section{Detection and analysis of waves and oscillations in time-series data}
\label{sec:detection_in_time_series}

The most popular traditional  techniques for the analysis of oscillatory signals in time-series data are Fast Fourier transform (FFT), Continuous Wavelet Transform (CWT), and least squares fitting.
All of these techniques are based on mode decomposition or some sort of matching of an analysed signal with a model or a set of models.
Least-squares fitting does this directly by fitting the data with a predefined model depending on a set of free parameters.
The model parameters are tuned until the best match (the lowest $\chi^2$-criterion) between the model and data is found.

FFT and CWT are based on the decomposition of an oscillatory signal into a set of modes using a predefined basis.
The Fast Fourier Transform decomposes the input signal into a set of harmonic functions  which are not modulated and therefore have constant frequencies and amplitudes.
The most valuable benefits from the FFT analysis are gained when the  FFT basis coincides with  a set of natural modes of the  physical system which is investigated.
A good example of such a system  is acoustic waves in the solar interior.
The natural modes of the Sun as an acoustic resonator are almost-pure, very narrow-band harmonic waves which can be observed for months and even years at the same frequencies.
Therefore, FFT has become the main tool in the seismology of the solar interior (helioseismology) and has allowed  a lot of valuable information about the internal structure of the Sun to be obtained.

In contrast to the solar interior, the characteristic evolution time of the structures observed in the solar corona and their life-time as well are around several tens of minutes or even less, which is comparable to typical periods of MHD oscillations observed in the corona.
Moreover, the number of observed natural modes (harmonics) of coronal oscillations is always very  limited.
Usually, only the fundamental mode and 1--2 of its overtones can be detected. 
Thus, the oscillatory processes in the solar corona are expected to be rather short-lived  and to exhibit a very limited number of harmonics (2--3) modulated in both frequency and amplitude due to fast evolution of oscillating plasma structures, even if their natural modes of  are pure sinusoids.

Since the classical FFT analysis does not have temporal resolution, it is not capable of resolving different physical modes having similar frequencies but separated in the time domain.
Since the coronal oscillations are observed as quasi-harmonic signals localised in both the time and frequency domain, a natural choice for their analysis is the CWT which provides a representation of the oscillating power of a 1D signal in a 2D time-frequency domain.
Like FFT, CWT is a decomposition of the signal into a set of oscillatory modes, but CWT modes are rather short wave-trains as opposed to the plane sinusoidal  waves used by FFT.

Despite the successful applications of FFT and CWT in coronal seismology, both techniques suffer from the same drawback: 
Their predefined modes not always match the physical modes of oscillating structures such as coronal loops.
These modes, in turn, are known only for just a few specific coronal magnetic plasma configurations (e.g. magnetic cylinders with a sharp boundary and plasma slabs with a smooth boundary determined by the symmetric Epstein profile). A possible first step to alleviate this problem is described by \citet{2019ApJ...876...86R} who considered how to determine the normal modes of a physical system by iterative application of time-dependent numerical simulations and the CEOF (Complex Empirical Orthogonal Function) analysis of simulation results. Provided the numerical models are realistic enough, this should produce normal mode characteristics comparable to observed loop oscillation events. 
Additionally, waves in coronal structures are subject to damping that introduces amplitude modulation, which is different from both the constant amplitude of the FFT modes and from the predefined symmetric shape of a wavelet.
Moreover, loop parameters such as length or temperature evolve in time causing corresponding evolution of the resonant frequencies.
These circumstances stimulated development and adaptation of new time-series analysis techniques, such as EMD, which do not use a predefined set of basis functions.
Instead,  the oscillatory modes are obtained empirically from the signal itself, potentially allowing for the extraction of true physical modes.
The detailed description of the EMD analysis is given in Section~\ref{sec:EMD}.

Below, we present  individual tools for time-series analysis including  EMD (Section~\ref{sec:EMD}) and software packages for automated time-series analysis AFINO (Section~\ref{sec:AFINO}) and Stingray (Section~\ref{sec:Stingarray}).
Following this, we discuss possible pitfalls of the time-series processing that may cause detection of false periodicities even in pure noise (Section~\ref{sec:time-series_pitfalls})  and summarise everything on the example of searching for quasi-periodic pulsations (QPPs) in flaring lightcurves (Section~\ref{sec:QPP}).

\subsection{Empirical mode decomposition for analysis of oscillatory processes on the Sun and stars}  \label{sec:EMD}

A novel promising method for the detection and analysis of quasi-periodic and non-stationary oscillatory patterns in astrophysical observations, in particular various oscillations in the solar and stellar atmospheres, is empirical mode decomposition \citep[EMD,][]{1998RSPSA.454..903H}. Originally designed for geophysical applications \citep[see e.g.][for review]{2008RvGeo..46.2006H}, its potential for astrophysical application was revealed rather recently, thus demonstrating an excellent example of scientific knowledge transfer. In this section, we briefly outline the basic principles of EMD, overview examples of its application for processing of solar, stellar, and magnetospheric observations, describe current trends in the further development of the method, highlight its advantages in comparison with more traditional and well-elaborated Fourier and wavelet approaches, and summarise known shortcomings and perspectives for improvement.

\begin{figure}
	\centering
	\includegraphics[width=0.93\linewidth]{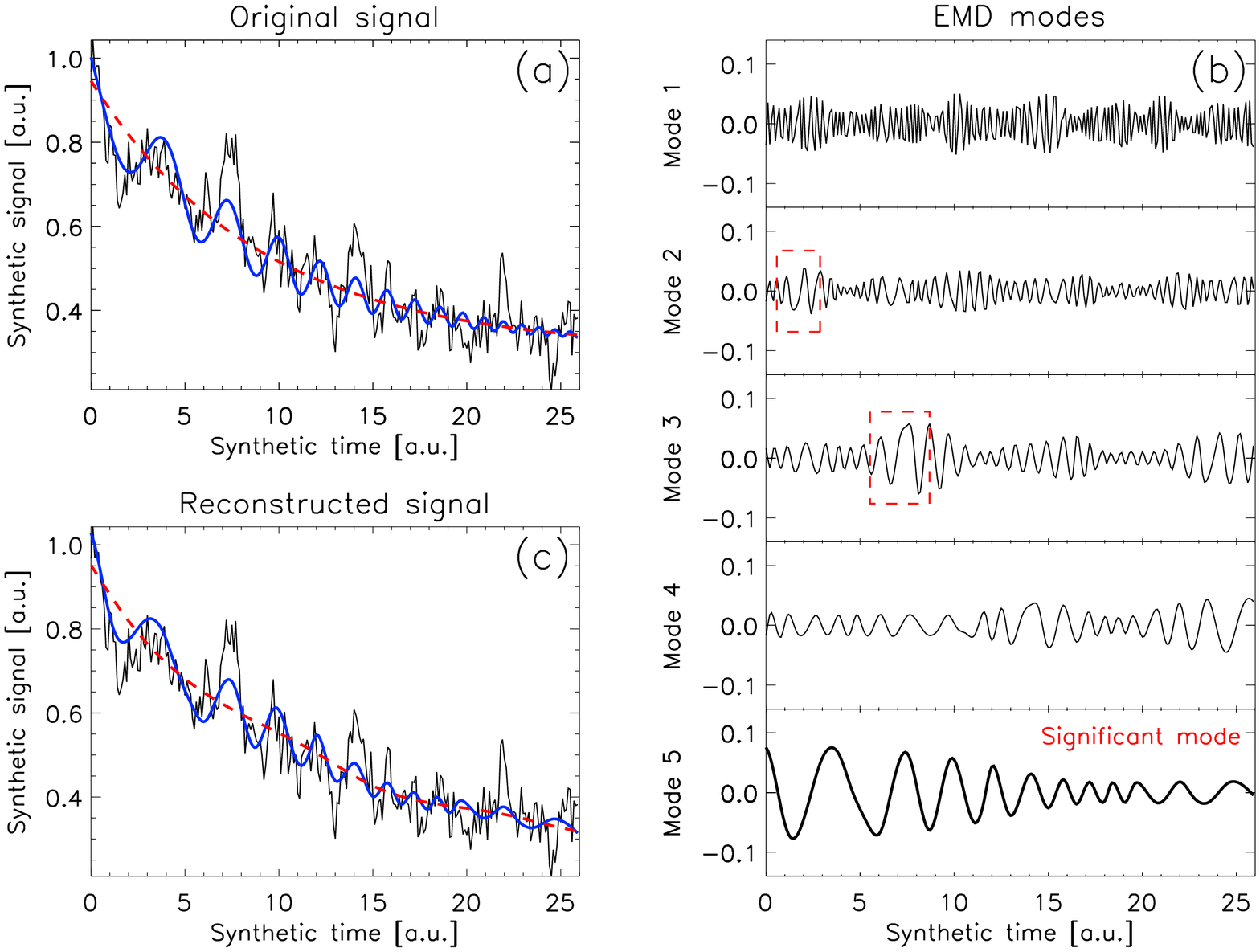}
	\includegraphics[width=0.93\linewidth]{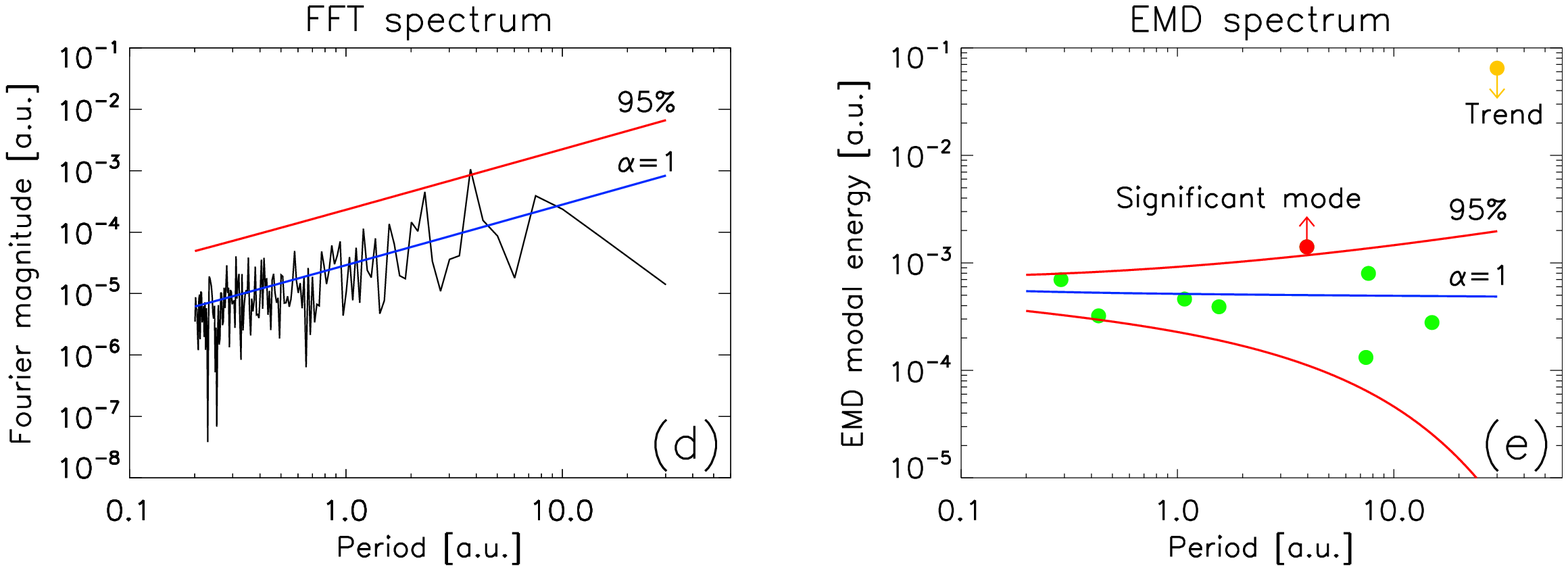}
	\caption{
		Application of the EMD and FFT methods to a synthetic signal, shown in black in panel (a) and consisting of the exponential trend (red dashed), a non-stationary oscillatory component with decaying amplitude and shortening period (blue), and coloured noise with the power-law index $\alpha=1$ (see Eq.~\ref{eq:emd_mean_spectrum}). Panel (b) shows first five EMD-revealed intrinsic mode functions of the original signal, including a statistically significant mode 5 which retains the non-stationary properties of the input oscillatory signal. The red dashed boxes illustrate the phenomenon of mode mixing in modes 2 and 3. Panel (c) shows the EMD-revealed significant mode 5 (blue) superimposed onto the EMD-revealed trend (red dashed) of the original signal. Panels (d) and (e) show the representation of the original signal with the EMD-revealed trend subtracted in the Fourier spectrum and in the EMD spectrum, with the latter defined by Eq.~(\ref{eq:emd_mean_spectrum}) as dependence of the EMD total modal energy on the mean modal period. The green, red, and yellow circles in the EMD spectrum correspond to the total energies and mean periods of each EMD-revealed mode, including one statistically significant mode and one mode associated with the aperiodic trend of the original signal. The blue solid lines in both panels (d) and (e) show best-fitting of the corresponding spectra by a power law function, see Eq.~(\ref{eq:emd_mean_spectrum}). The red solid lines in both panels show the 95\% confidence levels, above which the oscillatory components are considered as statistically significant \citep[see][for details]{2016A&A...592A.153K}.
	}
	\label{fig:emd_synth-example}
\end{figure}

Treating a signal of interest as an ensemble of various active time scales defined by the local extrema, EMD reduces the signal into a number of intrinsic mode functions (IMF)
which are smooth quasi-periodic functions with approximately zero mean and having slowly varying amplitude and period (see examples in Figure~\ref{fig:emd_synth-example}b). IMFs are  extracted from the original signal
through the so-called sifting process, iteratively searching for and extracting local time scales from the input time-series. More specifically, for each IMF the sifting procedure runs through:
\begin{itemize}
	\item Constructing the upper and lower envelopes of the input signal via connecting, for example by cubic spline interpolation, all the local maxima and minima;
	\item Obtaining the mean of these two envelopes and subtracting it from the input signal;
	\item Repeating the two previous steps until a condition determining the end of the sifting process for each individual intrinsic mode is met. This stopping criterion could be implemented via evaluation of the standard deviation between two consecutive sifting iterations, known as a shift factor, or by limiting the maximum number of sifting iterations by some reasonable value;
	\item Once this stopping condition is met, the resultant intrinsic mode is subtracted from the input signal and the whole procedure repeats for the next intrinsic mode, until a non-oscillatory residue is reached.
\end{itemize}
As a result of this sifting process, each detected IMF should approximately satisfy the following conditions:
\begin{itemize}
	\item The number of zero-crossings and extrema differ by no more than one;
	\item The local mean of the upper and lower envelopes is zero.
\end{itemize}
Thus, the basis of decomposition in the EMD analysis is derived locally from the data and is not prescribed \textit{a priori}. The local nature of EMD makes it adaptive and therefore entirely different and independent from other, more conventional Fourier transform based techniques. Namely, intrinsic modes are not necessarily harmonic, can be highly non-stationary, and their number is   small
($\sim\nobreak\log_2(N)$, where $N$ is the length of a time-series)
 in comparison with the number of harmonics revealed by the discrete Fourier transform.
For instance, EMD decomposes a time series shown in Figure~\ref{fig:emd_synth-example}a into eights IMFs (five of them are shown in Figure~\ref{fig:emd_synth-example}b) and a non-oscillitory residual or trend (dashed red line in Figure~\ref{fig:emd_synth-example}c).

 On the other hand, the completeness of the obtained empirical modes, i.e. recombining them to restore the initial signal, and their approximate orthogonality are discussed in detail in Section 6 of \cite{1998RSPSA.454..903H}. We also note here that in reality even two pure sinusoidal waves with different periods are not exactly orthogonal because of the finite data length. The detected empirical modes are typically characterised by increasing intrinsic time scales, with the last modes representing the longest-term variations of the input signal. Deduction of the shortest-period modes from the original signal is equivalent to its low-pass filtering or smoothing, while the aperiodic residue (or a combination of several longest-period modes) represent a slowly-varying trend, thus allowing one to use EMD for self-consistent detrending and smoothing. The analysis of instantaneous period and amplitude of each locally narrow-band intrinsic mode obtained with EMD can be performed, for example, with the Hilbert transform. A combination of EMD and the Hilbert transform for observational signal processing is known as the Hilbert--Huang transform approach.

An interesting experiment aiming at comparative analysis of the efficiency of various state-of-the-art methods, including EMD and the Fourier and wavelet transform based techniques, for detecting and analysis of quasi-periodic pulsations (QPPs) in solar and stellar flares was performed in \citet{2019ApJS..244...44B}. Due to their essentially irregular nature and relatively short lifetime (usually lasting for only a few oscillation cycles), the question of robust and reliable detection of QPPs remains open (see also Section \ref{sec:QPP}  of this review and  \citet{2021SSRv..217...66Z} for more details on the phenomenon of QPPs in solar and stellar flares). The best practice revealed by \citet{2019ApJS..244...44B} is that QPPs demonstrating different observational properties should be analysed with different methods. For example, EMD was found to perform very efficiently for detecting QPPs with non-stationary periods, which were not detected by the other methods based on the assumption of harmonic basis functions or their wavelets. A combination of several independent techniques can also help to increase the reliability of detection.
Figure~\ref{fig:emd_synth-example} demonstrates an example of applying the EMD analysis to a synthetic signal imitating a QPP in the decay phase of a solar flare and consisting of an monotonic trend and a non-stationary oscillation with varying period and amplitude, and contaminated by  coloured noise. EMD has successfully detected and recovered both the trend and oscillatory signal as the only significant modes in the decomposition.

Thus, due to its advantages in processing irregular and short-lived oscillatory signals, EMD has been extensively used for detection and analysis of solar and stellar oscillations on various time scales and of different physical natures. For example, the above-mentioned phenomenon of QPPs in solar flares with periods from a few seconds to tens of minutes was detected with EMD in e.g. \cite{2015A&A...574A..53K, 2019PPCF...61a4024N, 2020arXiv200802010K, 2020STP.....6a...3K}, including QPPs in the most powerful solar flare of cycle 24 \citep{2018ApJ...858L...3K}. Likewise, EMD was successfully employed for studying QPPs in stellar flares \citep[e.g.][]{2018MNRAS.475.2842D, 2019MNRAS.482.5553J} and for establishing the analogy between solar and stellar flares via QPPs \citep[e.g.][]{2016ApJ...830..110C}.
The use of EMD has also allowed for an advancement of our understanding of MHD waves and oscillations in solar coronal loops \citep{2004ApJ...614..435T} and at chromospheric and transition region heights \citep{2019FrASS...6...36N} with typical periods of a few minutes, and for identification of quasi-periodic oscillatory modes in long-lived solar facular regions with periods ranging from several minutes to a few hours \citep{2017A&A...598L...2K, 2018Ge&Ae..58..893S}.
Oscillatory variabilities in the longer-term solar proxies with periods from about a month up to the entire solar cycle were found with EMD in 10.7\,cm solar radio flux and sunspot records \citep{2007SoPh..243..193Z, 2018Ap&SS.363...84M}, coronal Fe XIV emission \citep{2015JASTP.122...18D}, flare activity index and occurrence rate of coronal mass ejections \citep{2019MNRAS.488..111D, 2012RAA....12..322G},  total and surface solar irradiance \citep{2012ApJ...747..135L, 2015JASTP.132...64L, 2018NPGeo..25...19B}, helioseismic frequency shift \citep{2015MNRAS.451.4360K},  spatio-temporal dynamics of the solar magnetic field \citep{2012ApJ...749...27V} and in the Sun-as-a-star observations of the solar mean magnetic field \citep{2016AJ....151...76X}, solar radius data \citep{2015RAA....15..879Q}, and also in direct numerical simulations of convection-driven dynamos \citep{2016A&A...589A..56K}.

Another important practical application of EMD draws on its ability to self-consistently detect aperiodic or low-frequency trends in observational time-series. For example,  \citet{2010PPCF...52l4009N} obtained solar flare trends with EMD which allowed for a simultaneous detection of highly anharmonic QPPs of a symmetric triangular shape in the microwave and hard X-ray flare fluxes. Likewise, \citet{2016PhRvL.117w5102H} used EMD to get rid of nonharmonic trends, associated with short large amplitude magnetic field structures (SLAMS) in the in-situ measurements of the magnetic field in the terrestrial quasi-parallel foreshock by the multi-spacecraft Cluster mission. This allowed for the first direct observational detection of nonlinear wave trains on scales smaller than those of SLAMS and numerical modelling of them in terms of the derivative nonlinear Schr\"{o}dinger equation. Thus, the application of EMD could provide an additional method for detrending in the areas of solar and heliospheric physics for which the presence of low-frequency background variations in observational time-series is crucially important \citep[see e.g. the problem of a long-term trend in the total solar irradiance measurments][]{2009A&A...501L..27F}.

Similarly to the conventional Fourier-based decomposition methods in which the presence of noise in the analysed signal leads to the appearance of spurious peaks in the power spectrum, whose significance can be assessed through the application of well elaborated and robust statistical techniques \citep[see Sec. \ref{sec:QPP} of this review and also][]{2005A&A...431..391V, 2017A&A...602A..47P}, not all the modes revealed by EMD necessarily represent statistically-significant oscillatory processes. In other words, significance of the EMD-revealed intrinsic modes must be tested in comparison with the background noise in a similar manner to the corresponding tests in the Fourier and wavelet methods. For example, properties of white noise in the EMD analysis were studied in \citet{2004RSPSA.460.1597W}. However, the solar and heliospheric observations are often contaminated by a combination of white and coloured noises \citep[e.g.][]{2015ApJ...798..108I, 2015ApJ...798....1I}, with white/coloured noise dominating at higher-/lower-frequency parts of the spectrum, respectively.
These noisy components could manifest different physical processes of natural or instrumental origin. For example, coloured, i.e. frequency-dependent, noise may be associated, for example, with spontaneous regimes of small-scale magnetic reconnection, for which the characteristic distribution of the released energy is found to obey a power-law dependence \citep[e.g.][]{2011ApJ...737...24B}.

This urgent need for incorporating statistics of coloured noise in the EMD analysis was recently addressed by \citet{2016A&A...592A.153K} through the development of a method for assessing the statistical significance of the intrinsic EMD modes in comparison with a power-law distributed background. Namely, the fact that EMD operates as an approximate dyadic filter bank \citep{2004ISPL...11..112F}, i.e. the frequency coverage in each intrinsic mode function decreases approximately by a factor of 2 with the mode number so that the higher-frequency modes occupy a broader range of frequencies, allowed for re-writing the distribution of the Fourier spectral power $S$ over frequencies $f$ for a power-law distributed noise, $S\propto 1/f^{\alpha}$ with typical $\alpha \gtrsim 0$, in terms of the total energy of each EMD mode $E_\mathrm{m} = \sum_i x_i^2$ and dominant modal period $P_\mathrm{m}$ as
\begin{equation}\label{eq:emd_mean_spectrum}
E_\mathrm{m}P^{1-\alpha}_\mathrm{m} = \mathrm{constant.}
\end{equation}
Equation (\ref{eq:emd_mean_spectrum}) can be referred  to as the mean EMD power spectrum of a power-law distributed noise. The particular forms of Eq.~(\ref{eq:emd_mean_spectrum}) for white, pink (flicker), and red noises can be obtained by substituting $\alpha=0$, $\alpha=1$, and $\alpha=2$ into it, respectively. A similar relationship was obtained by \cite{2009NPGeo..16...65F} in the application to oscillations in climatological data, but for the background noise modelled as an autoregressive process of the first order.

\begin{figure}
	\centering
	\includegraphics[width=0.8\linewidth]{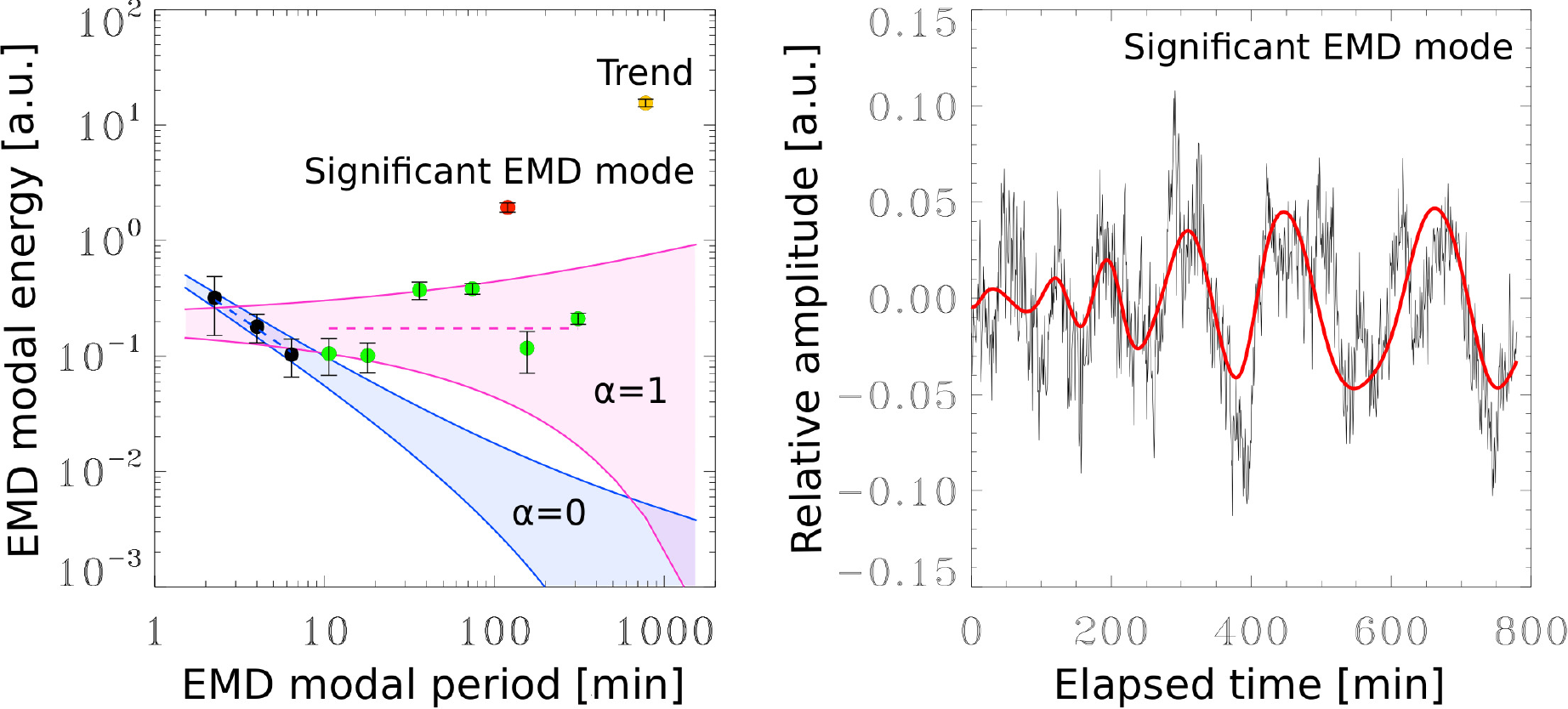}
	
	\caption{An example of the EMD analysis of oscillatory variations of the average magnetic field (the black line in the right-hand panel) in a solar facula \citep[adapted from][]{2017A&A...598L...2K}. The left-hand panel shows the EMD energy spectrum, that is dependence of the total energy of each intrinsic mode revealed by EMD (the black, green, and red circles) on the corresponding modal dominant period. The blue and pink shaded areas show the 99\% confidence intervals for while noise (with the spectral power-law index $\alpha=0$) and pink noise ($\alpha=1$), respectively, obtained with the method developed by \cite{2016A&A...592A.153K}. The dashed blue and pink lines show best-fits of the corresponding EMD modal energies by Eq.~(\ref{eq:emd_mean_spectrum}). The red circle indicates the EMD mode with statistically significant properties, that is shown in the right-hand panel in red. The yellow circle indicates a slowly varying trend that was subtracted from the original signal. 
	}
	\label{fig:emd}
\end{figure}

An additional fact about EMD is that the instantaneous amplitudes of intrinsic mode functions obtained from pure noise samples are normally distributed, that was demonstrated for the models of white \citep{2004RSPSA.460.1597W} and coloured \citep{2016A&A...592A.153K} noise. Combined with the definition of the total energy of each EMD mode as a sum of all instantaneous amplitudes squared, this allowed \cite{2016A&A...592A.153K} to show that this total EMD modal energy of coloured noise 
obeys a chi-squared distribution
with the mean value determined by Eq.~(\ref{eq:emd_mean_spectrum}). The number of degrees of freedom (DoF) in this distribution is greater than 2 providing the existence of the upper and lower confidence levels, and also the number of DoF was shown to depend on the modal oscillation period. In the method of \citet{2016A&A...592A.153K} this dependence was established heuristically, i.e. from numerical experiments. Thus, all the EMD modes with total energies situated outside this distribution possess properties statistically different to noise and hence can be considered as statistically significant. Revealing the properties of white and coloured noise in the EMD analysis and developing the noise testing scheme radically improves the ability of EMD to detect quasi-oscillatory signals in observations.
Figure~\ref{fig:emd_synth-example} demonstrates  detection a statistically significant non-stationary oscillation in a synthetic time-series by EMD and Fourier analysis.
An example of real data processing is shown in Figure~\ref{fig:emd} where out of ten oscillatory modes detected with EMD in the time-series of the average magnetic field in a solar facula only a single mode was found to posses statistically-significant properties, representing a non-stationary oscillation with the period increasing from about 80 min to 230 min.

Despite being developed just a few decades ago, EMD has already been unequivocally proven as an efficient and powerful method for revealing non-stationary and irregular quasi-periodic patterns in observational time-series of various natures. The test of statistical significance developed by \cite{2016A&A...592A.153K} enabled a more rigorous and thus meaningful use of EMD for solar, stellar, and magnetospheric observations. This opens up clear perspectives for exploiting the potential of EMD for processing the upcoming data from recently commissioned and future observational instruments, such as, for example, \emph{Parker Solar Probe} and \emph{Solar Orbiter}. On the other hand, the full-scale application of the method also requires understanding of the realm of its capability and pitfalls. Below, we summarise a few known shortcomings of EMD and provide suggestions for further potential improvements:
\begin{itemize}
	\item {Sometimes, for example due to poor time resolution or other peculiarities in the analysed signal or improper settings of the decomposition algorithm, EMD is unable to properly distinguish between local time scales, causing a so-called mode-mixing (also known as mode-leakage) problem. It manifests as either the appearance of widely disparate time scales in a single intrinsic mode or when a signal of a similar time scale resides in different intrinsic modes. An example of apparent mode-mixing can be seen 
	in Figure~\ref{fig:emd_synth-example}b where it is marked by red dashed rectangles.
	In particular, the mode-mixing problem can adversely affect the dyadic nature of EMD and thus corrupt the applicability of Eq.~(\ref{eq:emd_mean_spectrum}) and compromise the entire analysis. To suppress such inter-mode leakages, the noise-assisted ensemble empirical mode decomposition (EEMD) was proposed \citep[see][for details]{2008RvGeo..46.2006H}. Although EEMD was indeed was found to be able to cope with the mode-mixing problem, its application is time-consuming and would benefit from parallelising.}
	\item {The assessment of the statistical significance by the method developed by \citet{2016A&A...592A.153K} fails if there is an insufficient number of oscillation cycles in the EMD-revealed intrinsic mode \citep[see e.g.][]{2019ApJS..244...44B}. In this case, such low-frequency modes are better to be attributed to a slowly-varying trend of the original signal and excluded from the list of potential oscillations.}
	\item {The choice of the internal EMD parameter, a shift factor, determining the criterion for stopping the sifting process for each individual intrinsic mode and thus controlling the sensitivity of the decomposition is another currently unresolved issue of EMD.
	For example, if this shift factor is chosen too low, that corresponds to a large exponentially growing number of sifting iterations, then the decomposition reduces to the Fourier transform in the limit of infinite run time \citep{2010EJASP2010..145W}. In contrast, if the shift factor is too high then 
		the sifting process is stopped too early, and IMFs
		 appear to be undersifted, i.e. highly obscured by noise. Hence, an optimisation of the choice of the shift factor is currently needed.}	
\end{itemize}


\subsection{AFINO code}
\label{sec:AFINO}

Most studies that search for oscillatory phenomena in solar data make use of popular techniques such as wavelets or periodograms. The Automated Flare Inference of Oscillations (AFINO) method \citep{2015ApJ...798..108I, 2016ApJ...833..284I, 2018JGRA..123.6457M, 2020ApJ...895...50H} takes an alternative approach to this problem, instead searching for oscillations in time-series data by fitting and comparing various models to the Fourier power spectral density of the signal. AFINO is an open source tool written in Python and is publicly available.

The approach taken by AFINO is motivated by several factors. First, the realisation that many time-series in nature, and in particular in astrophysics, exhibit power laws in the Fourier power spectrum \citep{2006Natur.444..730M, 2010AJ....140..224C, 2011A&A...533A..61G, 2013ApJ...768...87H}, a feature that must be carefully accounted for when using techniques such as a periodogram or wavelet [e.g. Vaughan 2005, 2010]. Secondly, while it is common to prepare signals for analysis by performing running mean subtractions or pre-filtering of the initial time-series, such practices require arbitrary choices that can lead to inaccurate results and also cause a reproducibility problem (see also Section. \ref{sec:time-series_pitfalls} for a more detailed discussion of these issues). Finally, the AFINO algorithm was designed to enable quick analysis of a large number of time-series to enable statistical studies, something difficult to achieve with methods that require manual fine-tuning. Given these motivations, the AFINO method instead analyses the full Fourier spectrum of an input signal, avoiding manual intervention and providing fast, reproducible results. 

\begin{figure}
\begin{center}
\includegraphics[width=1.\linewidth]{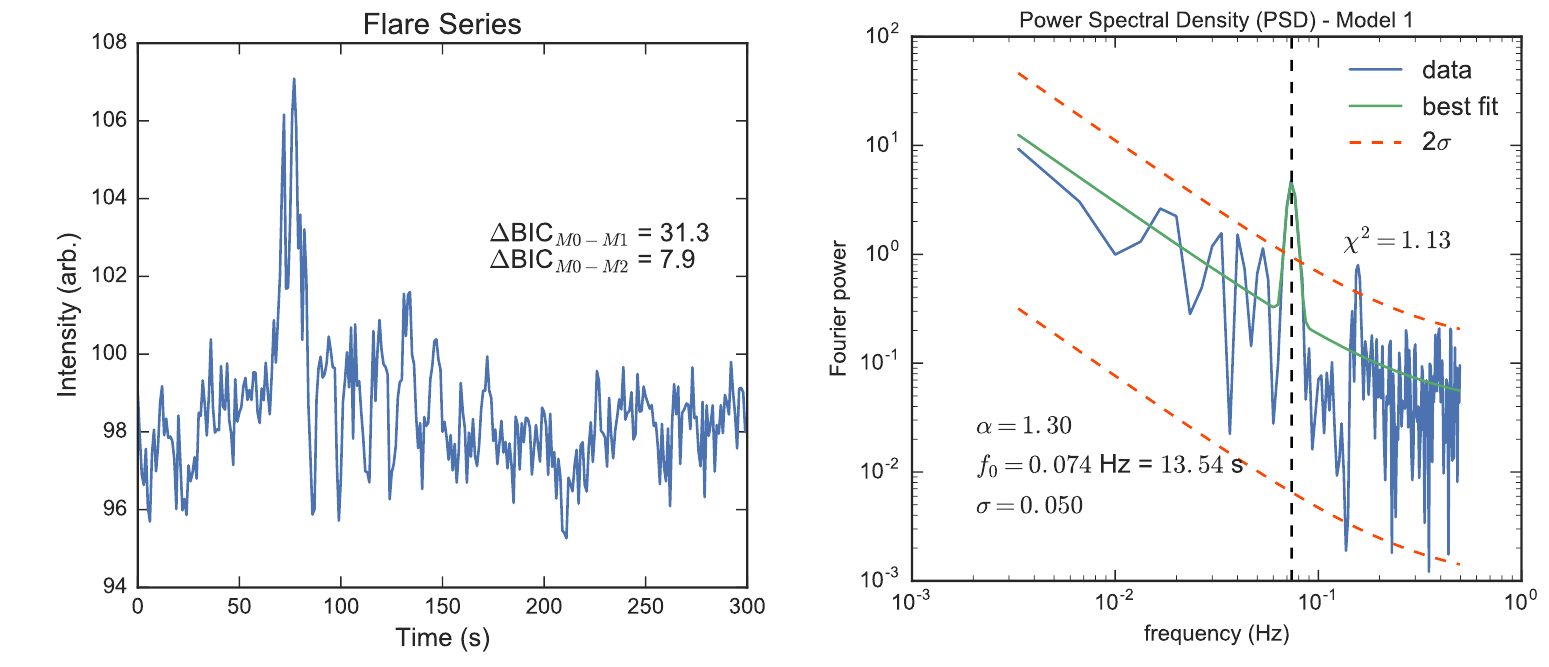}
\caption{The AFINO technique applied to a synthetic flare signal with a known oscillation period of ~13s. The flare input time series is shown in the left panel. The right panel shows the Fourier Power Spectral Density (PSD) in blue, with the best-fit model superimposed in green. Using the Bayesian Information Criterion (BIC) comparison, AFINO finds that the model 'M1' that contains a Gaussian enhancement is strongly preferred over the other candidate models, a single power law (model M1) and a broken power law (model M2). The oscillation period is correctly identified with a frequency of 0.074 Hz, corresponding to a period of 13.5s. Adapted from \citet{2019ApJS..244...44B}, Figure 5.  }\label{fig:afino}
\end{center}
\end{figure}

The AFINO methodology is described in detail in \citet{2015ApJ...798..108I, 2016ApJ...833..284I}. First, an evenly-sampled input time series is acquired, normalised and multiplied by a window function. The choice of window function is not critical, and a number of popular functions may be used (for example the Hanning window, or the Blackman-Harris window). Second, AFINO fits a set of candidate models to the Fourier power spectrum of the modified signal using the maximum likelihood method, using tools available in SciPy. In the \citet{2015ApJ...798..108I, 2016ApJ...833..284I} formulation, one of these models includes a Gaussian enhancement in frequency space, representing a quasi-stationary oscillation, while other ‘null hypothesis’ models include variations on a power-law in frequency space.  Third, these models are compared using the Bayesian Information Criterion (BIC) \citep{1978AnSta...6..461S}. The BIC is defined as $BIC$ = $-2 \ln (L) + k \ln(n)$, where $L$ is the maximum likelihood, $n$ is the number of spectral data points and $k$ is the number of free model parameters. Thus, BIC can be calculated directly from the maximum likelihood. The BIC includes a penalty term associated with adding extra free parameters, the idea being that a simpler model will be preferred unless there is substantial evidence in favour of a more complex alternative. The last major step is to perform an additional check to test whether any of the applied models were acceptable using a $\chi^2$-like parameter derived by \citet{2014ApJ...789..152N}. This is done since the BIC comparison itself does not make an explicit determination of whether any of the chosen models are an acceptable fit. Once this is complete, the evidence for an oscillatory signature in the data can be assessed based on the BIC values.
An illustrative example of the analysis of a synthetic signal mimicking a QPP in a stellar flare is presented in Figure~\ref{fig:afino}. The AFINO code has successfully detected a predefined oscillatory component and correctly identified its period. The presence of the oscillations were  assessed based on the BIC values computed for two competing models: an oscillatory signal superimposed with a power law noise and a power law noise alone.

One caveat of this approach is that the choice of models is empirical. In particular, a power law with a Gaussian enhancement is used to represent a discrete oscillation signal, while in reality this may be an incomplete or inaccurate model of the signal. Another limitation of this method is that many oscillatory signals on the Sun (and in other phenomena) evolve significantly over time. Such signals are likely to be overlooked by this approach that searches for localised frequency enhancements in the Fourier power spectrum of the entire signal. The performance of many QPP detection methods, including AFINO, was recently assessed by \citet{2019ApJS..244...44B} using test data sets, thus revealing the positives and negatives of this technique. On the positive side, AFINO was found to produce few false positive results, i.e. detections claimed by the method were genuinely present in the data. However, the downside was that this approach produced more false negatives than other methods, meaning a number of real signals were missed.

\subsection{Stingray code}
\label{sec:Stingarray}

Another recently released Python package that provides time-series analysis tools is Stingray \citep{2019ApJ...881...39H}. Stingray is a general-purpose tool that provides a range of Fourier analysis techniques used in astrophysical spectral-timing analysis. The package was written in Python, and is fully open source, version controlled and publicly available. The package has already been used in a variety of recent studies of astrophysical phenomena \citep[e.g.][]{2018ApJ...861L...7B, 2018MNRAS.477.1120K, 2019ApJ...875..144P, 2018ApJ...853L..21B}.

Stingray was developed as a means to unify the distribution and use of time-series analysis tools in astronomy, which historically has been fractured. It aims to provide the core tools needed to analyse time-series data in a wide variety of contexts. Stingray provides a framework for data manipulation and analysis in the form of a \verb|LightCurve| class, which can be constructed from arbitrary time-series data. Via this object, many basic tasks can be easily performed, including addition and subtraction, time-shifting, re-binning, truncation and concatenation of time-series, as well as functionality to save data in certain formats. Stingray also provides an \verb|Events| class, which is used for single-event measurements such as individual photon counts. 

From these objects, Stingray can generate cross spectra and power spectra for the desired time-series, and calculate confidence limits. Another tool generated by Stingray of particular interest in the study of solar pulsations and oscillations is the dynamic Fourier power spectrum. This spectrum can be used to study the variation in Fourier power of a signal over time, similar to a wavelet spectrum. In the dynamic power spectrum, a time-series is divided into a number of segments, and the Fourier power spectrum calculated for each one. These individual power spectra are then plotted sequentially in vertical slices to visualise the change in power over time, a common need when studying non-stationary signals. 

\begin{figure}

\begin{center}

\includegraphics[width=1.\linewidth]{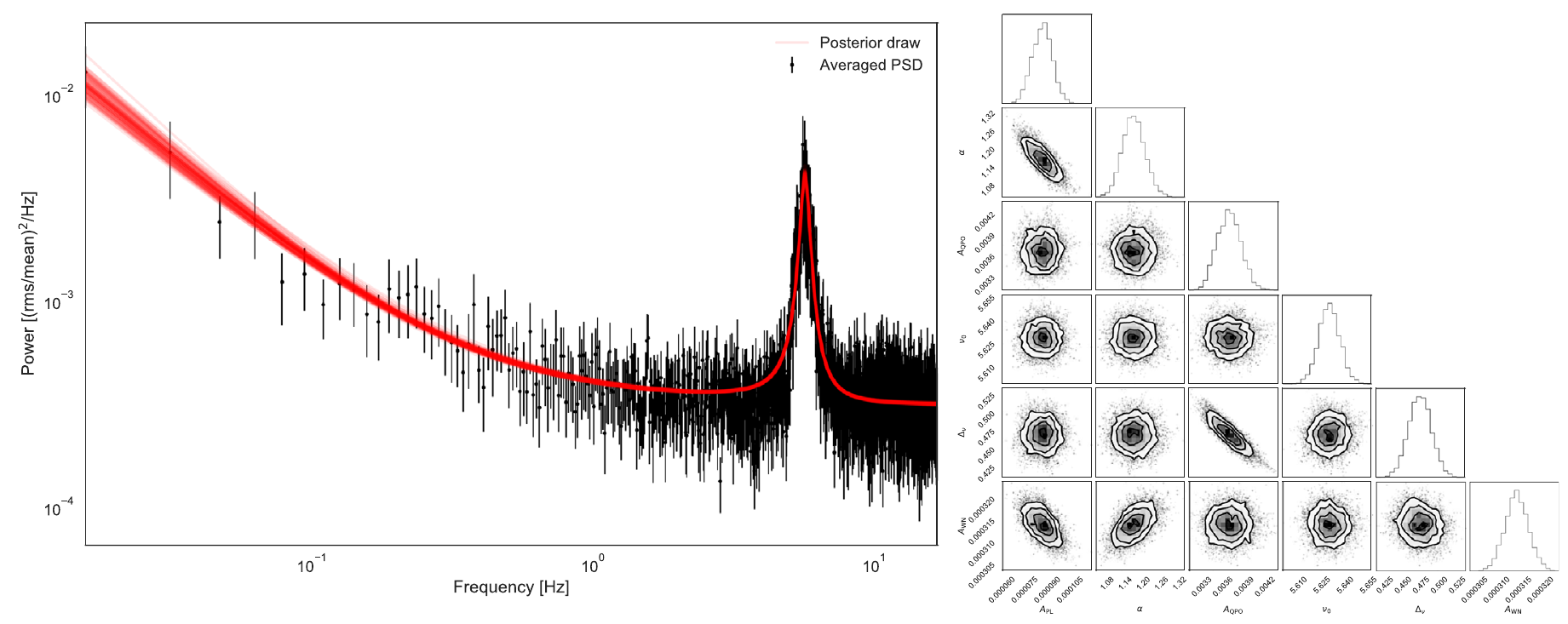}

\caption{An example of Stingray modelling tools being used to fit the Fourier power spectrum of X-ray binary GX 339-4 and observe a QPO signature. Left: The black data shows the averaged power spectrum from GX 339-4 from 15 different time segments. The red line shows the best fit model, consisting of a power law plus a Lorentzian and a constant. The right panel shows the marginal posterior distributions of the six model parameters. Reproduced from \citet{2019ApJ...881...39H}, Figure 4.}\label{fig:stingray}

\end{center}

\end{figure}

In addition to these core tools, Stingray includes additional features that allow for the modelling of time-series in a variety of ways. It can be used to fitting models to the Fourier power spectra of signals, a useful feature in the search for oscillations. A variety of comparison metrics are implemented, including the likelihood ratio test (LRT), the Akaike Information Criterion (AIC)\citep{1974ITAC...19..716A}, and the Bayesian Information Criterion (BIC)\citep{1978AnSta...6..461S}. For this model fitting, Stingray implements a full Bayesian Markov Chain Monte Carlo (MCMC) framework, allowing the full posterior probability distributions to be estimated (see Figure \ref{fig:stingray}). 

Stingray also includes a simulation package, which enables the creation of simulated time-series from an input power spectrum. This feature is particularly useful for validation and testing of analysis methods, such as that performed by \citet{2019ApJS..244...44B}. For example, using Stingray a user can quickly generate a sample time-series from a white-noise background spectrum and a red-noise background spectrum for comparison, and test the performance of their chosen analysis technique in both scenarios.


\subsection{Pitfalls of time-series pre-processing}
\label{sec:time-series_pitfalls}

In order to reduce noise, remove a trend, or stabilise the signal, researchers often perform  preliminary processing of a time-series before applying actual spectral analysis.
Such preprocessing may include the following steps:
\begin{itemize}
	\item Trend removal;
	\item Noise suppression;
	\item Various kinds of normalisation.
\end{itemize}

All these operations may  introduce new time scales to the time-series and, thus, may cause false positive detections of oscillatory components.
For instance, a common detrending procedure implies subtracting from the signal its version smoothed by a boxcar filter.
Here, the width of the boxcar filter defines a new time-scale which can appear in the result of the subsequent spectral analysis as a false periodicity.

To illustrate these caveats, we present a practical exercise demonstrating how smoothing and normalising the observational time-series modifies its Fourier power spectrum and can lead to the erroneous detection of oscillations in a pure noisy input signal.

\begin{figure}
	\centering
	\includegraphics[width=\linewidth]{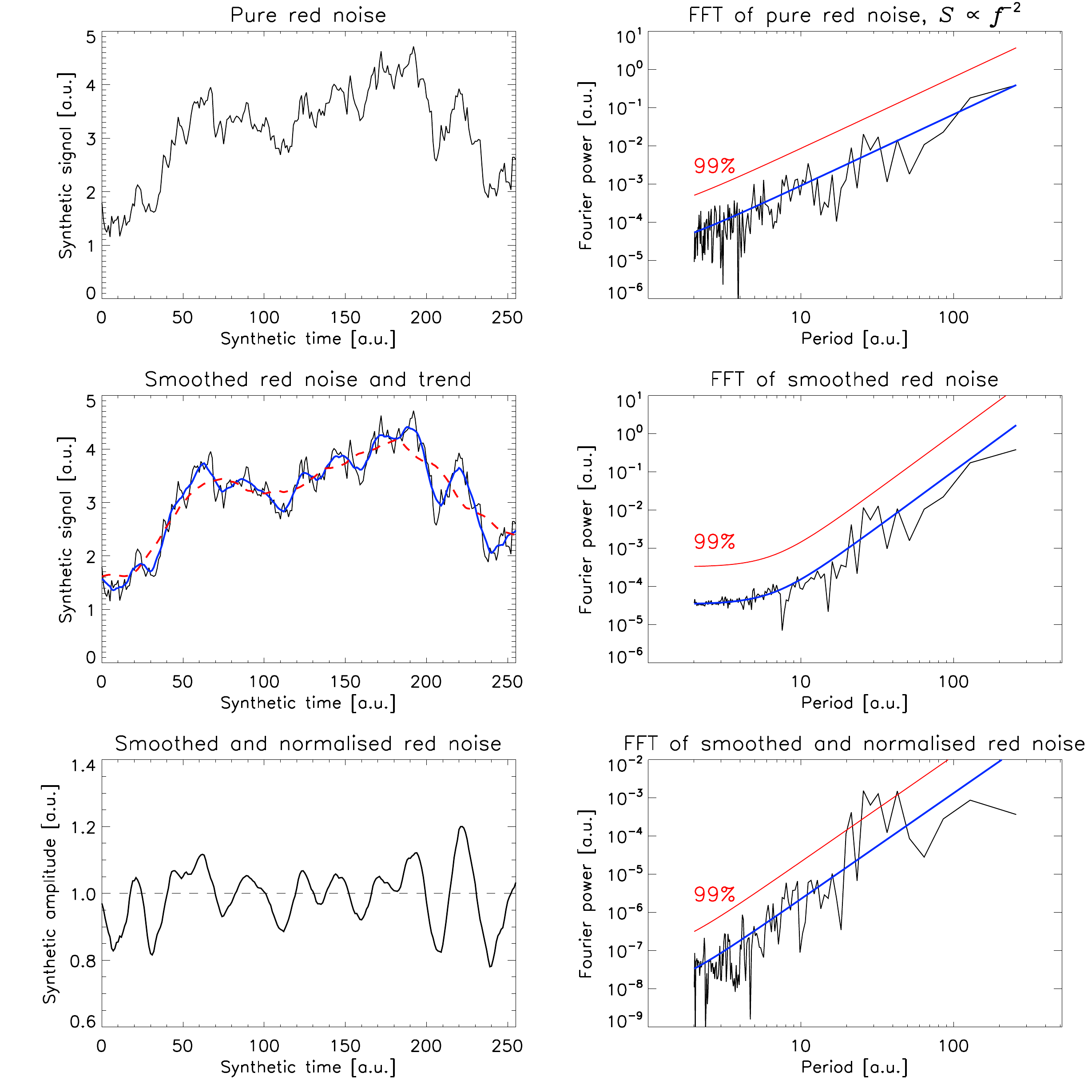}
	\caption{Top and middle rows: a synthetic red noise sample represented in the time and Fourier domains. The solid blue (dashed red) line in the time domain indicates the original signal smoothed over 10 (40) time units. The solid blue and red lines in the Fourier domain show the best-fit of the Fourier spectrum by a power-law model and the 99\% confidence level, respectively. The bottom row shows the red noise sample smoothed and normalised to its trend, also in the time and Fourier domains.
	}
	\label{fig:qpp_from_noise}
\end{figure}

Figure~\ref{fig:qpp_from_noise} demonstrates a sample of synthetic red noise with the Fourier spectral power $S$ distributed over frequencies $f$ as $S\propto f^{-2}$ and its Fourier power spectrum with the corresponding 99\% confidence interval estimated following the procedure analogous to that described in Section~\ref{sec:QPP} and references therein. The total length of the signal is 256 (in some arbitrary time units) and the amplitude is also set arbitrarily. As expected, neither of the Fourier peaks obtained from the pure red noise sample is seen to be statistically significant.

We now smooth the original signal over 10 time units using a basic boxcar smoothing technique. This is effectively equivalent to suppressing higher-frequency fluctuations in the original signal, which is evident from the Fourier power spectrum of a smoothed signal (see Fig.~\ref{fig:qpp_from_noise}). Indeed, the Fourier spectrum of a smoothed red noise changes its shape in the shorter-period (higher-frequency) part where the oscillation energy is filtered out by smoothing. As previously seen, all the Fourier peaks are detected well below the noise confidence level. We also apply the same smoothing method for obtaining a slowly-varying trend of the original signal. In this case, we smooth it over 40 time units.

As the next step, we normalise the smoothed red noise sample to its slowly varying trend. This causes the resulting signal to oscillate with respect to the unity level at a relatively stable period of about 30 time units and with no signatures of decay. In the Fourier domain, after the application of this data processing, the periods shorter than 10 time units and those longer than 40 time units are severely suppressed, which leads to the artificial enhancement of the group of Fourier periods at around 30 time units, well above the 99\% level of statistical significance. Thus, the resulting spectrum demonstrates disparately different behaviour in comparison with that of the original unprocessed pure red noise, and hence it cannot be used for detecting any potential oscillations.

We stress that the appearance of such a nice and apparently significant quasi-periodic pattern in a pure synthetic power-law distributed noise sample should not be associated with any oscillatory physical process or characteristic time scale, and is caused by a combination of the two data processing tricks which are smoothing and subsequent normalisation to the trend. As a power-law distributed noise is a common feature of solar and heliospheric observations (Sections \ref{sec:EMD}, \ref{sec:AFINO} and references therein), and, on the other hand, smoothing and various manipulations with the trend are rather common data processing approaches, through this quick example we illustrate that: 

\begin{itemize}
	\item It is fine to use smoothing procedures to get rid of high-frequency noise in the analysed observational signal.
	It may also be safe to use smoothing for obtaining its long-term trends if a sufficient diligence is applied to the choice of the smoothing window;
	\item However, normalisation of such a smoothed signal to its trend introduces an element of nonlinearity, i.e. interaction of the signal with itself;
	\item As the Fourier analysis strictly requires the oscillating system to be linear, the resulting signal no longer possesses the spectral properties of the original signal, which makes claiming significance of oscillations in the normalised signal meaningless;
	\item On the other hand, such a normalisation to the trend may be safely used for the physical interpretation of quasi-periodic patterns whose statistical significance was assessed independently.
\end{itemize}


\subsection{Detection of quasi-periodic pulsations in flaring light-curves}
\label{sec:QPP}

Quasi-Periodic Pulsations (QPPs) manifest as a pronounced oscillatory pattern within the electromagnetic radiation detected in solar and stellar flares. The oscillatory signals are detected over a range of periods and with period modulation, coupled with apparent amplitude modulation. The true physical mechanism(s) that underpin QPPs in flares is currently undergoing intensive theoretical study \citep[e.g., see reviews by][]{2018SSRv..214...45M,2020STP.....6a...3K,2021SSRv..217...66Z}. These specific properties lead to challenges in the detection and analysis of QPPs. Firstly, the low quality of QPPs should be mentioned, where the quality, or Q-factor, is defined as the number of full periods. Studying QPP signals observed in the same solar active region, it was found that typically QPPs have from two to ten full periods only \citep{2019PPCF...61a4024N}. The second problem relates to the parameters of the QPP (period, amplitude) which can vary significantly in time. This non-stationarity of the parameters could be caused both by multi-modality of the QPP or by the temporal evolution of the plasma parameters during the flare \citep{2019PPCF...61a4024N, 2020STP.....6a...3K}. Given these challenges, different methods are used for the detection and analysis of QPPs in flare emission. Among them, there are standard methods utilising the periodic basic functions (FFT and CWT) as well as fully empirical methods such as EMD described in Section~\ref{sec:EMD} (see  \cite{2019ApJS..244...44B} for a comprehensive review). Each method has its own advantages and disadvantages, and the question arises as to what specific technique is most appropriate for detecting  a quasi-periodic signal with a certain combination of observational properties. 

As an example, let us consider the FFT method applied to the full flare light curve, or a time-series containing a flare, with a low-frequency trend, a QPP and different kinds of noise. Here, the Fourier spectrum takes the form of a power-law function. In the spectrum, the lowest-frequency component (trend) has the maximum spectral power while the highest-frequency white noise has the minimum spectral power. In addition, the noise component is distributed according to power-law  $S \propto f^{-\alpha}$. In particular, a power-law index $\alpha = 0$ corresponds to the uncorrelated (or white) noise, $\alpha = 1$ denotes correlated flicker (or pink) noise, and $\alpha = 2$ means correlated red noise. Often, a combination of the different  $\alpha$ values at the lower and higher Fourier frequencies can form a broken power-law spectrum \citep{2015ApJ...798..108I, 2017A&A...608A.101P}. The colour of the noise in the full time series is defined empirically by fitting its Fourier spectrum and estimating its power-law index,  $\alpha$. The spectral peak in the Fourier spectrum is treated to be significant if its power is higher than the noise level (or significance level) which is defined performing the $\chi^2$ test \citep{2005A&A...431..391V, 2017A&A...602A..47P}. 

The FFT method of the full time-series is appropriate for analysing a high-quality QPP with a stable period or with slightly varying periods, and in these cases, there is a high probability of the  QPP being detected. However, if the period of the QPP is varying significantly with time, the FFT method can give spurious results. For example, the spectral power of a QPP with a significantly-growing period will be distributed over an interval between the start and end values of the period, instead of the formation of a localised peak in the periodogram. Therefore, the significance of such a QPP will be underestimated, and these QPPs will be mistakenly sifted out.

In contrast, if we consider a short time-series containing a low-quality QPP, the number of counts in the Fourier spectrum will be  small. This can lead to an incorrect estimation of the $\alpha$ value and, consequently, to overestimation or underestimation of the QPP power. For example, making the assumption of a white noise background spectrum for analysis of a time-series that actually contains red noise can lead to a false detection, i.e. the lower frequencies appear to be significant. Therefore, estimation of the correct statistical significance of the QPP is an obligatory task.  

However, the presence of a low-frequency trend adds uncertainty to this task. The shape of the simplest flare consists of a fast rise and the consequent gradual decay.  An empirical template of the simplest white light flare on an M-type star fits the decay flare phase with a broken exponential function \citep{2014ApJ...797..122D}. An analytical template of the solar flare in  soft X-rays represents the trend as a combination of a Gaussian function at the rise phase and exponential relaxation at the decay phase  \citep{2017SoPh..292...77G}. A curious result is that the Fourier spectrum of the flare trend obtained by \citet{2017SoPh..292...77G} has a similar slope as the Fourier spectrum of red noise \citep{2019PPCF...61a4024N}. Thus, it is difficult to determine whether the slope of the Fourier spectrum is caused by the flare trend or by background red noise. It is possible that the detrending procedure (the subtraction of the trend from the full time-series) could help to resolve this uncertainty.

Anyone who studies QPPs meets a conceptual problem: to detrend or not to detrend. On one hand, when analysing the full time-series (including trend) with the FFT method, a number of real QPPs occur to be below the significance level. The presence of a trend affects the slope of the Fourier spectrum, resulting in an incorrect estimation of the noise parameters and, as a consequence, in the incorrect estimation of the significance level. This problem can be resolved by analysing not the full time-series, but its high-frequency component only. The high-frequency component can be obtained with direct methods of Fourier filtration \citep{2009A&A...493..259I} or wavelet filtration \citep{2012ApJ...749L..16D} or with an indirect method, i.e. the detrending approach. There are several methods for defining the trend. In the ideal case, the analytical form of the trend is known (e.g. the previously mentioned exponential decay phase). In this case, parameters of the trend could be estimated, for example, with the least-squares technique. However, most flares have a more complicated shape which is difficult to fit with such an analytical function. Therefore, the trend could be defined with a smoothing procedure, such as smoothing with the running average or convoluting the time-series with a polynomial function (the IDL routine SAVGOL.pro).
The potential of the EMD method for obtaining the flare trends is discussed in Sec.~\ref{sec:EMD}.  

On the other hand, there is a danger that the trend could be determined incorrectly, e.g. if the smoothing was performed over an inappropriately large window. This could lead to the appearance of a false peak (or peaks) in the Fourier spectrum (see also Section \ref{sec:time-series_pitfalls}). For this reason, some studies do not apply a detrending procedure \citep[e.g.,][]{2017A&A...602A..47P, 2018SoPh..293...61D}.

Given these common pitfalls in QPP detection, including detrending, trimming, coloured noise, possible stationary or non-stationary periodicities, one must be careful to deploy appropriate techniques to reliably identify QPPs and minimise false detections. \cite{2019ApJS..244...44B} reviewed multiple techniques for detecting QPPs, including  Gaussian Process Regression \citep{foreman-mackey2017}, wavelet analysis, AFINO/Automated Flare Inference of Oscillations (see Section~\ref{sec:AFINO}), Smoothing and Periodogram (with a trimmed versus untrimmed signal), 
EMD/Empirical Mode Decomposition (see Section~\ref{sec:EMD}), forward modelling of QPP signals with the use of MCMC and Bayesian analysis, and Periodogram-based Significance Testing. Several recommendations were articulated to help avoid these key pitfalls, which include using simulations to test the robustness of the detection method, taking red noise into account during detections, being cautious with detrending, including only trimming the data around the QPPs if this benefits detection, using EMD for non-stationary signals, and then using the wavelet analysis on the detrended and EMD-decomposed signal.

\section{Methods for detecting and analysing periodic transverse displacements of coronal structures}
\label{sec:transverse_motions}

\subsection{Enhancing transverse motions with the DT$\mathbb{C}$WT-based motion magnification}
\label{sec:motion_magnification}

It was discovered \citep[e.g. ][]{2015A&A...583A.136A} that the  coronal loops and other structures seen in SDO/AIA   EUV images support omnipresent low amplitude transverse periodic motions.
In most cases, these motions have an amplitude less than the pixel size. Therefore, they are difficult to recognize by eye and require  advanced data analysis methods for processing.
A possible solution for the analysis of low amplitude motions in imaging data is a technique called \textit{motion magnification}.
Motion magnification  aims to magnify  low amplitude transverse motions to make them visible by eye and available for traditional data processing methods  such as investigating oscillatory patterns in time-distance plots.

The first implementations of  motion magnification were based on explicit estimation of the velocity field and subsequent wrapping of individual images \citep{Motion_mag_Liu_2005}.
However, such approaches were found to be computationally heavy and able to produce artefacts.
The further development of the motion magnification algorithms led to the  Eulerian video magnification method developed by \citet{emoRecognition:Wu_2012}.
Their approach eliminates explicit computation of the velocity field. Instead, the algorithm uses a temporal broadband band-pass filter to extract and amplify intensity variations associated with small transverse motions. The main limitations of this approach are small magnification factors, and image distortions due to amplification of the brightness noise and brightness variations not related to transverse motions.

The current generation of motion magnification algorithms is so-called phase-based motion magnification. It relies on the decomposition of the images into complex 2D wavelets or wavelet-like pyramids. The elements of such a decomposition are localised in space and wave-number domains. Thus, every element of such a decomposition is characterised by its location, spatial scale, orientation, and complex amplitude. The latter encodes the brightness and position of individual structures present in the original images.
The brightness  is related to the absolute value of complex amplitudes, while any transverse motions cause variations of the phase. In the case of small displacements these two dependencies are linear and decoupled, allowing for measuring and magnifying small transverse motions in time-resolved imaging data.
The first implementation of phase-based motion magnification was proposed by \citet{Wadhwa:2013:PVM:2461912.2461966}. Their algorithm decomposes a sequence of images into Complex Steerable Pyramids \citep[see][for the detailed description]{citeulike:3723307,citeulike:3808781}, which are  a complex wavelet-like 2D spatial decomposition. 

\citet{2016SoPh..291.3251A} developed an original algorithm of the phase-based motion magnification tuned for analysis of EUV images of the solar corona.
Their algorithm uses the two-dimensional dual tree complex wavelet transform (DT$\mathbb{C}$WT) developed by \citet{2005ISPM...22..123S}. The distinct features of DT$\mathbb{C}$WT are perfect reconstruction, good shift invariance, and computational efficiency.
Moreover  DT$\mathbb{C}$WT is implemented in several programming languages and  is published as an open source software \citep[e.g.][]{rich_wareham_2014_9862}.
Since DT$\mathbb{C}$WT is a 2D wavelet decomposition, its individual elements are localised both in space and in wave number domains allowing for independent processing of structures in images with different locations (several loops in an active region), spatial scales (individual thin strand in a thick multi-stranded loop) and orientations (overlapped loops with different orientations).

Like 1D wavelets, the 2D wavelet transform has a trade-off between spatial and wave-number resolutions. The limiting cases of such a trade-off are spatial Fourier transform (perfect wave-number resolution, no spatial resolution) and the image itself (perfect spatial resolution, no wave-number resolution).
The lower spatial resolution of individual wavelet elements allows for higher magnification coefficients, but sacrifices the ability to resolve different motions in neighbouring structures. The motion magnification code developed by \citet{2016SoPh..291.3251A} follows the opposite approach. The algorithm uses basic DT$\mathbb{C}$WT elements with higher spatial and lower wave-number resolution to allow accurate resolution of motion in neighbouring structures, which is more important than the high magnification coefficient while analysing complex images of solar corona with an active background and overlapping coronal loops.

\begin{figure*}
	\centering
	\includegraphics[width=0.75\textwidth]{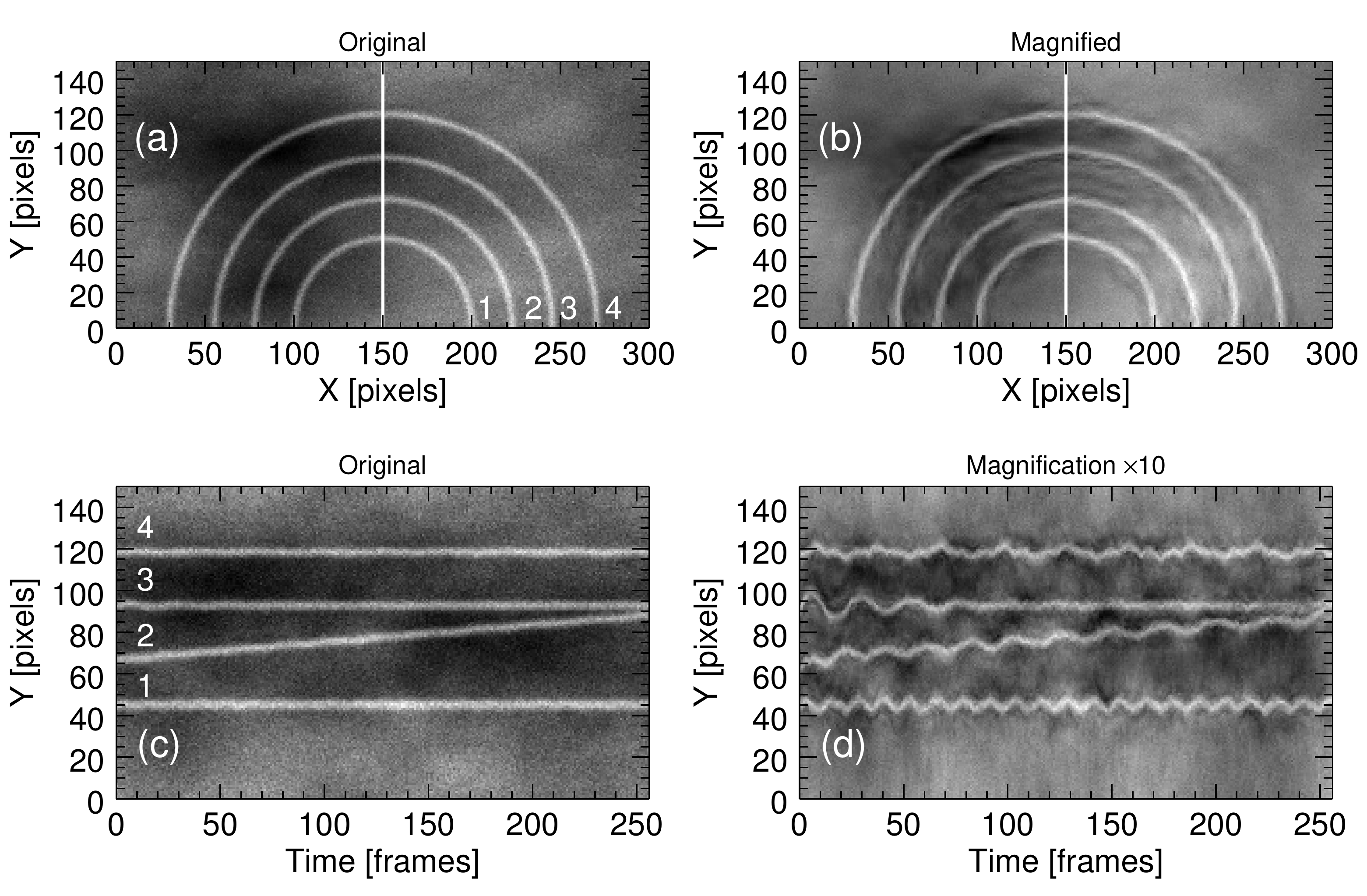} \caption{Detection of synthetic signals by the method of DT$\mathbb{C}$WT-based motion magnification. Panel (a) shows a frame with four loops superimposed with a time varying spatially non-uniform background. Loop 1 oscillates harmonically with a constant period of 30 frames and amplitude 0.2~px. Loop 2 oscillates with constant amplitude of 0.2~px; its major radius increases with time, and the oscillation period increases with the increase in the loop length. Loop 3 performs a decaying oscillation with the initial amplitude of 1~px, and the damping time of 40 frames. Loop 4 performs a two-harmonic oscillations, with the period ratio of 3, the amplitude of the longer period oscillation is 0.2~px, and of the shorter period is 0.1~px. The white vertical line shows the location of the slit for time-distance maps.
		Panel (b) shows the same frame as (a) but after the application of the DT$\mathbb{C}$WT-based motion magnification. Panel (c) gives the time-distance plot made along the slit shown in panel (a) for the original synthetic data. Panel (d) shows the time-distance plot after the application of the DT$\mathbb{C}$WT-based motion magnification. The figure is taken from \citet{2016SoPh..291.3251A}.
	}
	\label{fig:motion_mag_model}
\end{figure*}

\begin{figure*}
	\includegraphics[width=0.39\textwidth]{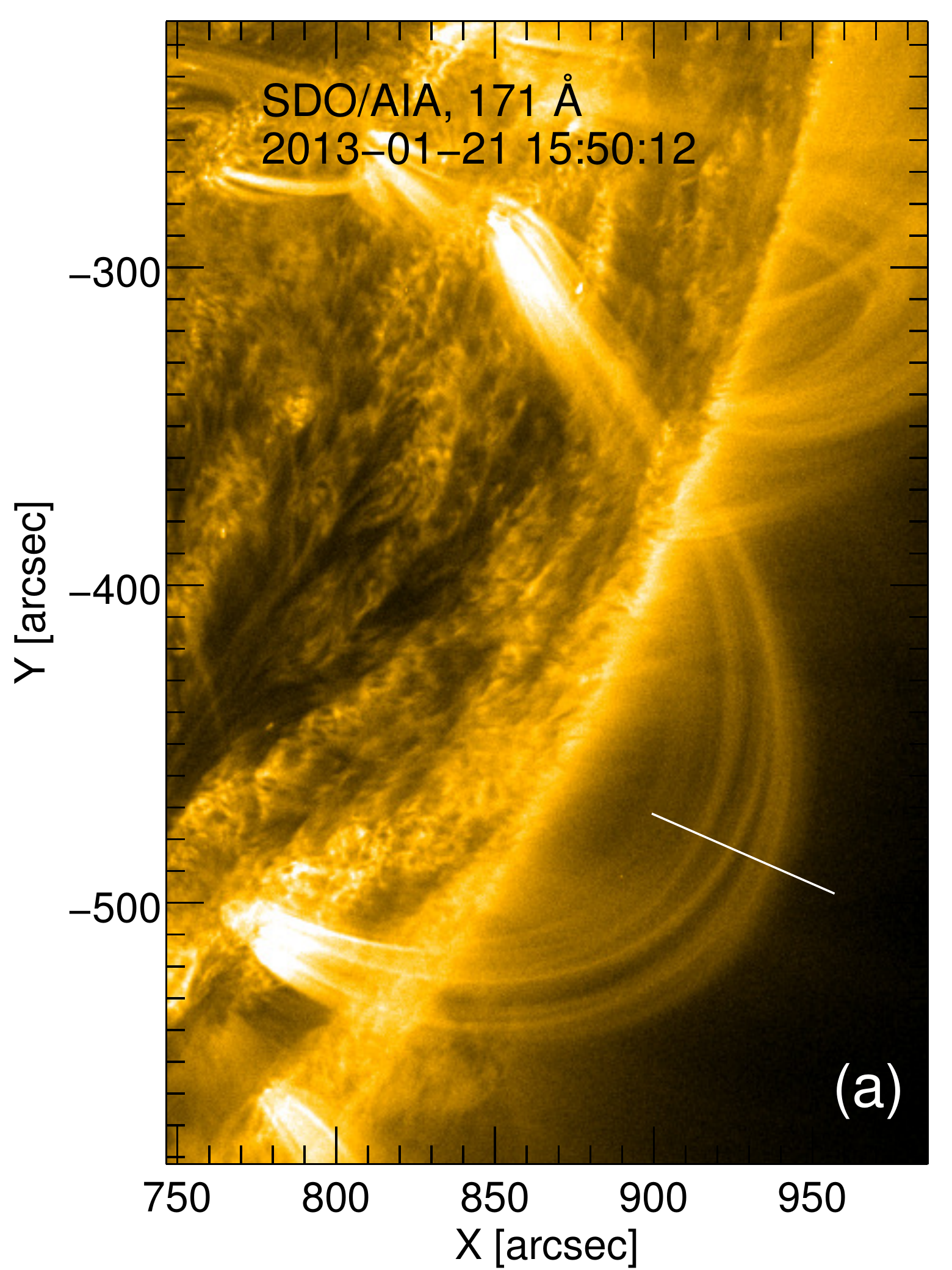}\includegraphics[width=0.61\textwidth]{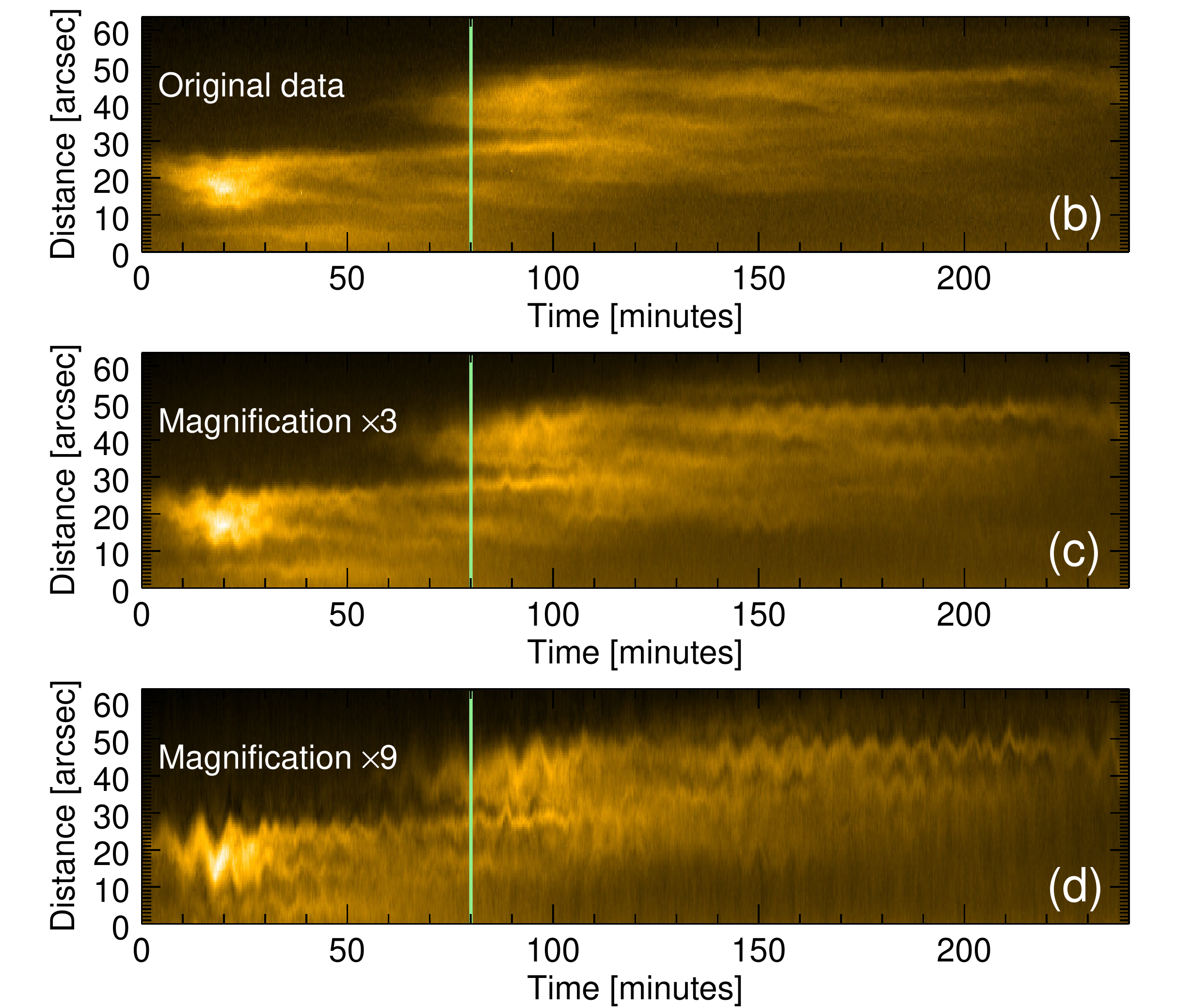} \caption{EUV image (a) of the coronal loop system observed in 171~\AA~ SDO/AIA channel on 21 January 2013. Right panels show time-distance plots made with the use of the original data (b) and processed with the motion magnification technique with the magnification factors $k = 3$ (c) and $k = 9$ (d).  The vertical green line in plots (b--d) indicates the instant of time when the image (a) was taken. The artificial slit used for the time-distance plot construction is indicated by the straight white line in panel (a). The distance is measured along the slit, starting at its left edge. The figure is taken from \citet{2016SoPh..291.3251A}.}
	\label{fig:motion_mag_real_data} 
\end{figure*}

The motion magnification algorithm \citep{2016SoPh..291.3251A} was tested on  synthetic data sets in order to reveal its potential usage in coronal seismology. The synthetic sequence of images has four distinct loops: a steady loop oscillating at a single frequency, a loop slowly moving and oscillating at the same time, a loop exhibiting exponentially decaying transverse oscillations, and a fourth loop oscillating at two different frequencies simultaneously. The data was superimposed with a randomised non-uniform background and contaminated with photon noise. Figure \ref{fig:motion_mag_model} illustrates the synthetic data sets and the results obtained by motion magnification with the magnification factor of 10.
All transverse motions that were put in the model are clearly seen in the time-distance map produced from the magnified data (Figure \ref{fig:motion_mag_model}d), unlike in the original data (Figure \ref{fig:motion_mag_model}c), where they are visually not recognisable.
An example of detecting decayless kink oscillations in a coronal loop observed by SDO/AIA in 171~\AA\ channel at 2013-01-21 15:50:12 UT  is shown in Figure \ref{fig:motion_mag_real_data}.
The oscillating patterns are clearly visible in magnified time-distance plots, unlike in the original data where they are barely noticeable. 

The DT$\mathbb{C}$WT-based motion magnification  algorithm is aimed at analysing different kinds of oscillatory signals including non-harmonic and non-stationary quasi-periodic displacement.
Therefore it is crucial to provide linear and uniform magnification with respect to the amplitude and period of the original signal. A set of synthetic tests revealed that the DT$\mathbb{C}$WT-based motion magnification demonstrates linear dependence of the magnified amplitude versus the original one for the cases when the periodic displacements in the original data do not exceed 2--3 pixels.
However  displacements larger than 1--2 pixels do not  need any magnification to be analyzed with  traditional techniques.
It was also found that the oscillation frequency has no effect on the actual magnification factor   in a broad  range of periods, allowing for reliable analysis of multi-modal and non-harmonic transverse oscillations in coronal loops and other structures.

The motion magnification method has several limitations. First of all, the method can be applied to only steady or slowly moving structures, since it implies gathering information from a fixed time-window covering several subsequent images, and an oscillating structure should persist during this time window.
The width of this window is a parameter of the algorithm called the smoothing width and should be tuned to be shorter than the characteristic evolution time of an oscillating structure (i.e. coronal loop) but longer than the longest timescale of investigated motions.
Secondly, the projected amplitude of transverse motions should not exceed 1-2 pixels, otherwise the motion-magnified data will have significant distortion.
Also, the background brightness noise which is always present in the data will be magnified alongside the real transverse motions and may cause distortion in the processed images.
Keeping in mind these limitations, \citet{2016SoPh..291.3251A} suggest the following strategy for data analysis using motion magnification.
\begin{itemize}
    \item Process the data cube of interest with a magnification coefficient of 10 and a smoothing width corresponding to the expected period of the oscillation.
    \item Make several time-distance plots from the magnified and original data.
    \item If there is a significant distortion in the magnified data in comparison to the original one, reduce the magnification coefficient and try again.
    \item Examine the time-distance plots and estimate the main (longest) oscillation period of interest.
    \item Set the smoothing width parameter to be slightly larger (for example, by 10\,\%) than the main oscillation period and process the original data again, obtaining the final result.
\end{itemize}
By following this strategy  one can get reliable results and avoid adverse effects causing possible distortions \citep{2021SoPh..296..135Z}.
The practical application of the motion magnification method to analysing transverse oscillations in coronal loops made it possible to obtain new seismological information from the low amplitude kink oscillations of coronal loops.
In particular, the presence of a second harmonic was detected for the first time in the decayless regime of kink oscillations \citep{2018ApJ...854L...5D}.
The  analysis of omnipresent decayless kink oscillations simultaneously observed in several loops of a single active region allowed for mapping the Alfv\'en speed (and potentially the magnetic field) in the corona \citep{2019ApJ...884L..40A}.
\citet{2020ApJ...893L..17L} discovered low amplitude transverse  oscillations with growing period in a diffuse loop observed by SDO/AIA 171\,Å\ associated with the QPP event observed during the  rising phase of a solar flare by Nobeyama radioheliograph at 17\,GHz.
Thus, motion magnification allows for seismological analysis of the solar corona by the low-amplitude transverse motions in coronal loops in non-flaring active regions and for investigating weak oscillatory responses of coronal structures to the eruptive events in nearby active regions.

\subsection{Automatic detection of transverse motion with the NUWT code}
\label{sec:NUWT}

\begin{figure}[t]
\includegraphics[width = 0.47\textwidth]{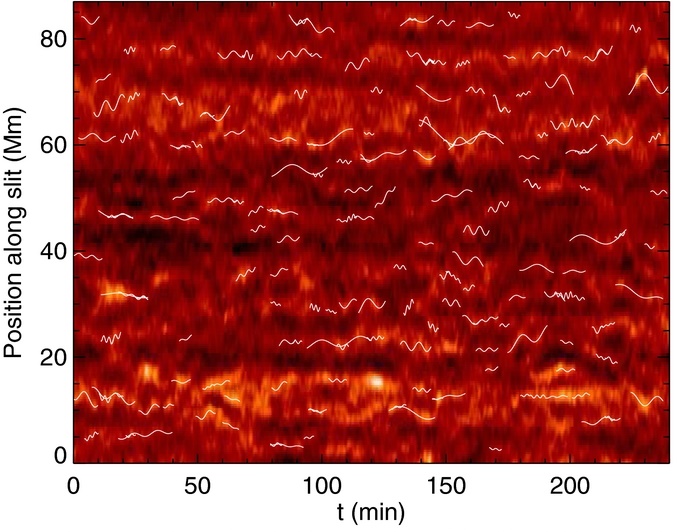}
\includegraphics[width = 0.47\textwidth]{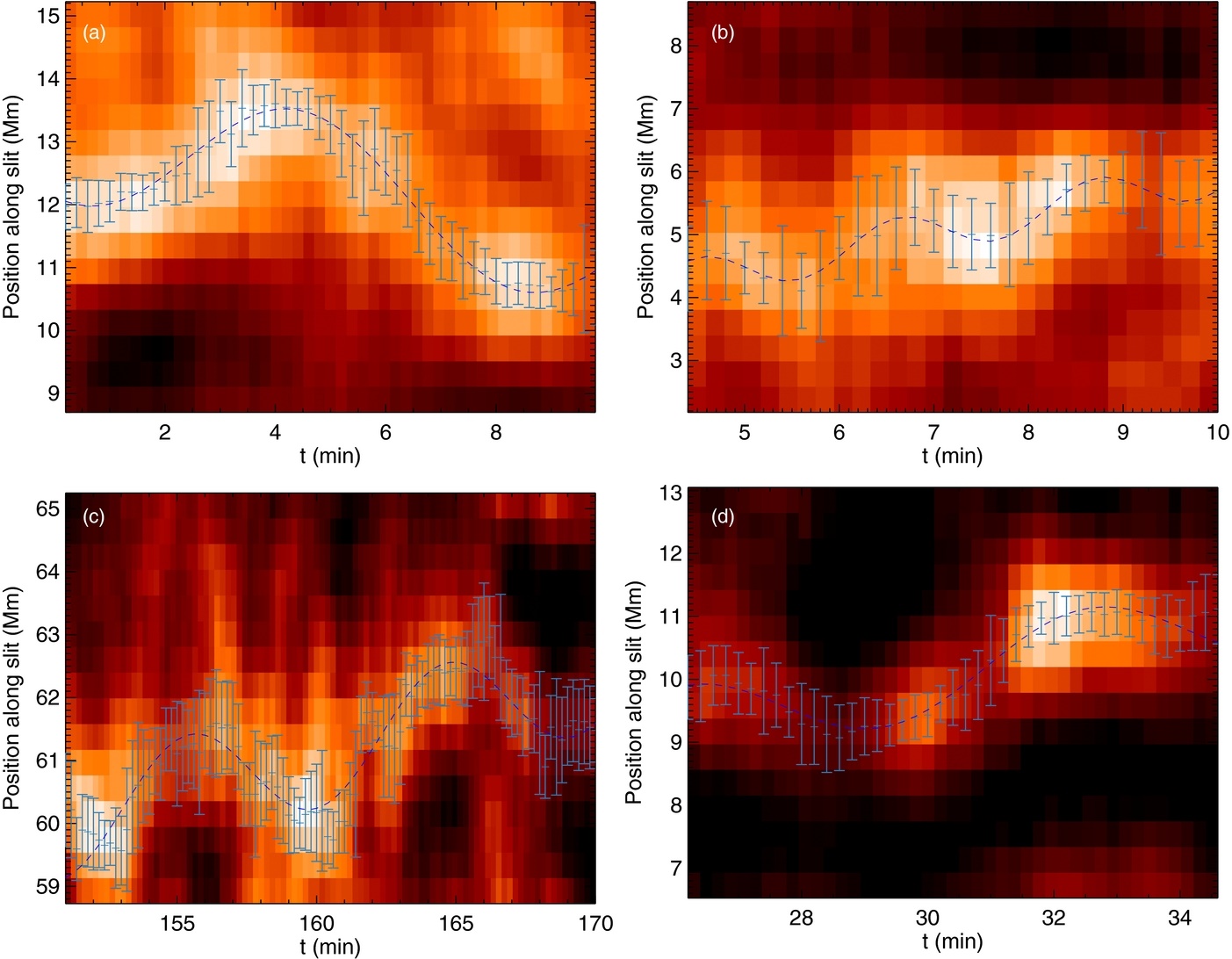}
      \caption{{(Left)} Time-distance map made from slit at 8.71 Mm for the whole time series. A selection (i.e. not full sample) of longer-period fits that were made to the oscillating features are overlaid in white. (Right)  Close-up view of fitted oscillations. Centre of the blue vertical error bars show the feature centre from time-to-time as determined by Gaussian fitting routine and $\pm \sigma$ uncertainty on that position. Blue-dashed curve through the feature the best sinusoidal fit to the feature centre, from which the wave parameters are derived.
   }
\label{Figure_McLaughlin_One}
\end{figure}

NUWT (Northumbria University Wave Tracking) is an image-processing algorithm, specifically an automated Gaussian-fitting method, that can accurately identify individual brighter/darker features in time-distance diagrams and track their evolution through time. The fundamental framework of NUWT was reported in \citet{2013A&A...553L..10M} - who compared and contrasted Hi-C \citep{2013Natur.493..501C} and SDO/AIA observations of transverse MHD waves in active regions - and \citet{2013ApJ...768...17M}, who demonstrated the connection between convectively-driven photospheric flows and incompressible chromospheric waves. NUWT has also been used to provide an analysis of the fine-scale structure of moss in an active region \citep{2014ApJ...789..105M}, including making the first direct observation of physical displacements of the moss fine structure in a direction transverse to its central axis. NUWT was used by \citet{2014ApJ...784...29M} for statistical studies in the highly-dynamic solar chromosphere, which allowed for the determination of the chromospheric kink wave velocity power spectra. \citet{2015NatCo...6.7813M} used NUWT to reveal that counter-propagating transverse waves exist in open coronal magnetic fields using the CoMP instrument \citep{2008SoPh..247..411T}.

\citet{2014ApJ...790L...2T} used NUWT to report the first direct measurements of transverse wave motions in solar polar plumes and evaluate their energy contributions. The authors considered the solar north pole as seen by SDO/AIA in 171\AA\ on 6 August 2010 at 00:00 UTC. A slit of 200 pixels (87 Mm) at an altitude of 8.71\,Mm was used to create a time-distance diagram. The time-distance diagram was characterised by intermittent, bright streaks (corresponding to the locations of the over-dense, plume structures) which exhibited clear signs of transverse motion. NUWT extracted the transverse displacements, periods and velocity amplitudes of 596 distinct oscillations observed in these time-distance diagrams. The measurements made for the slit at an altitude of 8.71\,Mm are shown in Figure \ref{Figure_McLaughlin_One} (left) and, as a typical example, four fits are shown in further detail in Figure \ref{Figure_McLaughlin_One} (right).

The NUWT code can be downloaded via \citet{2016zndo.....54722M} and NUWT has been further automated and improved in \citet{2018ApJ...852...57W}, who also tested and calibrated the algorithm using synthetic data, which included noise and rotational effects. The calibration indicated an accuracy of 1\%-2\% for displacement amplitudes and 4\%-10\% for wave periods and velocity amplitudes. Here, we briefly review the basic operation of the code \citep[see][for details]{2018ApJ...852...57W}. There are six steps in the NUWT data processing pipeline:
\begin{enumerate}
\item{\emph{Data acquisition and preprocessing.} After acquired the data, relevant instrument-specific corrections and processing should be performed, which could include rotating, rescaling, co-alignment, de-spiking and suppression of random noise.}
\item{\emph{Slit Extraction.} A two-dimensional time–distance diagram is constructed by extracting the data from a virtual slit. The intensity uncertainties in the time–distance diagram are also collated, including expected contributions from standard sources (e.g. photon noise, dark current). These uncertainties (data errors) influence the uncertainties on model parameters in stage 3, and so it is important that accurate values are extracted or estimated.}
\item{\emph{Feature Identification.} Local maxima  are located in the time–distance diagram by comparing values to their nearest neighbours, via  a crawling algorithm. Then,  gradients either side of the determined location are checked against a user-adjustable threshold gradient. If the gradient is sufficient, the point is considered a local maximum. This step also removes spurious peaks due to noise. Once pixels containing local maxima are determined, their position is refined to sub-pixel accuracy by fitting the local neighbourhood with a Gaussian model, using a nonlinear least-squares fitting method \citep{2009ASPC..411..251M} which takes into account the intensity uncertainties. For  SDO/AIA 171\AA, the fitting is weighted by intensity errors taken as $\sigma _{\rm{noise}}(F) \approx \sqrt{2.3 + 0.06F} /\sqrt{5}\: (DN)$ as per \citet{2012A&A...543A...9Y}, where $F$ is the pixel-intensity of the unaltered Level 1 data and  $\sqrt{5}$ arises because the time-distance diagrams are constructed from an average of 5-neighbouring slices. The uncertainty on the position of the local maxima is taken as the $\sigma$ estimate on the position of the apex of the fitted Gaussian and the uncertainty on position due to sub-pixel jitter (added in quadrature).}
\item{\emph{Thread tracking.} Local intensity maxima are  traced through the time series by connecting the maxima into \lq{threads}\rq{}. This is achieved using a nearest-neighbour method that scans a user-adjustable search box in both time and space. A given peak cannot be assigned to more than one thread. Threads containing less than a minimum number of data points are rejected. For SDO/AIA, 20 data points is found to be reasonable cut-off, which means we reject threads that persist for less than 240\,s. Fits dependent on a set of points that include a jump of more than 2.5 pixels from point-to-point are discounted since, for SDO/AIA, this would constitute an instantaneous transverse velocity of greater than 100\,km\,s$^{ - 1}$ which is an order of magnitude greater than the representative velocity amplitudes measured for all samples (even when this criterion is relaxed). Such jumps are artifacts from the thread-following algorithm.}
\item{\emph{Application of FFT.} A split cosine bell windowing function is applied to the time series and then FFT is applied. We correct the output power spectrum to account for signal lost due to windowing. If required, gaps within each thread can be filled using linear interpolation.}
\item{\emph{Filtering Waves and Calculating Observables.} The significant wave components of the FFT power spectrum are selected. Every peak with power greater than the significance threshold are identified as different waves that are  propagating on the same structure. The tracking-routine picks up threads of variable lengths and so longer threads can often be fitted with different oscillations during different stages of their lifetime. In addition, it is often the case that shorter-period oscillations can be seen super-imposed upon longer period trends. Where multiple fits to different subsections of a longer thread are possible, all such fits are taken and contribute to the sample. Wave displacement amplitudes,  $\xi$, and period, $P$, are calculated from the power spectrum and the velocity amplitudes are calculated using $v = 2\pi \xi / P$.}
\end{enumerate}

The NUWT automated algorithm unlocks a wide range of statistical studies that would be impractical to conduct by hand (i.e. extremely labour-intensive). For example, NUWT has been used by \citet{2019NatAs...3..223M} to reveal  that the Sun's internal acoustic modes (p-modes) make a basal contribution to the Alfvénic wave flux in the solar corona, delivering a spatially-ubiquitous input to the coronal energy balance that is sustained over the solar cycle. \citet{2020ApJ...894...79W} used NUWT to make direct measurements of key parameters of transverse waves in a coronal hole, and reported how those measurements change with altitude through the low corona. Interestingly, this enabled them to derive a relative density profile for the coronal hole environment but, crucially, without the use of spectroscopic data.

Although, the NUWT code allows for automated detection of periodic transverse motions in the corona and quick estimation of their parameters,  this algorithm has its limitations as any other method.
First of all, the algorithms used by NUWT for detecting  and  testing significance of oscillatory events are simplified in favour of high performance.
For instance, the significance of a peak in the Fourier power spectrum is tested against white noise,  whereas the background noise in the corona is known to have power law spectrum \citep{2016ApJ...828...89M, 2016A&A...592A.153K}.
Thus, the algorithm may produce false detections of oscillatory events in some cases (see Sections \ref{sec:EMD}, \ref{sec:time-series_pitfalls}, and \ref{sec:QPP}  for a detailed discussion on testing significance of oscillatory signals in the presence of coloured noise with the power law spectrum).
Also, the NUWT algorithm searches for oscillations in well distinguished non-overlapping structures,  that is often not the case in the EUV image.
Thus, NUWT is a good tool for quick detection and statistical analysis of large amount of transverse oscillations, whereas the detailed analysis of individual events  should be performed with other, more comprehensive methods.
One of these methods aimed at the precise investigation of transverse oscillations in multiple overlapping loops is discussed below in Section~\ref{sec:multiple_loops}.

\subsection{Analysis of multiloop active regions}
\label{sec:multiple_loops}

\begin{figure*}
\centering
\includegraphics[width=0.45\textwidth]{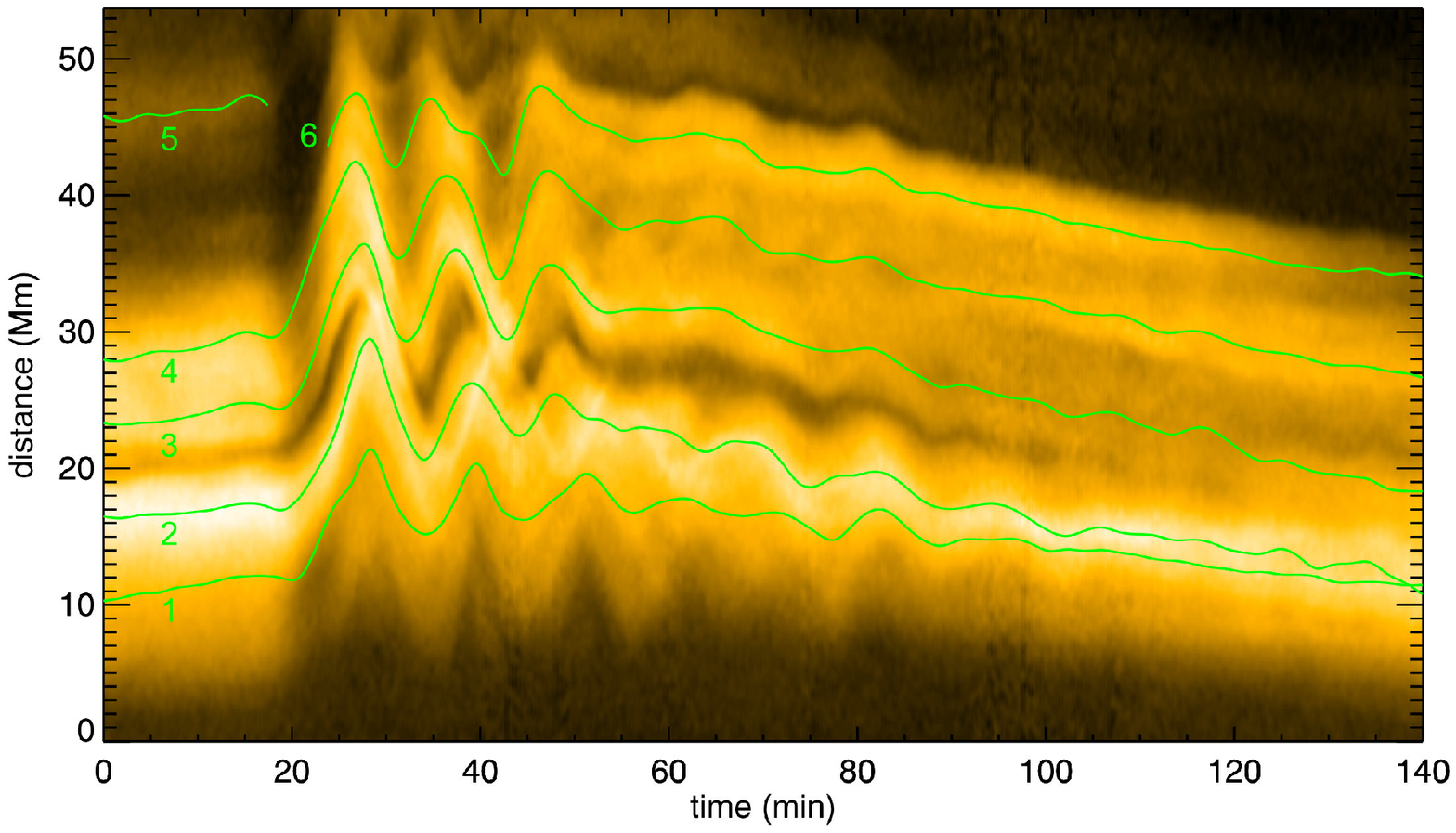}
\includegraphics[width=0.45\textwidth]{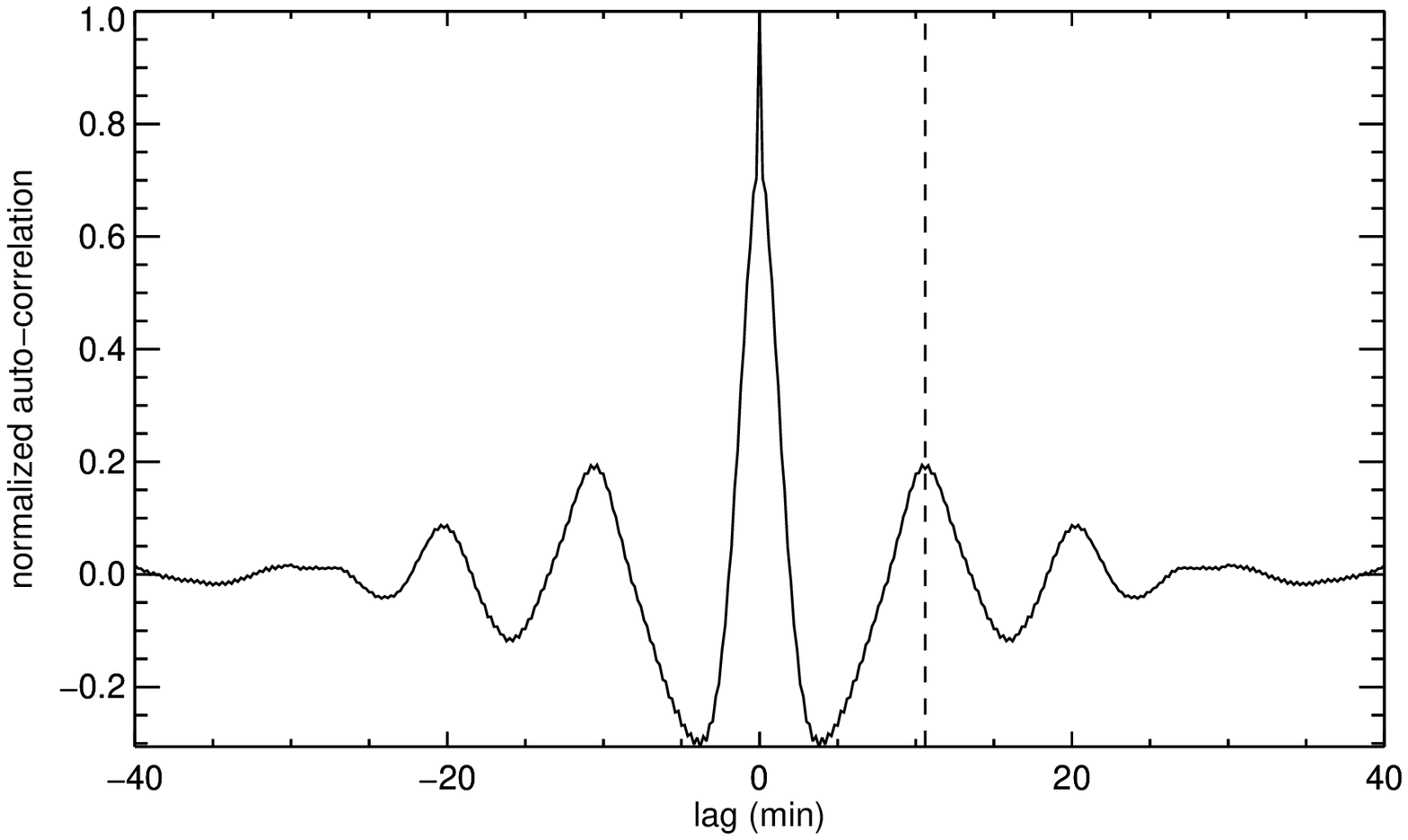}
\includegraphics[width=0.45\textwidth]{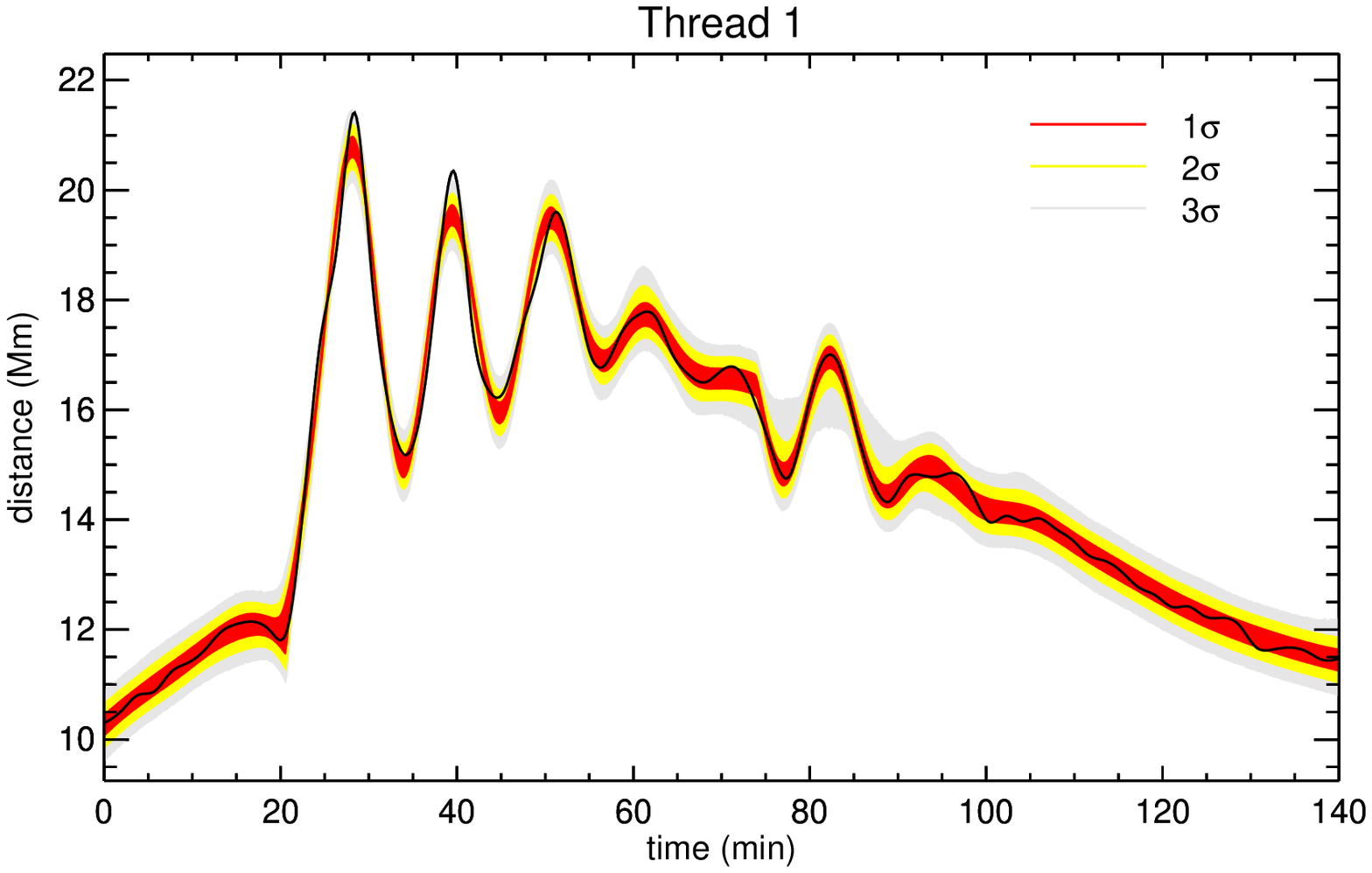}
\includegraphics[width=0.45\textwidth]{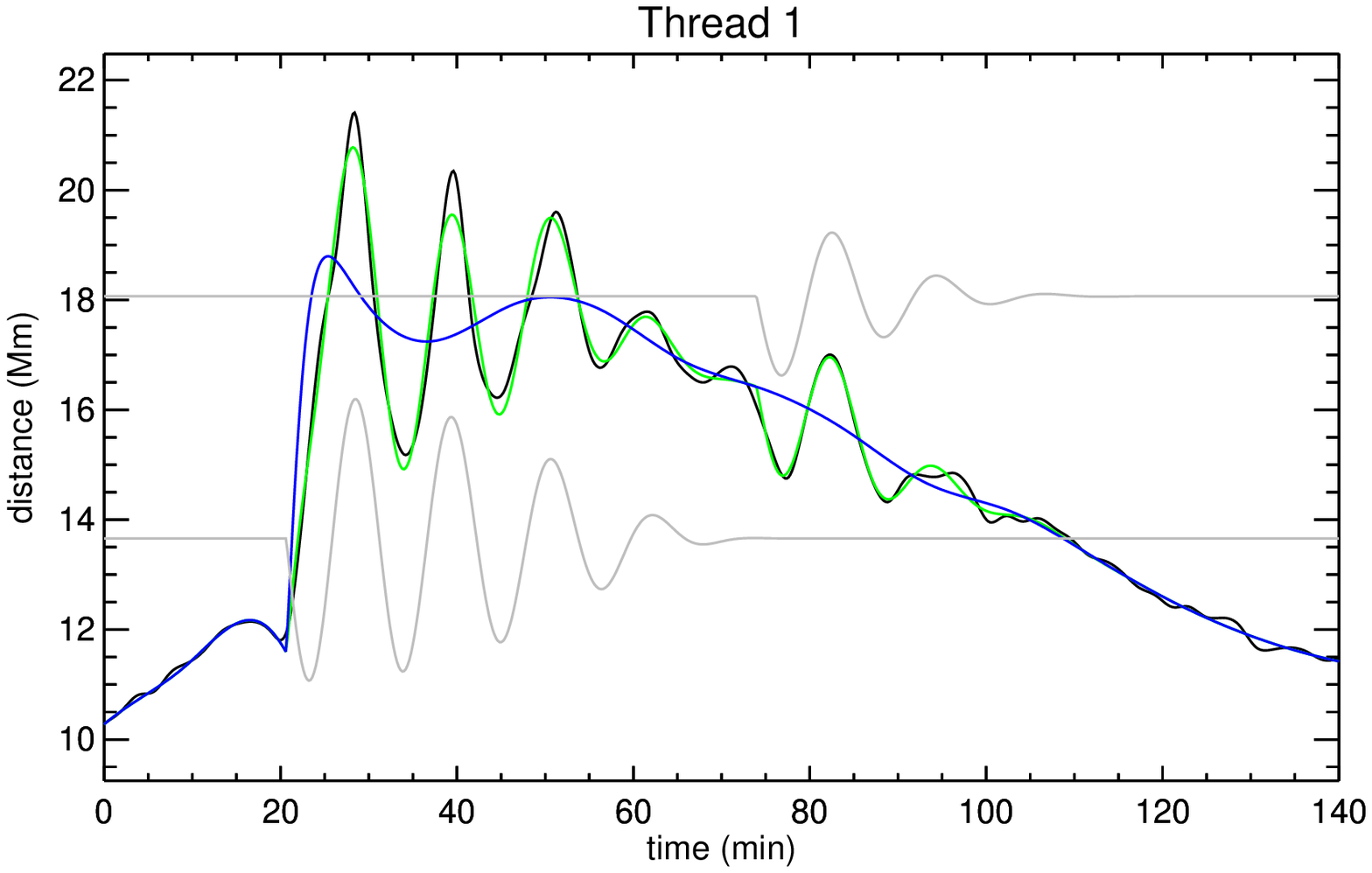}
\caption{
Tracking of six threads in a time-distance map for AR11967 using the method of \cite{2020ApJ...898..126P} (top left).
The typical period of oscillation (dashed line) may also be estimated for the whole time-distance map using the autocorrelation method of \cite{2019ApJ...880....3A} (top right).
The bottom left panel shows the tracked position of thread 1 (black line) and posterior predictive distribution based on $10^6$ MCMC samples of a model containing two separately excited oscillations, with colour contours representing  $1\sigma$ (red), $2\sigma$ (yellow), and $3\sigma$ (grey) confidence levels.
The bottom right panel shows the fit based on the maximum a posteriori values of the model parameters (green) with the background (blue) and oscillation (grey) components shown separately.
}
\label{fig:djp_waves}
\end{figure*}

Decaying standing kink oscillations are commonly observed in coronal loops.
Since they cause periodic displacements of the loop axis they are usually measured by placing an observational slit perpendicular to the loop axis which is used to construct a time-distance  map from a sequence of EUV images.
The use of multiple slits along the loop axis may also be used to demonstrate the spatial dependence of kink oscillations and hence confirm they are longitudinal harmonics of standing modes \citep[e.g.][]{2016A&A...593A..53P,2018ApJ...854L...5D,2019A&A...632A..64D}.

Analysis of coronal loops and their oscillations is complicated by the fact that active regions typically include multiple loops and, since coronal EUV emission is optically thin, the contribution from each loop is integrated along the observational line-of-sight.
Loops being in close proximity in EUV images can make it difficult to choose a slit which isolates a single loop.
Recently, two new techniques to measure the periodicities in active regions have been demonstrated.
These two methods are demonstrated in Figure~\ref{fig:djp_waves} for the analysis of coronal loops in active region AR11967 on 27 January 2014.

\citet{2019ApJ...880....3A} developed a method based on the autocorrelation function for the perturbations in the time-distance map.
The smoothed time-distance map is subtracted from the data to produce the detrended map $\tilde{I}$ for which the 2D autocorrelation function is calculated as

\begin{equation}
c \left( \Delta t, \Delta x \right) = \iint \tilde{I} \left( t, x \right) \tilde{I} \left( t + \Delta t, x + \Delta x \right) dt dx
\label{eq:auto}
\end{equation}
where $\Delta t$ is the time lag and $\Delta x$ is the spatial offset.
The autocorrelation plot for $\Delta x = 0$ (top right panel in Figure~\ref{fig:djp_waves}) reveals the characteristic period of oscillation as the first nonzero lag for which the function is maximal (dashed line).

The standard technique of tracking a loop in a time-distance map  is based on fitting each frame separately with a function, such as a Gaussian, which describes the intensity enhancement due to the loop (this technique is also used by the NUWT code described in Section \ref{sec:NUWT}).
It is simple to extend this method to consider multiple intensity enhancements but in practice the fitting may be difficult due to issues such as the proximity of the loops, spatial resolution, and noise limiting the information which can be reliably extracted from individual frames.
The technique used by \cite{2020ApJ...898..126P} is instead based on fitting the entire time-distance map with a number of smoothly moving and evolving intensity structures representing the coronal loops.
This allows the properties of the loop to be constrained using multiple frames.
The smoothness of each loop parameter (position, radius, intensity enhancement) can be varied separately depending on the expected behaviour. For example, kink oscillations cause displacements of the loop axis, do not perturb the loop radius, and changes in intensity are expected to have longer timescales than the period of oscillation.
The fit in the top left panel of Figure~\ref{fig:djp_waves} is therefore based on 701 frames being modelled by threads each using 75 parameters for position, four for intensity, and one for the radius.
This use of physical interpretation to reduce the number of fitted parameters means that, for this case, each parameter uses approximately 30 times as much observational data to constrain each parameter than an equivalent fit based on analysing each frame of the time-distance map separately.
A benefit of obtaining the time series for the loop position is that not only can the period of oscillation be calculated but other properties can be tested using advanced seismological methods such as Bayesian analysis (see Section~\ref{sec:bayesian_analysis}).
In this case, Bayesian model comparison demonstrated that the damping profiles of the oscillations were better described as Gaussian rather than exponential, consistent with the mechanism of resonant absorption \citep[e.g.][]{2012A&A...539A..37P,2021SSRv..217...73N}, and that there were two perturbations rather than one, coinciding with two solar flares.
For Thread 1 (bottom panels of Figure~\ref{fig:djp_waves}) it was also found that the damping rate increased in time which may be indicative of loop evolution due to the Kelvin–Helmholtz instability \citep{2020FrASS...7...61P}.


\section{Methods for analysing propagating longitudinal waves in open structures}
\label{sec:longitudinal_waves}

\begin{figure}
\includegraphics[width=0.98\textwidth]{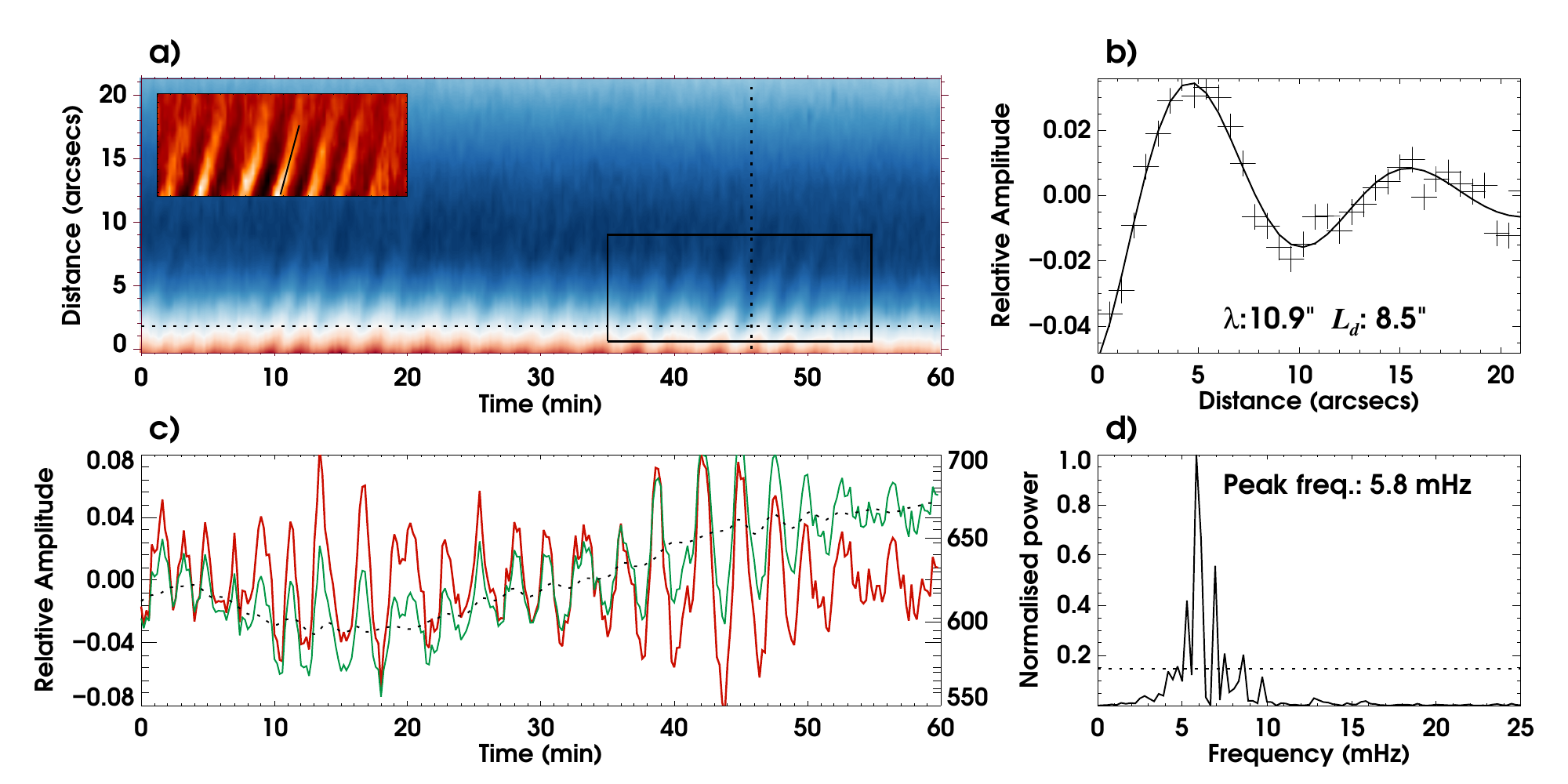}
\caption{Analysis of propagating longitudinal waves. a) A sample time-distance map constructed from a sunspot-loop structure. The region outlined by the solid black box is processed to enhance the ridges and shown in the inset panel. The slanted solid black line in the inset panel marks the ridge used in the propagation speed estimation. b) Intensity profile along the loop at a fixed temporal location marked by the vertical dotted line in panel a. The solid curve represents the best fit damping sinusoid to the data. The obtained wavelength, $\lambda$, and damping length, $L_d$, from the fit are listed in the plot. c) The temporal intensity profile at the spatial location marked by a horizontal dotted line in panel a. The green solid, black dotted, and red solid curves correspond to the original, background, and the filtered (background subtracted and normalised) intensities respectively. The scale for the original and the background intensities are shown on the right. d) Fourier power spectrum for the filtered light curve shown in panel c. The peak frequency is listed in the plot. The dotted line corresponds to a significance level of 0.01 (99{\,}\% confidence level). }
\label{kpfig1}
\end{figure}
Propagating longitudinal waves are commonly observed in extended coronal structures such as  sunspot fan loops, polar plumes and other on-disk plume-like structures, and interpreted in terms of slow magnetoacoustic waves \citep{2020arXiv201208802B}. Because of their compressive nature, they can be readily detected using imaging observations alone. The high-resolution, high-cadence, multi-wavelength data provided by SDO/AIA over the entire solar disk have vastly enlarged our observation sample of these waves. Similar to the analysis of transverse kink oscillations, discussed in Sec. \ref{sec:transverse_motions}, an important first step in the analysis is to construct time-distance maps from the image sequences (three-dimensional data). This involves extracting a one-dimensional slice from each image by manually tracing pixel locations along a target structure and stacking similar slices from successive frames adjacent to each other. Intensities across the loop are typically averaged either over a uniform or varying (following the expanding loop) width to improve the signal to noise \citep[e.g.,][]{2000A&A...355L..23D, 2011A&A...528L...4K}. The propagating longitudinal waves appear as slanted ridges of alternating brightness in the time-distance maps. One such map is shown in Fig.~\ref{kpfig1}a. A number of important oscillation properties such as the amplitude, period, propagation speed, damping length etc., can be extracted from these maps. 

\subsection{Extraction of basic oscillation parameters}
\subsubsection{Oscillation amplitude}
The apparent ridges in the time-distance maps can be further enhanced by subtracting a background level. One may also normalise the resulting intensities with the same background to obtain relative amplitudes. A section of the time-distance map shown in Fig.~\ref{kpfig1}a, outlined by a solid black box, is processed in this fashion and displayed in an inset panel. As can be seen, the brightness ridges appear much more sharp and clean. The background in this case is constructed from a running average at each spatial position over 30 data points effectively filtering any oscillations with periods longer than 6 minutes. Note, that the filtering cut-off period was selected to be twice as large as the main period in the power spectrum to keep all oscillations with periods around 3 minutes untouched. The original lightcurve, background level and the resultant filtered light curve after normalisation are shown in Fig.~\ref{kpfig1}c in green solid, black dotted, and red solid curves, respectively. Note the scale for the original and the background curves is shown on the right side of this panel. It may be noted that the oscillation appears quite regular with its amplitude varying between $\pm$8{\,}\% of the background. A root-mean-square (rms) value may be obtained from the filtered light curves to get an estimation on the oscillation amplitude.
\subsubsection{Oscillation period}
The spacing between the ridges in time-distance maps gives the periodicity of the oscillation. The exact period may be obtained by applying Fourier analysis techniques over the time series. The normalised light curve shown in Fig.~\ref{kpfig1}c is subjected to Fast Fourier Transform (FFT) and the resulting Fourier power spectrum is displayed in panel d) of the same figure. Here, the peak power is found to be at 5.8{\,}mHz corresponding to an oscillation period of 172{\,}s. The dotted line in this panel corresponds to a significance level of 0.01 implying for any random time series there is less than 1{\,}\% chance to produce Fourier peaks above this level. The application of the FFT method to a time series normalised to its trend is discussed in detail in Section~\ref{sec:time-series_pitfalls}.
\subsubsection{Propagation speed}
The speed at which the longitudinal waves propagate is another key parameter and it is directly related to the temperature of the plasma within the supporting structure. This is generally measured from the slope of the ridges in time-distance maps. Although the positional coordinates from only two spatio-temporal locations along the ridge are enough to measure a linear slope and thereby get an estimation on the propagation speed, the associated uncertainties are often substantially large. To alleviate this problem, \cite{2012SoPh..281...67K} identified temporal coordinates of the local maxima at each spatial position along a preselected ridge and fitted them with a linear function. The inverse of the  slope obtained from the fit in this case gives the propagation speed. The bright ridge marked by a solid black line on the inset panel of Fig.~\ref{kpfig1}a gives a speed of 48.0$\pm$2.9{\,}km{\,}s$^{-1}$ according to this method. Another method that is applicable here employs a cross-correlation technique. In this, the time series at each spatial position is cross-correlated with a reference series to obtain the respective time lags which can be further fitted with a linear function, the slope of which gives the propagation speed \citep[see for e.g.,][]{2009ApJ...697.1384T}. Since the slope is estimated from a larger set of points and both these methods involve less manual intervention, the obtained speed values are naturally more reliable and result in considerably lower errors. Also, these methods can be easily adapted to cases where the waves display non-constant speeds by simply changing the linear fit function to a more appropriate one. As can be seen from Fig.~\ref{kpfig1}a, some ridges are prominently bright while some are faint and barely visible. Often we find very few ridges with reasonably good amplitude. It is also quite possible to find changes in the slope of the ridges with time. In such cases, the method by \cite{2012SoPh..281...67K} is more preferable as it does not involve the full time series.
\subsubsection{Damping length}
One of the puzzling properties of propagating longitudinal waves is their rapid damping. They are visible over very short distances (of the order of a wavelength) near the foot point of the supporting structure. The background subtracted and normalised intensity profile at a particular time instant (marked by a vertical black dotted line in Fig.~\ref{kpfig1}a) is shown in Fig.~\ref{kpfig1}b using '+' symbols. The black solid line in this figure represents the best fit damping sinusoid curve defined by the function $I(x)=A_{0} e^{-x/L_{d}} \mathrm{sin}(k x + \phi)+B_{0}+B_{1}x$. Here $x$ is the distance along the loop, $L_d$ is the damping length, $k = 2\pi/\lambda$ is the angular wave number where $\lambda$ is the wavelength, and $\phi$ is the initial phase. $A_0$, $B_0,$ and $B_1$ are appropriate constants. The obtained wavelength and damping length for this particular case are 10.9$\arcsec$, and 8.5$\arcsec$, respectively as listed in the plot. It may be noted that because these waves are propagating, the wavelength is not necessarily a constant but instead can vary with distance depending on the changes in propagation speed \citep[see for e.g.,][]{2017ApJ...834..103K}.
\subsection{Other characteristic properties}
\subsubsection{Amplitude modulation}
The amplitude of oscillations is usually not constant in time (see Fig.~\ref{kpfig1}c). Indeed, it has been found that the amplitude varies with a period between 20-30 min, for the short-period oscillations detected in sunspot-related loop structures \citep{2015ApJ...812L..15K, 2020arXiv200405797S}. \cite{2015ApJ...812L..15K} found similar amplitude fluctuations in multi-instrument, multi-wavelength data spanning different atmospheric layers including the photosphere. It has been previously suggested that this amplitude modulation is due to a beat-like phenomenon produced by a set of closely spaced frequencies. Evidently, multiple peaks can be seen surrounding the main peak in the power spectrum displayed in Fig.~\ref{kpfig1}d. However, since the Fourier power spectrum does not have temporal information, in order to study the amplitude modulation, one needs to apply a wavelet transform to the light curves. The temporal variation of the wavelet power averaged over a desired frequency band could then be used to deduce the associated amplitude variation. The amplitude curve, thus obtained, can be further subjected to Fourier analysis to calculate the modulation period.  
The Hilbert-Huang transform approach, combining the EMD method and the Hilbert transform as described in Sec. 2.1, could also be used for the analysis of the instantaneous amplitude behaviour \citep[see e.g.][]{2019ApJ...884..131R}.
\subsubsection{Frequency-dependent damping} 
\begin{figure*}
\centering
\includegraphics[width=1.0\textwidth]{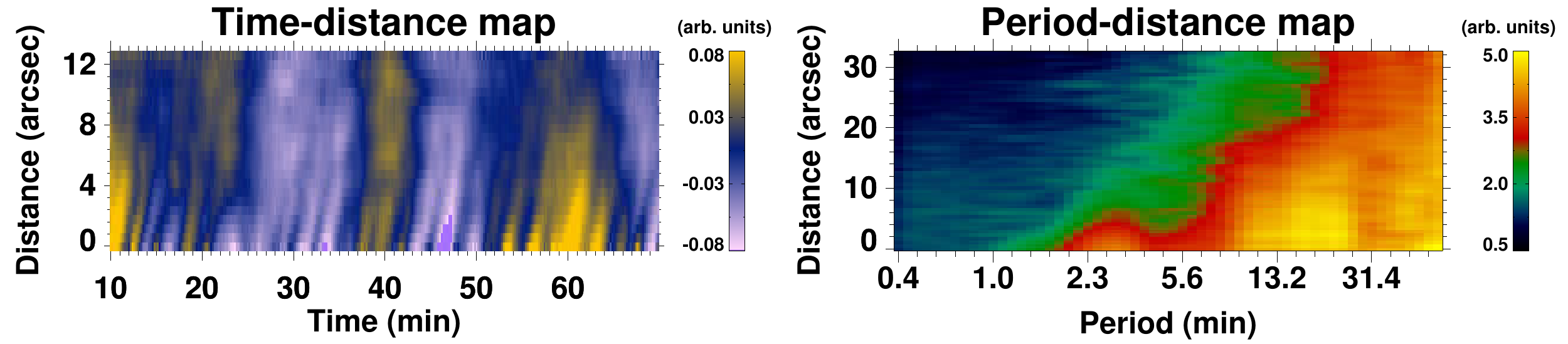}

\caption{
\textit{Left}: A section of a time-distance map depicting the simultaneous existence and propagation of waves with multiple periods. \textit{Right}: Corresponding period-distance map obtained by replacing the intensity time series at each spatial position with the respective Fourier power spectrum. The data are shown here for larger distances to highlight the decay at longer periods. Note the scale for periods is not linear and is truncated to show periods shorter than 60 minutes. Adapted from \cite{2014ApJ...789..118K}.}

\label{fig:kpfig2}
\end{figure*}

The damping of propagating longitudinal waves has been found to be frequency-dependent \citep{2014ApJ...789..118K, 2014A&A...568A..96G}.
In order to study the frequency dependency, \cite{2014ApJ...789..118K} constructed period-distance maps from the time-distance maps by replacing the time-series at each spatial position with the corresponding Fourier power spectrum (see Fig.{\,}\ref{fig:kpfig2}).
These maps allow one to extract the spatial variation of oscillation amplitude ($\propto$ square root of Fourier power) at multiple frequencies and thereby study the associated damping lengths as a function of frequency. By grouping the results from similar structures together, \cite{2014ApJ...789..118K} found that the dependences obtained for the open structures in polar regions do not concur with the existing linear wave theory. In polar regions, the long-period waves appear to display stronger damping compared to their short-period counterparts. The behaviour observed in sunspot loops and other on-disk plume-like structures also had some discrepancy with the theory, a resolution to which was suggested later through numerical modelling \citep{2016ApJ...820...13M} .

\section{Estimating the 3D geometry of oscillating coronal loops}
\label{sec:3D_shape}

Observations of loops in the solar corona are affected by projection effects. A loop with a given orientation in three dimensional (3D) space may appear different according to the location on the Sun: elongated in the proximity of the solar limb with a poor view of the position of the footpoints, or flattened when viewed on the solar disk with strong uncertainties about the height.
In the context of coronal seismology, the  loop length is a crucial parameter for the determination of the kink speed,  and consequently for the estimation of reliable values of the Alfv\'en speed and the magnetic field in the loop. The global kink speed for the fundamental harmonic of the transverse oscillations in loops is given as $C_\mathrm{K}=2L/P$, with $L$ the length of a loop and $P$ the period of the transverse oscillations. 
Oscillation periods are usually  obtained with good precision  by fitting tracks of loops in time-distance maps, resulting in uncertainties of the order of a few per cent. On the contrary, uncertainties of the loop lengths are much larger, affecting kink speed estimates with errors of the order of 10\%.

Finding the length of a loop with small uncertainties is related to the problem of defining the geometry of a curvilinear feature in 3D space. The exact geometry of a coronal loop is also important for the correct interpretation of the polarisation mode of kink oscillations: if vertical, horizontal or a composition of both (elliptical polarisation) \citep{Aschwanden2009,Verwichte2009,White2012, 2017A&A...607A...8P}. Possible deviations of the loop shape from the elliptical one are also of importance. Recent numerical MHD simulations by \citet{2020ApJ...894L..23M} show that the writhe-type deformations of a coronal loop can lead to the deviations up to 20\,\% in inferring magnetic field from kink oscillations. While for the plain elliptical case, \citet{2020ApJ...894L..23M} found almost exact correspondence between the actual magnetic field value and its seismological estimate.

\subsection{Geometry and loop coordinate system}

A coronal loop can generally be described as an elliptical curve depending on six parameters: the length of the footpoint baseline, the position of the midpoint (i.e., the point between the footpoints) on the solar surface (heliographic longitude and latitude), the inclination angle with respect to the solar surface normal, the azimuthal angle with respect to the meridian passing through the midpoint, and the height of the loop. The parametric equations are cumbersome if the loop coordinates are taken in a reference frame relative to the Sun (e.g., the Heliocentric Earth Equatorial (HEEQ) coordinate system).

The description gets simplified if a local reference frame for a loop is adopted \citep{Verwichte2010}, with the origin coinciding with the loop midpoint. The \lq\lq natural\rq\rq\ way to take the  coordinate axes in the local frame is as follows: two orthogonal axes lying in the loop plane, the third one perpendicular to the loop plane. Therefore, the loop with a general elliptical profile is parametrised {as a function of $\eta$} by the following equations:
\begin{eqnarray}
x_\mathrm{L} & = & a \cos\eta, \nonumber \\
y_\mathrm{L} & = & b \sin\eta, \nonumber\\
z_\mathrm{L} & = & 0, \nonumber \\
\label{eq_loop}
\end{eqnarray}
with $x_\mathrm{L}$, $y_\mathrm{L}$ coordinates of the loop points that lie on a common plane ($z_\mathrm{L}=0$). The coefficients $a$ and $b$ are the minor and major radii of the ellipse (a circle is trivially obtained when $a=b$) and represent the axes of the loop. The parameter $\eta$ varies between two extrema (ideally, between $0$ and $\pi$).

\begin{figure}
	\centering
	\includegraphics[width=0.7\textwidth]{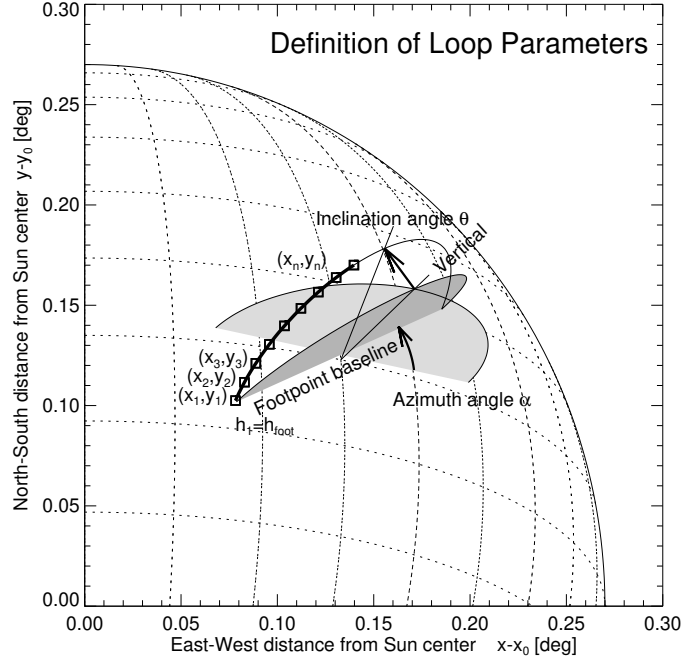}
	\caption{Geometry of a coronal loop with the definition of some loop parameters, as the azimuth angle $\alpha$, the inclination angle $\theta$, and the loop footpoint baseline. Adapted from \citet{Aschwanden_1999ApJ...515..842A}}.
	\label{fig_loop_geometry}
\end{figure}

The conversion of the coordinates from the local loop  reference frame to a solar reference frame, and vice-versa, is laborious but is a crucial step in the reconstruction of the geometry of coronal loops. Coordinate transformations between the local loop plane coordinate system and the heliographic coordinate system are given in Sect. 3.4.4 of \citet{2004psci.book.....A}.

\subsection{Techniques of 3D reconstruction of coronal loops}

Observations  of a coronal loop from a single point of view pose some difficulties in the context of 3D geometric reconstruction. Loop length estimates have been considered by simply approximating the loop shape with a semi-circle, with the radius determined as half of the distance between the loop footpoints for  those loops observed on the solar disk, or as the distance of the apex from  the footpoint baseline for those observed  off-limb \citep[e.g.,][]{Nakariakov_1999Sci...285..862N,2015A&A...583A.136A}.

Geometric triangulation through stereoscopy is a more robust approach for retrieving the 3D shape of coronal loops and of any feature appearing in the corona. The principles of stereoscopy are presented in \citet{2006astro.ph.12649I}. For example, stereoscopy was applied to reconstruct the geometry of coronal loops in active regions by using  the only point of view offered by SoHO/EIT. The rotation of the solar disk was exploited to observe the targeted loops at two different times, hence from two different perspectives \citep{Aschwanden_1999ApJ...515..842A}. This approach, referred to as non-simultaneous or \lq\lq dynamic\rq\rq\ stereoscopy, implies the drastic assumption that the active region and the loops do not change dramatically during the time interval between the two observations. The launch of the twin spacecraft of the Solar Terrestrial Relations Observatory mission \citep[STEREO,][]{Kaiser_2008SSRv..136....5K} opened to the possibility of simultaneous stereoscopy \citep{Aschwanden2008}.

Another method was also developed and presented in \citet{Verwichte2009,Verwichte2010}. The technique, consisting in a forward-modelling of a coronal loop, was applied  with appropriate variations to single-view and stereoscopic observations. It is suitable for loops that are well observed on the solar disk and that can be fully traced along their entire length with a finite number of points defined by the pair of coordinates projected onto the plane-of-sky of the image. The third coordinate of each point, which denotes the height above the solar surface, is added by assuming that the loop is planar or non-planar. Therefore, the problem consists in finding the best values of a few parameters defining the loop geometry, like the inclination angle with respect to the surface normal for a planar loop. If a second view is available, the loop is forward modelled in this field-of-view for a wide range of inclination angles. The best value of the inclination angle is found when the model fits the loop profile in the image. If the second view is unavailable, the 3D shape of the loop is found by adopting dynamic stereoscopy or by minimising the variation in the curvature.  

Other techniques are based on cubic B\'ezier curves to trace a loop in the EUV images and finding the most appropriate loop shape by comparing different realisations of the 3D curve with different magnetic field models \citep{Gary2014a, Gary2014b}.

\begin{figure}
	\centering
	\includegraphics[width=1.\textwidth]{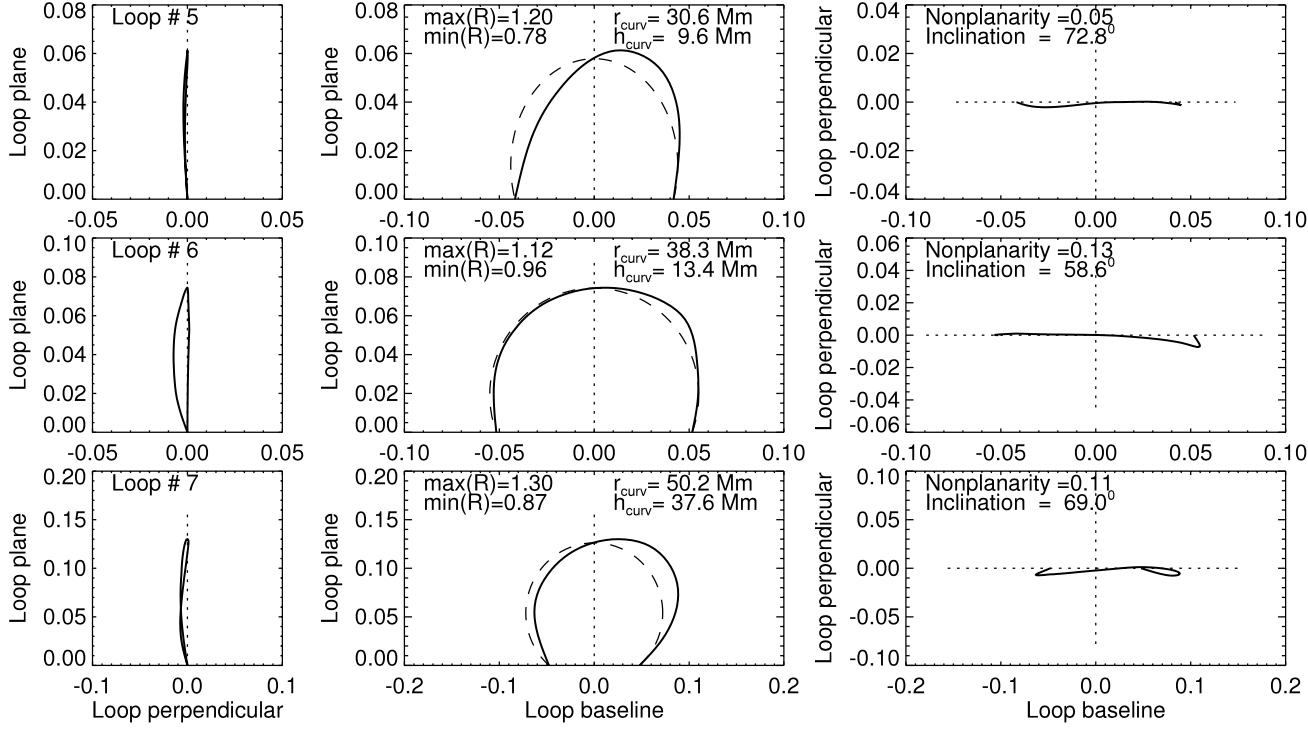}	\caption{Example of 3D reconstructed coronal loops observed with STEREO on May 9, 2007, and projected in the loop plane (middle panels) and in the orthogonal directions: edge-on view (left panels) and a top-down view (right panels). Adapted from \citet{Aschwanden2008}.}
	\label{fig_from_Aschwanden}
\end{figure}

A tool for 3D geometric triangulation is available in the SolarSoftWare (SSW) package based on the IDL language with the routine {scc\_measure.pro}\footnote{\url{https://hesperia.gsfc.nasa.gov/ssw/stereo/secchi/idl/display/scc_measure.pro}}. It is a widget application that allows a user to select a coronal feature appearing in a pair of stereoscopic images and deriving the 3D coordinates of the feature by stereoscopy triangulation. Therefore, concerning loops, it is possible to sample their path with a reasonable number of 3D points (typically ~20 tie-points). To reconstruct the loop profile, spline interpolation was used to connect the 3D tie-points \citep{Aschwanden2008}. The usage of interpolation, however,  can result in spurious effects which cause the curves to be quite irregular, especially when the number of tie points is limited (e.g., see Fig. 7 in \citet{Aschwanden2009}), even though, in general, the shape resembles that of a circle or an ellipse. Some examples are presented in Fig. \ref{fig_from_Aschwanden}). The upper panel of Fig. \ref{fig_from_Aschwanden} shows the reconstruction of a loop that has a shape more similar to an ellipse with the major axis tilted with respect to the normal to the surface. Therefore, based on these studies, a method to properly fit the 3D points was required.

\citet{Nistico2013} proposed the principal component analysis (PCA) to fit the 3D tie-points obtained from the routine scc\_measure.pro. 
The theoretical fundamentals of PCA can be found in \citet{Jolliffe1986}. It is also known as minimum and maximum variance analysis and has found wide application in science, such as in the analysis of the solar wind magnetic field data \citep{Sonnerup1998}. The peculiarity of PCA is that of reducing the dependencies and correlations between data, which are given in a specific reference frame or \lq\lq vector basis\rq\rq, and project them into a new one, whose basis vectors constitute the principal components. Therefore, the set of original data can be written as a linear combination of the principal components. Mathematically, this is equivalent to a change of basis in a vector space, where the new vector basis is associated with the maximum variability of the dataset. In the context of 3D loop reconstruction, the data points are interrelated between each other because of the common parameters that define the loop geometry in the Sun's coordinate system. PCA  removes any correlation in the data and projects them into a new reference frame, whose basis vectors represent the axes of the loop and the normal to the loop plane. These vectors are obtained from the eigendecomposition of the covariance matrix of the data points.

 Given a set of $N$ 3D coordinates $(x, y, z)$ sampling a loop in a certain Sun's reference frame, the covariance matrix is constructed as:

	\begin{equation}
	 S=
	\begin{bmatrix}
	\sigma^2_{x,x} & \sigma^2_{x,y} & \sigma^2_{x,z} \\
	\sigma^2_{x,y} & \sigma^2_{y,y} & \sigma^2_{y,z} \\
	\sigma^2_{x,z} & \sigma^2_{y,z} & \sigma^2_{z,z}
	\end{bmatrix}.
	\end{equation}
	
The diagonalisation of the covariance matrix is performed to maximise the variances of the data (the diagonal elements of $S$) and to minimise the covariances (the matrix elements out of the diagonal).
	
The calculated three eigenvalues are sorted in ascending order. $\lambda_n \le \lambda_a \le \lambda_b$, and the corresponding  eigenvectors define the vector normal to the loop plane ($\hat{ e}_n)$, and the vectors along the minor and major axes of the ellipse ($\hat{ e}_a, \hat{ e}_b$), respectively.
The lengths of the ellipse axes are related to the variances and are equal to
\begin{equation}
a=\sqrt{{2 \lambda_a}} ~~~~~~~~~~~~~~~~~~b=\sqrt{{2 \lambda_b}}
\label{eq:radii}
\end{equation}
The eigenvalue $\lambda_n$ is very small, in practise equal to zero. From a geometric point of view, this means that the data points lie on a same plane (i.e., the loop plane).

\begin{figure}[htpb]
	\center
	\begin{tabular}{c}
		\includegraphics[width=15 cm]{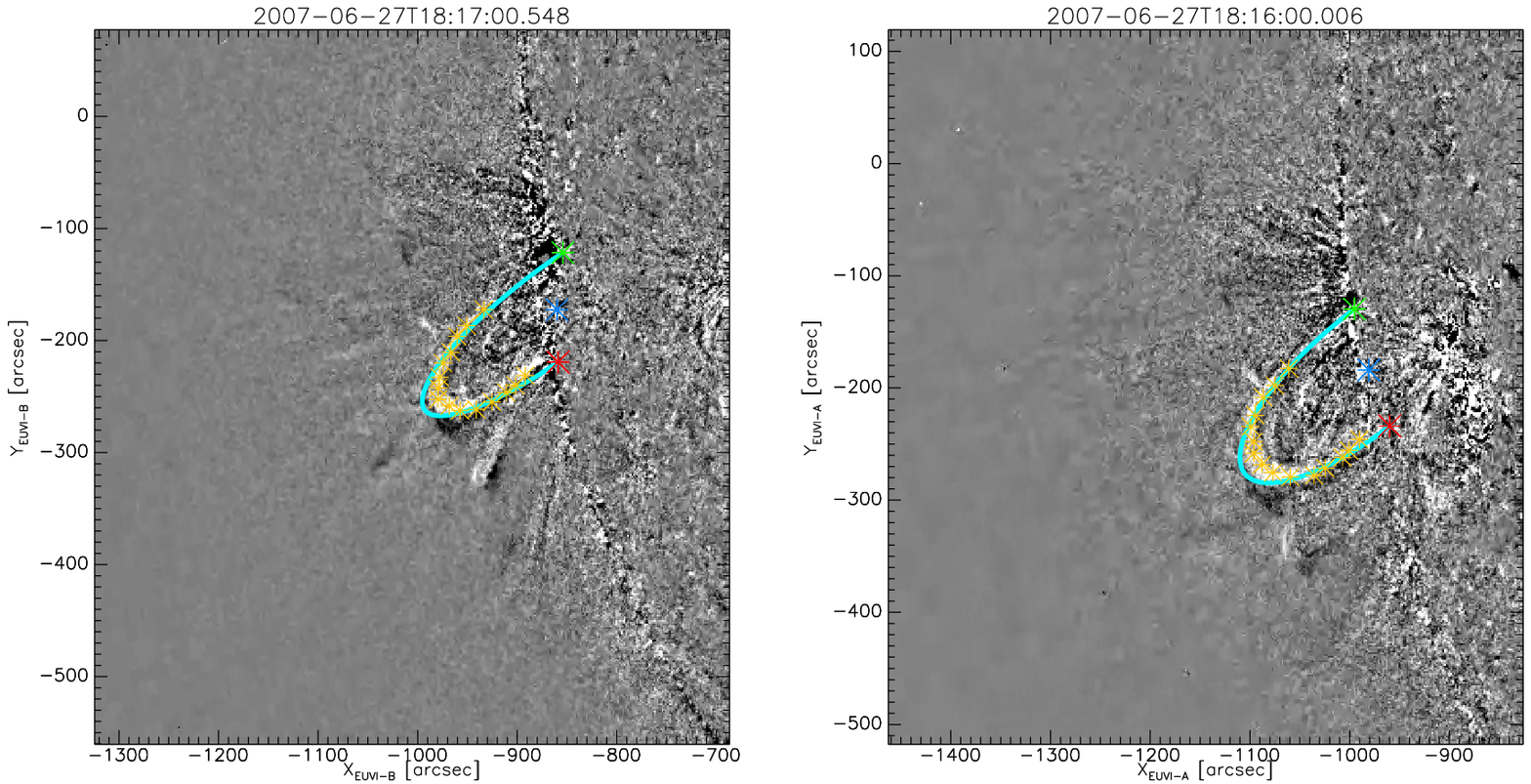}\\
		\includegraphics[width=15 cm]{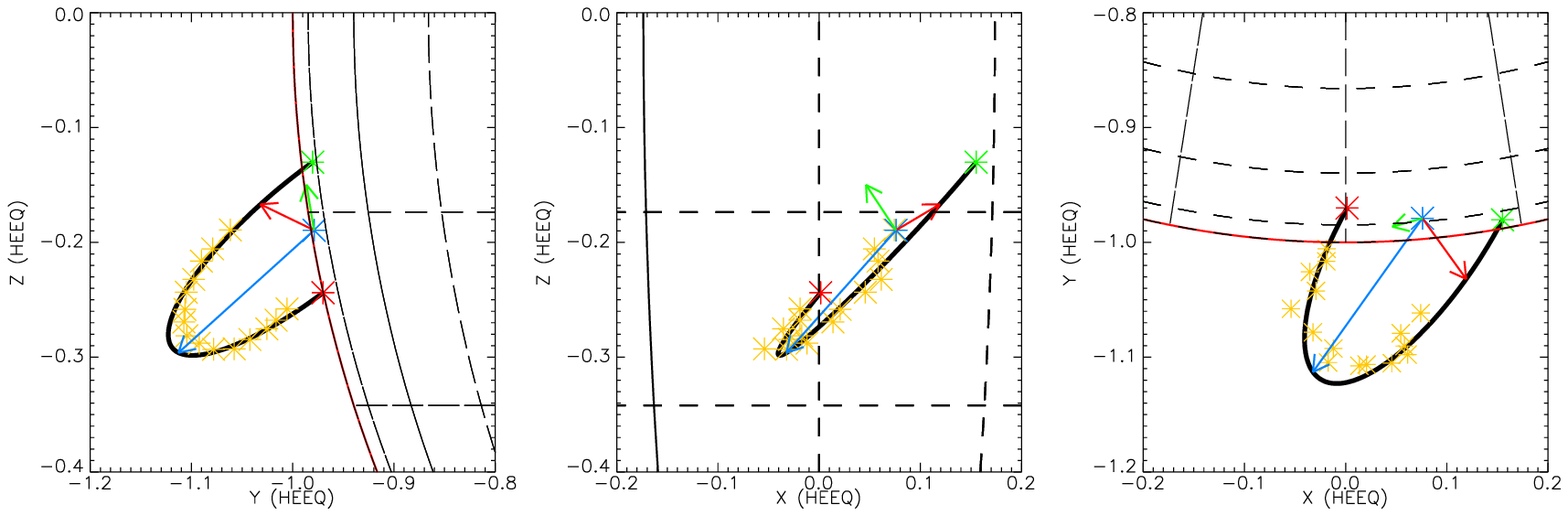} \\
		\includegraphics[width=15 cm]{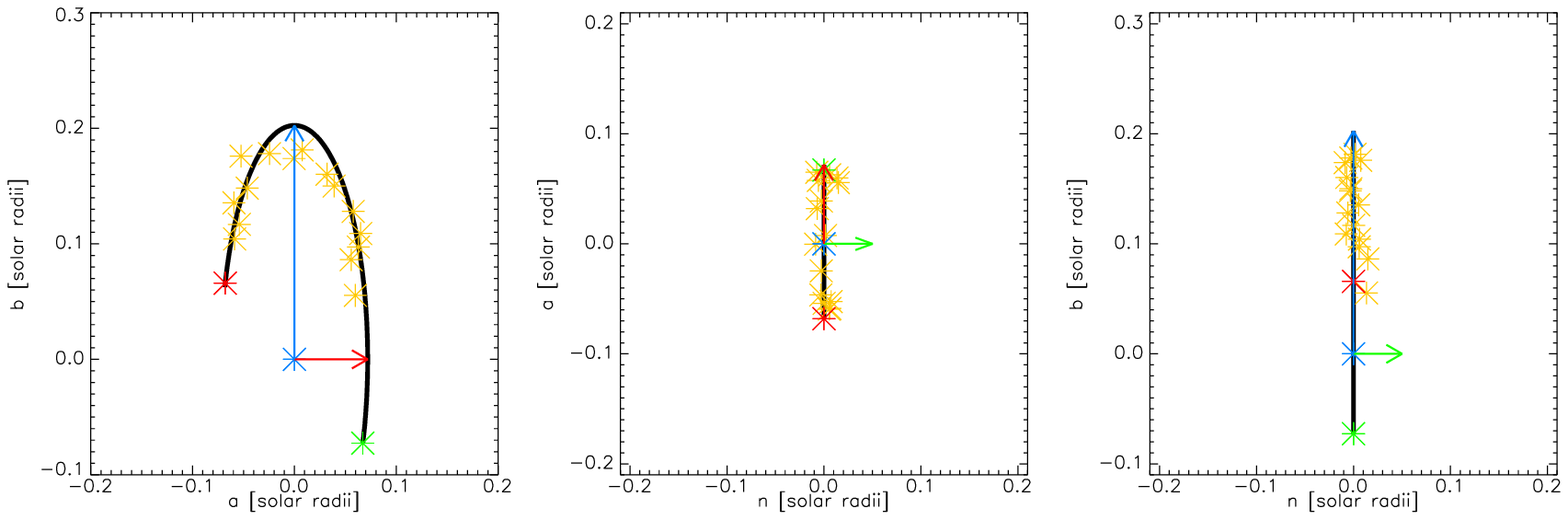} 
	\end{tabular}
	\caption{A coronal loop on the east solar limb during a flare event on June 27, 2007. Top: field-of-views from STEREO-B (left) and STEREO-A (right). The yellow symbols are the 3D data points measured with the routine {scc\_meaure.pro}. The light-blue curve fits the loop; at the base, the midpoint (i.e. the loop centre at the baseline) is in blue, and the footpoints inferred from the reconstruction are in red and green colours. Middle: plots of the reconstructed loop against the yellow data points, for the different orientations of the HEEQ coordinate system. The coloured vectors represent the three axes of the loop inferred with PCA: in red and blue, the minor and major axes; in green, the normal to the loop plane. Bottom: plots of the loop in the local reference frame for different orientations. Adapted from \citet{Nistico2013}.}
	\label{fig_Nistico_fitting}
\end{figure}

The ellipse fitting the loop can be traced by defining a grid of points and by using the parametric equations in \eqref{eq_loop} for an ellipse with $\eta = [0,2\pi]$. The curve is defined in the local reference frame and its coordinates can be converted into the Sun's reference frame by the change-of-basis matrix constructed from the eigenvectors. Then, the transformed points whose distance from the Sun's centre is $\geq 1.0 R_\odot$, defines the curve fitting the loop in the Sun's reference frame.  
Finally, the length of 3D reconstructed loops can be found numerically, by summing all the distances between the grid points defining the fitting curve.
 An example of the fitting method with PCA is given in Fig. \ref{fig_Nistico_fitting}. It shows a loop observed by STEREO on June 27, 2007. The event was also studied by \citet{Verwichte2009} and \citet{Aschwanden2009}, and offers a term of paragon for the different approaches in the 3D reconstruction of loops. By applying the forward-modelling technique with a circular and planar loop approximation, \citet{Verwichte2009} found a length for this loop of 346 Mm, while the 3D reconstruction by \citet{Aschwanden2009} resulted in a non-cicurlar, non-planar and highly asymmetric loop with a length of 311 Mm.
The PCA-based method results in an elliptical and planar loop with an intermediate estimate of the loop length of about 325 Mm. The strong asymmetry in the loop shape is due to a tilt of the loop axes with respect to the normal to the solar surface.
 
\section{Forward modelling}
\label{sec:forward_modelling}
As the name suggests, `forward modelling' consists of anticipating how physical data (i.e. temperature, density, velocity and so forth) obtained from specific numerical simulations would look if it  corresponded to solar data observed with a specific instrument and under specific observing conditions.
The step of synthesising observations based on numerical simulation data is crucial for proper comparison with solar observations. This process allows us to to predict specific features of modelled physical mechanisms (such as resonant absorption) that one may want to look for in observations, assuming that the numerical modelling is correct. At the same time, the analysis of synthetic observations also allows us to identify what may be missing from the numerical modelling in order to match real observations. 

Forward modelling therefore consists of various steps that can be listed as follows:
\begin{enumerate}
	\item Identify numerical simulation data,
	\item Identify a target solar instrument to compare the results with,
	\item Identify imaging and/or spectroscopic spectral line synthesis,
	\item Identify a specific spectral line or an imaging channel ,
	\item Use an atomic database (e.g. CHIANTI for optically thin case) and/or a radiative transfer code (for optically thick case) for calculating the line emission,
	\item Identify a specific observing mode, such as a LOS, a raster with specific number of slits etc,
	\item Identify spatial and spectral resolution for the results exceeding the capabilities of the target instrument,
	\item Degrade the calculated emission map(s) to the desired resolution assuming a specific spatial point-spread function (PSF) of the target instrument, rebin your results to a specific platescale and convert the intensity into the specific intensity flux detected by the instrument (taking into account the effective area and photons per data number of the CCD). 
\end{enumerate}
For example, if we have 3D simulation data and we want to compare with AIA 171\,\AA\ observations, then we want to perform imaging in the AIA 171\,\AA\ channel, which is dominated by (but not limited to) the \ion{Fe}{IX}~171.073\,\AA\ spectral line. Since it corresponds to EUV optically thin emission we can use CHIANTI \citep{Dere_1997AAS..125..149D}. If we want to compare with on-disk data we may want to choose a LOS of $90^{\circ}$ to the solar surface. The target temporal cadence and spatial resolution (incl. PSF effect) of AIA are 12\,s and $\approx1.2\arcsec$ with a platescale of $0.6\arcsec$. The intensity will then need to be multiplied by the effective area of AIA at 171.073\,\AA\ ($\approx2.9$~cm$^{-2}$) and a factor close to 1 to convert to photons/DN for that particular wavelength. Assuming the same settings but targeting Hinode/EIS spectral data we would need to degrade the spatial resolution to $3\arcsec$, with $1\arcsec$ platescale and a spectral resolution corresponding to $\approx36$\,km\,s$^{-1}$. The effective area in this case is $\approx0.3$~cm$^{-2}$ and the number of photons/DN is between $2-3$ depending on the wavelength. Long exposure times need also to be accounted for since they define a time over which our synthesised results need to be summed. An example application of this process is provided in Fig.~5 of \citet{2020SSRv..216..140V}.

\subsection{The FoMo code}

The FoMo code \cite{VanDoorsselaere_0.3389/fspas.2016.00004} is designed to forward model the emission of optically thin plasmas under specific observing conditions and instrumentation, thereby covering all the points in the list above\footnote{In addition, a FoMo version for calculating the gyrosynchrotron emission from non-thermal particles is also available}. FoMo uses emissivity tables pre-calculated from the CHIANTI atomic database, and is written in such a way as to be independent from the CHIANTI package. 

\begin{figure}
	\centering
	\includegraphics[width=0.98\textwidth]{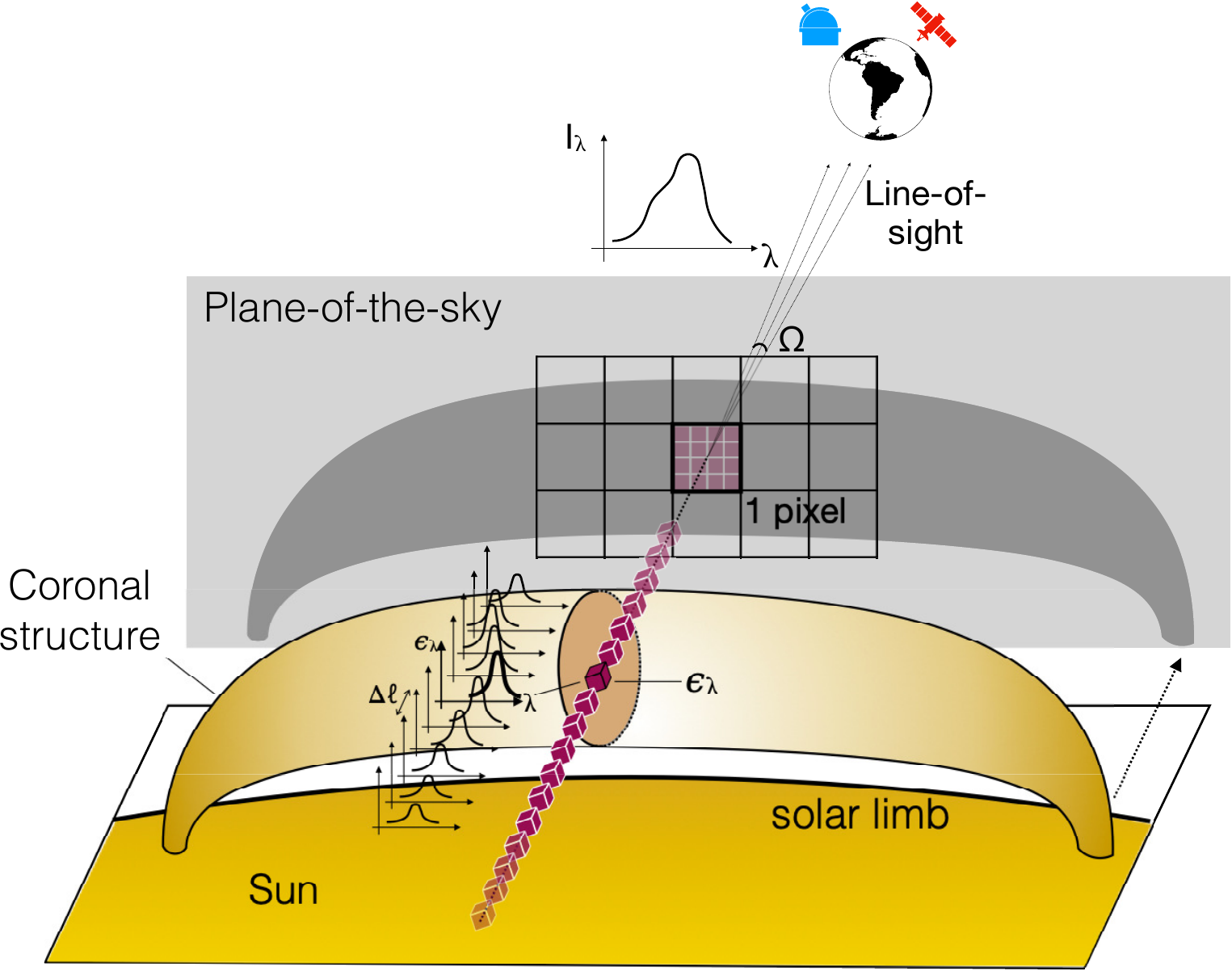}
	\caption{Sketch of optically thin line formation and measurement. A coronal structure at the limb of the Sun emits photons corresponding to a given spectral line. The emerging intensity $I_{\lambda}$ captured by a telescope on Earth is formed by the sum of emission from elementary plasma volumes (defined by constant temperature and density) along the line-of-sight, each emitting an amount $\epsilon_{\lambda}$ (see Eq.~\ref{Eq:emission}) and separated by a distance $\Delta \ell$ along the line-of-sight. The light going into a single CCD pixel in the telescope is defined by the solid angle $\Omega$ (set by the aperture of the telescope). This in turn defines an area of emission in the plane-of-the-sky for each CCD pixel (corresponding to the spatial resolution of the telescope). FoMo simulates these steps of emission and produces the specific intensity taking into account the properties of the observing instrument.}
	\label{fig:fomo_sketch}
\end{figure}

Specifically, FoMo calculates the sum:
\begin{equation}\label{Eq:emission}
I(\lambda,x^{\prime}_i,y^{\prime}_j)=\sum_k\epsilon(\lambda,x^{\prime}_i,y^{\prime}_j,z^{\prime}_k)\Delta\ell,
\end{equation}
where $I(\lambda,x^{\prime}_i,y^{\prime}_j)$ is the specific intensity (in ergs~cm$^{-2}$~s$^{-1}$~sr$^{-1}$~\AA$^{-1}$) at the wavelength $\lambda$, $\epsilon(\lambda,x^{\prime}_i,y^{\prime}_j,z^{\prime}_k)$ is the emissivity for a specific voxel at position $(x^{\prime}_i,y^{\prime}_j,z^{\prime}_k)$ in the observation frame of reference (see Fig.~\ref{fig:fomo_sketch}), which is related to the simulation grid $(x_i,y_j,z_k)$ by some specific transformation defined by the LOS angle along the direction $z^{\prime}$. Lastly, $\Delta\ell$ denotes a uniform resolution step along the LOS, defined by the numerical resolution. 

The emissivities $\epsilon(\lambda,x^{\prime}_i,y^{\prime}_j,z^{\prime}_k)$ are calculated from the temperature and density in each voxel, and a spectral profile is assigned to the voxel assuming a Gaussian distribution, with a width defined by the local temperature (i.e. a thermal line width) and Doppler shifted according to the local velocity. Details of the process and on the various FoMo versions for coding languages can be found in \cite{VanDoorsselaere_0.3389/fspas.2016.00004} and is maintained at \url{https://github.com/TomVeeDee/FoMo}.

\subsection{Applications of the FoMo code}

\begin{figure}
	
	\includegraphics[width=1.\linewidth]{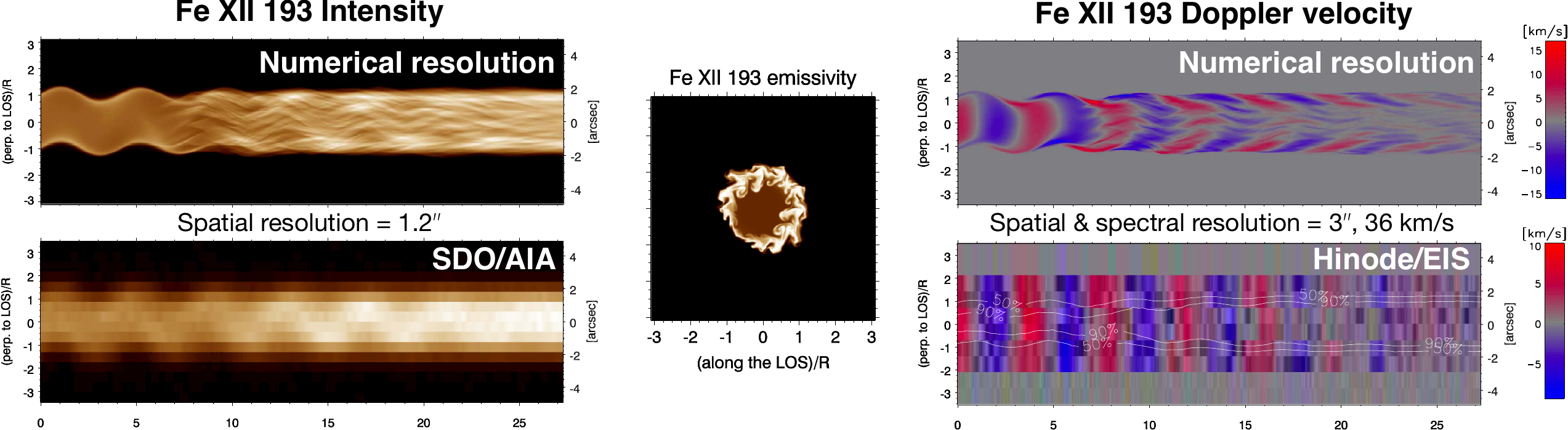}
	\caption{Forward modelling in the Fe XII 193\,\AA\ line (and corresponding AIA 193\,\AA\ passband) of an impulsively excited standing kink mode oscillation in a coronal loop. The simulation and forward modelling procedure are described in \citet{Antolin_2017ApJ...836..219A} and the figures are adapted from there. The two panels on the left show the intensity at numerical (top) and AIA spatial resolution (bottom). Note that the periodic brightening from the Kelvin-Helmholtz instability (KHI) results in an apparent decayless oscillation at AIA resolution. The two panels on the right show the Doppler velocity in the same line at numerical (top) and Hinode/EIS resolution (bottom). Note the absence of the characteristic amplitude increase and herringbone pattern from resonant absorption and phase mixing (respectively) at the loop boundary at EIS resolution. The contours with percentages in the panel denote intensity contours relative to the maximum intensity. The panel in the middle is adapted from Antolin et al. (2016) and denotes the loop cross-section of the emissivity in the same line. Note that this line is particularly sensitive to the boundary dynamics where the KHI occurs. Adopted from \citet{Antolin_2017ApJ...836..219A}.}
	\label{fig:FoMo_KHI}
\end{figure}

FoMo has proved especially popular for the forward modelling of wave phenomena in 3D MHD simulations (or 3D numerical boxes constructed from the expected perturbations based on wave dispersion relations).
In particular FoMo has allowed for the characterisation of observational signatures of various MHD modes in different coronal conditions.
This includes slow modes along flaring coronal loops \citep{Yuan_2015ApJ...807...98Y}, sausage modes in quiescent and flaring loops \citep{Antolin_VanDoorsselaere_2013AA...555A..74A,2018ApJ...863..167G, Shi_2019ApJ...883..196S}, gyrosynchrotron emission from sausage modes \citep{2014ApJ...785...86R}, torsional Alfv\'en waves \citep{Guo_2019ApJ...870...55G}, and kink modes in monolithic \citep{Yuan_2016ApJS..223...23Y, Antolin_2017ApJ...836..219A} and  multi-stranded coronal structures \citep{Magyar_2016ApJ...823...82M, Pant_2019ApJ...881...95P} and spicules \citep{Antolin_2018ApJ...856...44A}.
An example of forward modelling impulsively excited kink waves and observational effects associated with the Kelvin Helmholtz instability (KHI) are shown in Figure \ref{fig:FoMo_KHI}.
FoMo has also been used in other contexts, such as flux rope eruption \citep{Zhao_2019ApJ...872..190Z}, associated formation of KHI and Alfv\'enic vortex shedding during eruption \citep{Syntelis_2019ApJ...884L...4S} and in characterising the EUV signatures of heating associated with braiding \citep{Pontin_2017ApJ...837..108P}.
FoMo has also been adapted into a method to calculate the density and temperature profiles of the quiet corona and coronal holes by \cite{2019ApJ...884...43P}.

\section{Bayesian analysis}
\label{sec:bayesian_analysis}

\subsection{The concept of the Bayesian approach}

Any measurement of a physical quantity implies uncertainty.
Even direct measurements have uncertainties caused by a limited accuracy of our measurement tools and noise which is always present in the data.
In the case of coronal seismology, most of our measurements are indirect and require comparison of the results predicted by theoretical models with actual observational data.
This is a source of additional uncertainties associated with incomplete and uncertain observations, limited knowledge of additional parameters that are required by theoretical models but could not be measured directly. 
Theoretical models themselves are the source of uncertainty since there  could be several different models explaining the same observation.

Inferring physical parameters from the comparison between observed and theoretically predicted wave properties is not straightforward.
We are limited in our access to direct information on the physical properties of the magnetic and plasma structures of interest.
Our information from observations is incomplete and uncertain.
Because of this our assessment of physical parameters and plausible models has always a probabilistic nature, and our conclusions can at best be probability distributions \cite{gregory05}.
Therefore, it is natural to express any data and information used in the analysis in the form of probability distributions, including measurement data, theoretical models, inference results, as well as any prior knowledge about unknown model parameters.

The interpretation of a probability in the Bayesian paradigm is different from the common frequentist approach.
The probability is interpreted rather as a degree of belief of what the quantity value could be than as the frequency of an outcome of multiple identical experiments.
Note, that the Bayesian interpretation ideally suits the parameter inference problem, when one has to make conclusions about unknown parameters using only a single measurement. 

The central point of the Bayesian inference is Bayes' theorem,
\begin{equation} \label{eq:Bayes theorem}
P(\theta|D,M) = \frac{P(D|\theta,M)P(\theta)}{P(D)}.
\end{equation}
All quantities used in \eqref{eq:Bayes theorem} are probability distributions representing information about unknown model parameters $\theta$:
\begin{enumerate}
	\item \textit{The prior probability distribution} $P(\theta)$ represents the information about model parameters $\theta$ before considering observational data $D$. It could be information from previous measurements, a researcher's personal belief of what parameter values are realistic, or even the statement that nothing is known about $\theta$.
	\item \textit{The likelihood }$P(D|\theta,M)$ represents the measurement results by combining observational data $D$ and corresponding uncertainties with a theoretical model $M$ explaining these data.
	\item \textit{The posterior probability distribution} $P(\theta|D,M)$ represents the updated knowledge after considering new data $D$ under condition that the model $M$ correctly describes the investigated physical system.
\end{enumerate}

The normalisation constant $P(D|M)$ is the \textit{Bayesian Evidence} or marginalised likelihood
\begin{equation} \label{eq:Evidence}
P(D|M)=\int P(D|\theta,M)P(\theta|M) d\theta.
\end{equation}
It is used in quantitative model comparison by computing the Bayes factor if two models are compared, or by quantifying probability of a model $M$ to be most relevant to the observations in the case of multiple competing models.
Thus, Bayes' theorem describes how the prior knowledge $P(\theta)$   is transformed to an updated knowledge $P(\theta|D)$ by adding new information $P(D|\theta)$ from the measurements $D$.

The expected values and credible intervals of a selected model parameter $\theta_j$ can be estimated from marginalised posterior distribution which is of the posterior probability density function \eqref{eq:Bayes theorem} integrated over the whole parameter space by all parameters with except of $\theta_j$:

\begin{equation} \label{eq:marginal_posterior}
P(\theta_j|D,M) = \int P(\theta|D,M)d\theta_{i\ne j}.
\end{equation}

Probabilistic inference considers any inversion problem as the task of estimating the degree of belief in statements about parameter values or model evidence in view of observed data. It is based on Bayes' theorem and considers that our state of knowledge is a combination of what we know a priori independently of data and the likelihood of obtaining a given data realisation actually observed as a function of the parameter vector. Their combination leads to the so-called posterior distribution that accounts for what can be said about our parameters, conditional on data and the assumed model. 
Bayes' rule can then be applied to parameter inference by computing the posterior for different combinations of parameters and then marginalising. To model comparison, by computing posterior ratios and Bayes factors or to model averaging, by computing a combined posterior, weighted with the evidence for each model.

Thus, the core of Bayesian analysis is Bayes' theorem, which describes how a priori knowledge is updated by new observational data.
Since Bayes' theorem operates with probabilities and probability densities, the data analysis problem should be expressed in the language of probability distributions, where the prior distribution represents the knowledge about model parameters before observations, the likelihood function represents observational data and a model connecting observables with free parameters, whereas the inference result (i.e. the model parameters and their credible intervals, best-fitting the observations) will be obtained as a posterior distribution.
Thereby, data analysis within the Bayesian approach should follow general steps listed below:
\begin{itemize}
	\item Identify available observables and physical quantities to be inferred.
	\item Construct a model (or models) connecting observables with  the physical parameters of interest.
	\item Express observational information and model as a likelihood function.
	\item Express available knowledge about free model parameters  as prior probability density function.
	\item Use Bayes' theorem to obtain updated knowledge about free parameters as a posterior probability density function.
	\item Compute marginal posteriors of free parameters using analytical integration, direct numerical integration or MCMC sampling of the posterior probability density function.
	\item Estimate expected values and credible intervals from marginal posteriors.
\end{itemize}

Bayesian inference tools have been commonly used in several areas of astrophysics since decades \citep[see e.g.][]{loredo92}. It began to be applied in coronal seismology almost a decade ago \citep{arregui11} and is now being incorporated to an increasing number of studies \citep[see][for a review]{arregui18}.

\subsection{Model comparison}
\label{sec:model_comparison}

Bayesian analysis allows for quantitative comparison of two models $M_1$ and  $M_2$ by calculating the \textit{Bayes factor} \citep{Jeffreys1961}, defined as
\begin{equation}
\label{eq:Bayes factor}
B_{12} = \frac{P(D|M_1)}{P(D|M_2)},
\end{equation}
where the evidences $P(D|M_1)$ and  $P(D|M_1)$ are calculated according to Eq. \ref{eq:Evidence}.
Traditionally, the doubled natural logarithm of this factor is used, i.e.
\begin{equation}
\label{eq:Bayes_K}
K_{12} = 2\ln B_{12},
\end{equation}
where  values  of $K_{12}$ greater  than  2,  6,  and  10  correspond  to \lq\lq positive\rq\rq , \lq\lq strong\rq\rq, and \lq\lq very strong\rq\rq\  evidence for model $M_1$ over model $M_2$, respectively \citep{Kass1995}.

For further understanding of the Bayesian evidence \eqref{eq:Evidence}, let us use Bayes' theorem to compute the posterior probability of a model $M_i$ in the case of $N$ competing models $M_1, M_2 ... M_N$:

\begin{equation}\label{eq:model_probabilities}
P(M_i|D) = \frac{P(D|M_i) P(M_i)}{P(D)},
\end{equation}
where $P(M_i)$ is the prior probability of model $M_i$, the likelihood term $P(D|M_i)$  is a Bayesian evidence for model $M_i$ defined by \eqref{eq:Evidence}.  $P(D)$ is again a normalisation constant and can be computed as
\begin{equation}\label{eq:model_normalisation}
P(D) = \sum_{i=1}^N P(D|M_i) P(M_i).
\end{equation}

Thus, Equations \eqref{eq:model_probabilities} and \eqref{eq:model_normalisation} allow us to compute the posterior probability of a model $M_i$ to be the most relevant to observational data $D$ among a set of competing models $M_1 .. M_N$.
In the case of $N=2$ and  $P(M_1) = P(M_2)$, there is a direct correspondence between the model probability \eqref{eq:model_probabilities} and the Bayes factor \eqref{eq:Bayes factor} $B_{12} =\frac{P(M_1|D)}{P(M_2|D)}$ and hence
\begin{equation}\label{eq:Bayes_factor_2_probability}
P(M_1|D) = \frac{B_{12}}{1 + B_{12}},\; P(M_2|D) = \frac{1}{1 + B_{12}}.
\end{equation}
For instance,  $B_{12} = 20$ (strong evidence for model $M_1$) corresponds to $P(M_1) =20/21 \approx 0.95$ and  $P(M_2) =1/21 \approx 0.05$.

The posterior model probabilities \eqref{eq:model_probabilities} can be used to select the most appropriate model if one of the models has a probability  close to unity.
Otherwise, the posterior distributions computed using different models can be weighted to make a combined distribution accounting for several possible models:
\begin{equation} \label{eq:combined_distribution}
P(\theta|D) = \sum_{i=1}^N P(M_i|D)P(\theta|D,M_i).
\end{equation}

\subsection{Applications in coronal seismology}

Bayesian inference began to be applied in coronal seismology almost a decade ago.
\citet{arregui11} reanalysed  measurements of the periods and damping time of large-amplitude kink oscillations measured by other researchers.
Thanks to the Bayesian approach, they managed to find well-localised solutions and obtain reliable error bars for the Alfv\'en travel time and transverse inhomogeneity length scale, despite the ill-posed nature of the problem and the infinite number of solutions.
Though a model parameter (e.g. Alfv\'en travel time) has an infinite number of solutions matching observational data, individual values of free parameters are not equally probable.
Thus,  calculation of the marginalised probability density allows for estimation of the expected values of unknown parameters and corresponding credible intervals.

The very first models used for Bayesian inference of coronal loop parameters using observations of kink oscillations \citep{arregui11, 2012ASSP...33..159A,2013ApJ...765L..23A, 2013ApJ...769L..34A,2014A&A...565A..78A,2015A&A...578A.130A,2015ApJ...811..104A} were rather simple and had only 3 unknown parameters: Alfv\'en travel time, density contrast, and transverse inhomogeneity length scale.
The low number of free parameters allowed the authors to compute marginal posterior probability density by direct numerical integration of Eq. \eqref{eq:marginal_posterior}.

\citet{2015ApJ...811..104A} calculated the Bayes factors by direct numerical integration to compare the plausibility of different models of the transverse density profiles: linear, sinusoidal and parabolic.
\citet{2017ApJ...846...89M} applied Bayesian analysis to infer physical parameters of kink oscillating coronal loops such
as the density contrast, the transverse density inhomogeneity length-scales, and the aspect ratio of coronal loops. Also, they calculated Bayes factors to compare three models of damping for kink oscillations of coronal loops.
A similar study has been performed for transverse oscillation in prominence threads \citep{2019A&A...622A..88M}.
The authors used Bayesian analysis to infer physical properties of prominence threads, including the magnetic field, and quantified the plausibility of alternative damping mechanisms of transverse prominence oscillations by calculating the Bayes factor.
\citet{2019A&A...625A..35A} collected measurements of oscillation period, loop length and damping time reported in six publications by other authors for more than 150 oscillating loops.
They  investigated how the selection of priors for  density contrast and loop density affects the results and found significant dependence of the inferred magnetic field upon  the prior information on plasma density  while selection of different priors for density contrast barely influences the results.

The increase of model complexity and the number of free parameters quickly makes direct numerical integration of \eqref{eq:marginal_posterior} practically impossible when the number of free parameters exceeds 4--5, since every additional dimension in the parameter space increases the computation time by 2--3 orders of magnitude.
Thus, a large multi-parametric model requires an alternative method for evaluating \eqref{eq:marginal_posterior}. The most popular approach is to generate large number of samples from the posterior distribution using Markov Chain Monte-Carlo (MCMC) method \citep[first published in][]{1953JChPh..21.1087M} and subsequent construction of histograms  approximating marginalised posterior distributions of the parameters of interest.
The collected samples allow for estimation of the expected values and credible intervals by simple averaging and computing percentiles.

We should note that, despite its random nature, MCMC as with other Monte-Carlo techniques is a precise method, meaning that any given accuracy (in computing marginal posteriors) can be achieved by increasing the number of samples.
Due to this feature, a data analysis workflow may include several considerably fast \lq\lq quick look\rq\rq\ MCMC computations of low precision    followed by a time demanding final precise calculation with a much higher number of samples.

For the first time, the MCMC approach in coronal MHD-seismology was applied by \citet{arregui11,2013A&A...554A...7A}, who used their own implementation of  Metropolis-Hastings algorithm \citep{{1953JChPh..21.1087M}} which is the most basic and popular MCMC algorithm to sample proposal distribution.

An implementation of the same Metropolis-Hastings algorithm lies in the core of the Solar Bayesian Analysis Toolkit (SoBAT).
Thanks to SoBAT, the use of Bayesian inference in solar data analysis has dramatically increased in recent years.
SoBAT provides a framework for Bayesian parameter inference written in the  IDL programming language.
It was initially developed for analysing solar data but is applicable to a wide range of problems.
SoBAT is available at \href{https://github.com/Sergey-Anfinogentov/SoBAT}{https://github.com/Sergey-Anfinogentov/SoBAT} and described in detail in \cite{2020arXiv200505365A}.
SoBAT uses Markov chain Monte Carlo (MCMC) sampling with a Metropolis-Hastings acceptance algorithm to compare a user-provided model with data.
Another advantage of this approach is the accurate estimation of parameter uncertainties.
SoBAT is designed for the analysis of observations and so by default includes an estimation of the level of (normally distributed) noise in the data unless otherwise provided by the user.
The code can also calculate the Bayesian evidence for a model which is particularly useful for quantitative comparison of two or more competing models using Bayes factors given by Equation~\ref{eq:Bayes_K}.

MCMC sampling is useful for high-dimensional parameter inference problems i.e. ideal for complex models with numerous free parameters.
Unlike other approaches, for example least-squares fitting, it also does not require convergence to a unique solution.
This is useful for problems such as the seismological inference of the transverse density profile of corona loops using the damping of kink oscillations, for which the constraint of model parameters is highly sensitive to the level of noise in the data \citep[see Figure~3 of][]{2018ApJ...860...31P}.
On the other hand SoBAT does require the user to provide explicit prior information about the model parameters. This can take the form of strictly defined limits, information from previous analysis, or simply a range of values considered reasonable for the problem.
Also, we need to warn the reader that the Metropolis-Hastings  algorithm  used by SoBAT  has its own limitations. In particular, it becomes less effective for really high-dimensional problems, when the number of free parameters reaches several hundreds and more. In such cases, we advise to try alternative approaches such as Gibbs sampling or Hamiltonian Monte Carlo \citep[see][for a detailed discussion]{2017arXiv170102434B}.
Below we consider some recent examples of Bayesian inference and model comparison in solar physics using SoBAT.

The SoBAT code was first applied by \cite{2017A&A...600A..78P} who considered the damping of kink oscillations of coronal loops.
The method is based on using the shape of the damping profile of standing kink oscillations due to resonant absorption \citep[e.g. review by][]{0741-3335-58-1-014001,2021SSRv..217...73N} to infer information about the transverse density profile of the loop.
The non-exponential damping profile had been established through numerical simulations \citep{2012A&A...539A..37P,2013A&A...551A..40P,2015A&A...578A..99P}, analysis \citep{2013A&A...551A..39H}, and observations \citep{2016A&A...585L...6P}.
The use of Bayesian inference to determine the density profile using the shape of the damping profile was demonstrated by \cite{2013ApJ...769L..34A}.
The first application of the seismological method to observational data by  \cite{2016A&A...589A.136P} used least-squares fitting of the model and so was limited to three coronal loops with density profile parameters.

In the absence of accurate information about the shape of the profile, the problem becomes ill-posed and the solution is represented by an inversion curve in the parameter space \citep[e.g.][]{2019A&A...622A..44A}.
Bayesian inference is ideal for this problem since it can readily return marginalised posteriors rather than rely on convergence to a unique solution as with least-squares fitting.
Subsequent applications of the seismological method using SoBAT \citep{2017A&A...600A..78P,2017A&A...607A...8P,2018ApJ...860...31P,2019FrASS...6...22P} allowed less constrained examples to be considered.

Bayesian analysis is often combined with forward modelling allowing us  to easily incorporate our physical understanding into the model used to generate synthetic data.
This provides additional constraints which can assist in extracting weak signatures in the data, though the results are obviously contingent on the model applied being appropriate for the particular problem.
This was used by \cite{2017A&A...600A..78P} to test for the presence of higher longitudinal harmonics of standing kink oscillations.
Damping of kink oscillations by resonant absorption is frequency-dependent such that higher harmonics damp below a detectable amplitude more rapidly than the fundamental mode does.
\cite{2017A&A...600A..78P} demonstrated that including this frequency-dependent behaviour into the applied model allowed the second and third longitudinal harmonics to be detected, whereas their rapid damping would lead to minimal signatures in Fourier or wavelet analysis (see their Figure~18).

\begin{figure*}
	\centering
	\includegraphics[width=0.45\textwidth]{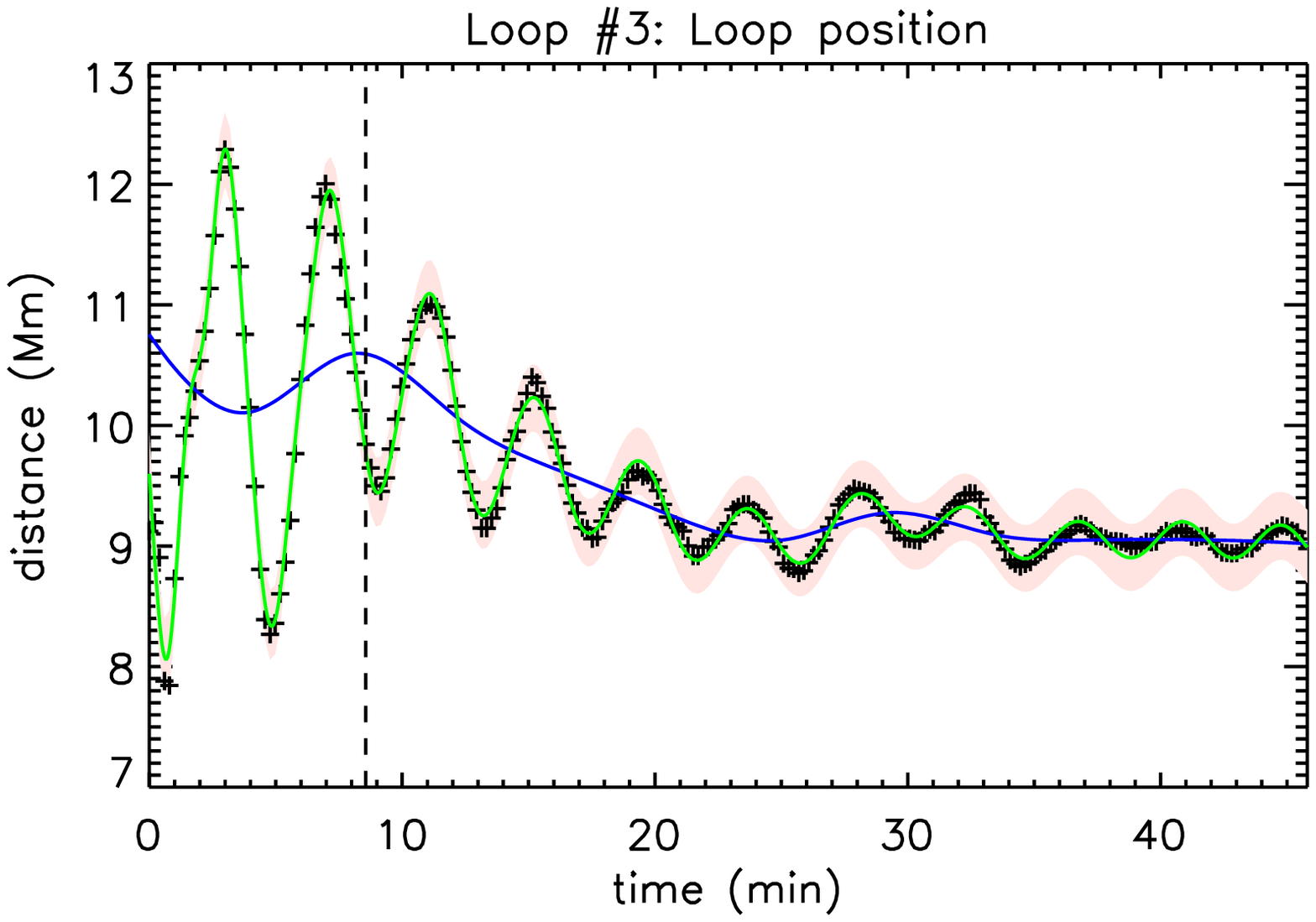}
	\includegraphics[width=0.45\textwidth]{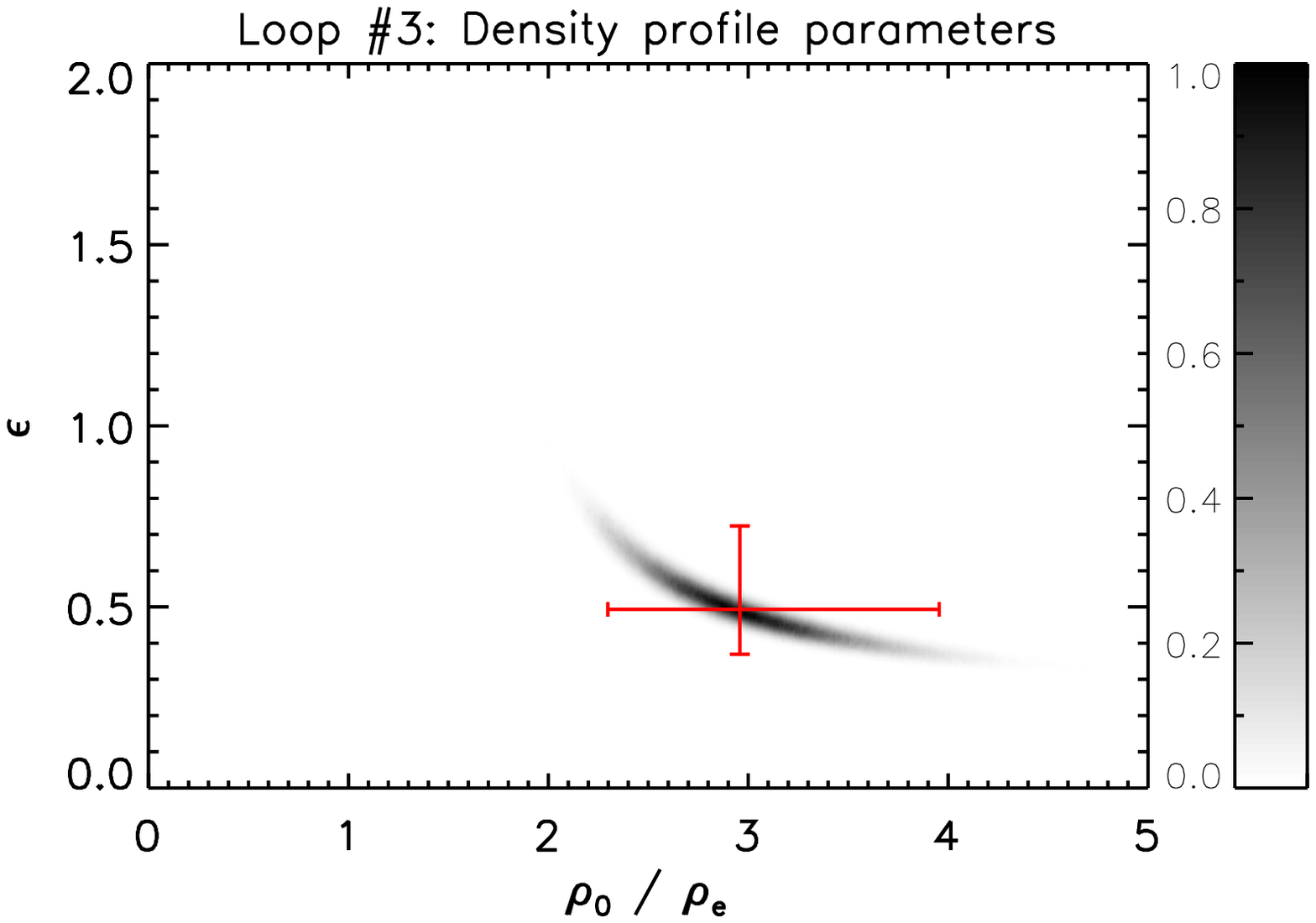}
	\includegraphics[width=0.45\textwidth]{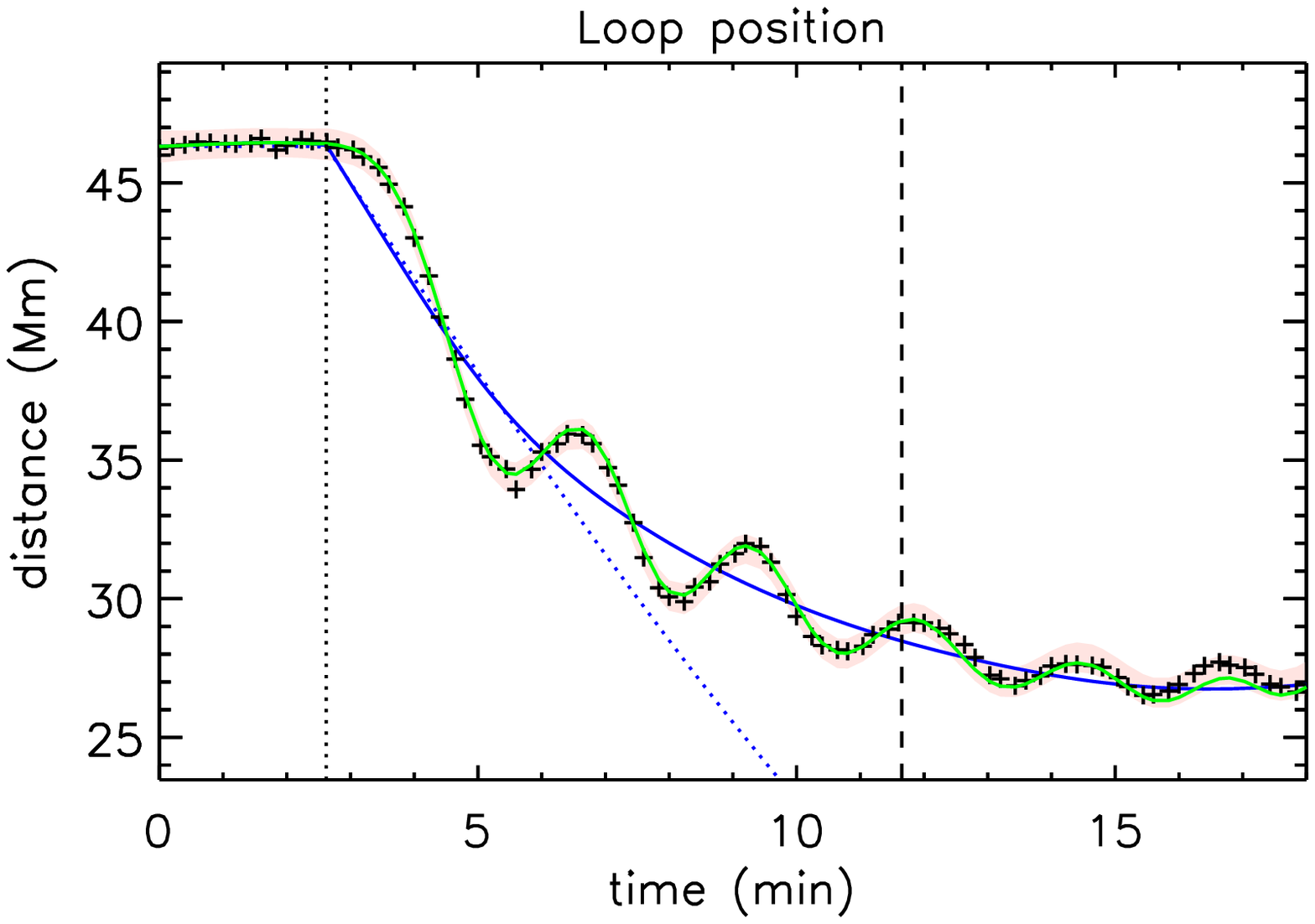}
	\includegraphics[width=0.45\textwidth]{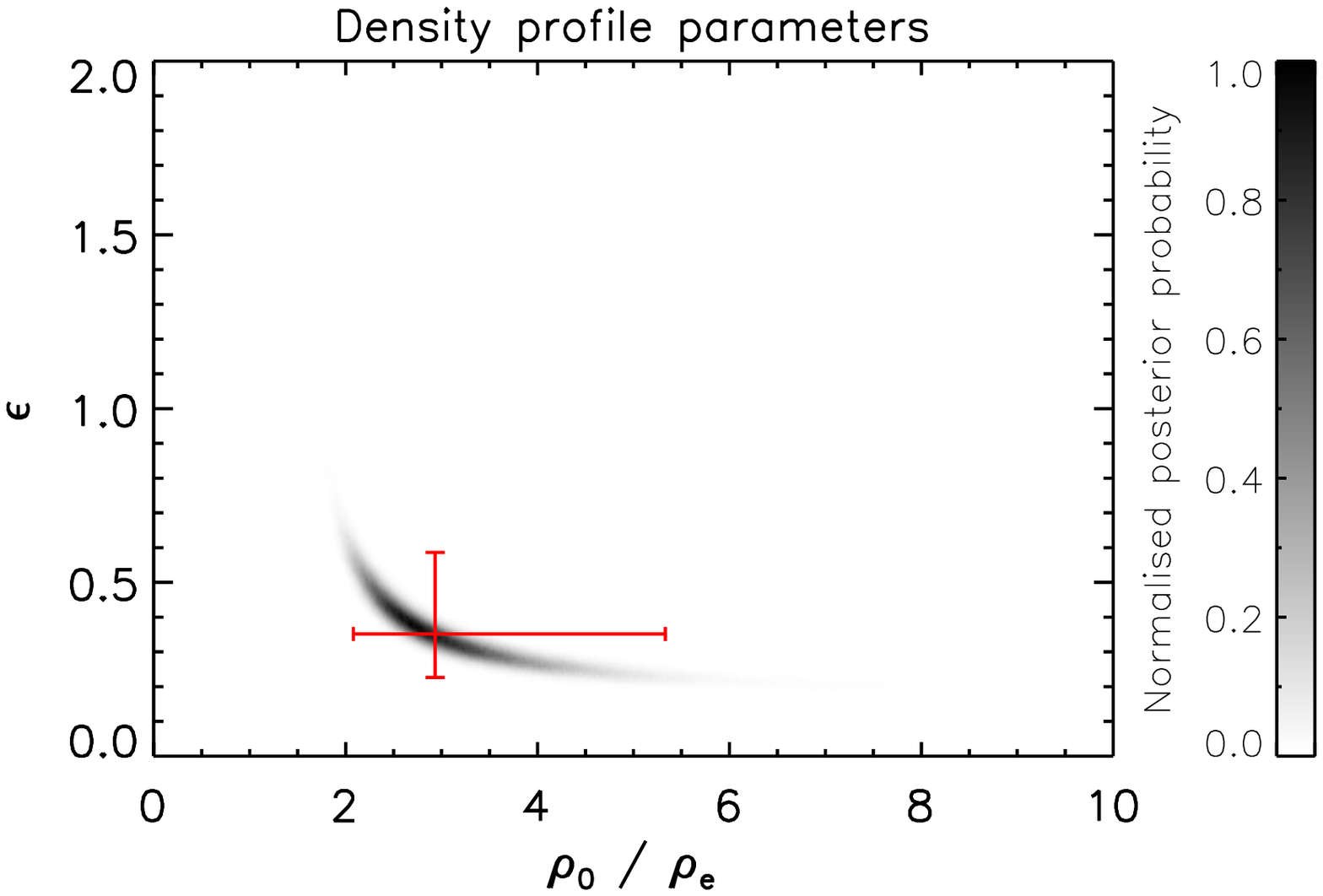}
	\caption{
		Examples of Bayesian analysis of kink oscillations adapted from \cite{2017A&A...600A..78P,2017A&A...607A...8P}.
		Left panels show the loop position (symbols) and the kink oscillation model based on median values of parameters (green lines).
		Background trends are shown by blue lines.
		The rose shaded region represents the 99\% credible intervals for the loop position predicted by the model, including an estimated noise.
		Vertical dashed lines correspond to the switch between Gaussian and exponential damping regimes \citep{2013A&A...551A..40P}.
		Right panels show 2D histograms approximating the marginalised posterior probability density function for the transverse density profile parameters estimated using the damping profile.
		The red bars indicate the median values and the 95\% credible intervals.
	}
	\label{fig:djp_seis}
\end{figure*}

Figure~\ref{fig:djp_seis} shows examples of Bayesian analysis of kink oscillations.
The top panels show analysis of a loop by \cite{2017A&A...600A..78P} which includes a decayless component \citep{2013A&A...552A..57N} and additional longitudinal harmonics generated by external excitation \citep[e.g.][]{2009A&A...505..319P,2014ApJ...784..101P}.
The bottom panels show an example from \cite{2017A&A...607A...8P} of an oscillation generated by loop contraction \citep{2013ApJ...777..152S,2015A&A...581A...8R} after a flare, without additional harmonics excited.

SoBAT has also been used to estimate the transverse density profile of coronal loops by forward modelling the EUV intensity profile to compare with SDO/AIA 171 data \citep{2017A&A...600L...7P,2017A&A...605A..65G,10.3389/fspas.2020.00061}.
In these studies different shapes were considered for the transverse density profile such as step and Gaussian functions, and a profile with a linear transition layer between the loop core and the background.
The studies found that the Bayesian evidence favoured the linear transition model over the step function, supporting the conclusion that coronal loops have inhomogeneous layers necessary for damping of kink oscillations by resonant absorption.
Furthermore, the Bayesian evidence also generally favoured the linear transition model over the Gaussian profile, consistent with the inhomogeneous layer having a finite spatial scale rather than loops being continuously non-uniform.

Bayesian inference has also been used to consider the problem of identifying quasi-periodic pulsations (QPPs) in solar and stellar flares (see e.g. \citet{2020ApJ...905...70p}) and for simultaneous multi-band detections and analysis of stellar superflares (\citet{2021ApJ...912...81K})).
\cite{2019ApJS..244...44B} performed an exercise which tested the capabilities of several different methods in confidently detecting the presence of QPPs in the time series of flux observations such as the white light flares detected by the Kepler space telescope \citep[e.g.][]{2016MNRAS.459.3659P}.
QPPs can be difficult to detect in Fourier-based methods due to their finite duration, similar to the problem of detecting higher longitudinal harmonics of kink oscillations discussed above.
Bayesian inference was found to be particularly useful for QPPs with non-stationary periods of oscillation due to the simplicity of incorporating this behaviour directly into the applied model.
Likewise, \citet{2018ApJ...861...33K} applied Bayesian inference and MCMC sampling for interpreting a quasi-periodic fine structure observed in the dynamic spectrum of a type III solar radio burst, in terms of a propagating fast magnetoacoustic wave.  

In \cite{2020ApJ...898..126P} the positions of several loops in an active region were tracked over a period of time which included two solar flares.
It was evident that the first flare excited kink oscillations in the loops but less clear if the second flare did.
Bayesian model comparison was used to investigate this question by comparing the evidence for two models; one with a single large amplitude perturbation and another with two distinct perturbations.
In three of the four loops considered the Bayesian evidence was very strong for the model with two perturbations while for the fourth the evidence was positive.

\subsection{Advantages and limitations}

Thus, Bayesian inference is a universal data analysis approach aimed at inferring model parameters, reliable estimation of the uncertainties and quantitative model comparison.
It becomes extremely useful when available observational data is limited or incomplete and is suitable even for solving ill-posed problems.

Despite being a complete data analysis solution, Bayesian inference has some limitations.
It is very computationally intensive and, therefore, is less suitable for fast processing of large amounts of data. 
Thus, the main area of usage where Bayesian inference shines is reliable estimation of uncertainties and extracting all information from limited or incomplete observational data sets.
Most of the data analysis problems in the field of coronal MHD-seismology are of this kind and, therefore, can be efficiently solved with Bayesian analysis,
provided that there is a reliable model (or several alternative models) connecting observables with unknown physical parameters of coronal plasma.
These models can be either analytical expressions or numerical forward modelling tools.

\section{Conclusion}

In this review, we presented novel data analysis techniques used in the context of coronal seismology.
The  tools presented here were developed as a response to challenges induced by the quick development of coronal MHD-seismology and new observational capabilities in recent decade.
New challenges include the problems of reliable detection of oscillatory signals including non-stationary ones in time-series and imaging data, automated processing of numerous observational data sets, coping with limited spatial resolution of existing EUV instruments, and reliable seismological inference of coronal plasma physical parameters when available observations are limited and even incomplete.

The MHD-seismology of the solar corona  utilises a lot of data analysis techniques from traditional Fourier transform to novel insights such as motion magnification.
We focused mainly on  methods developed in the recent decade  targeted at the  analysis of solar observations in the EUV range.
Also, we have discussed the most common pitfalls causing  artefacts and false detections for both conventional and well-established methods and novel techniques reviewed  here.
Therefore, any data analysis tools should be always used with caution and diligence.

There is no doubts that new advanced data analysis methods will be developed or adopted from other fields in the near future.
Some of the techniques presented in this review were adopted from other fields and lead to a numerous new observational results.
Good examples of such a scientific knowledge transfer are EMD, motion magnification and Bayesian analysis.
As for the possible future insights, the most promising prospective approaches are those based on machine learning and artificial intelligence, including image recognition and classification.
We also expect tight integration and complementary usage of methods developed in other branches  of solar physics, including data-driven modelling and inferring plasma parameters from observations in X-ray and radio bands.
As an ultimate task, these developments will lead us to the routine time-dependent 3D modelling of solar active regions consistent with all available observations from photospheric magnetograms to EUV images and X-ray light curves. Coronal MHD-seismology will be an  intrinsic and irreplaceable element in this modelling, providing crucial information about the coronal plasma conditions which could not be measured by any other means.

\section{Acknowledgements}
S.A. acknowledges support from the Ministry of Science and Higher Education of the Russian Federation and from the Russian Foundation of Basic Research  (projects 18-29-21016 mk and 19-52-53045 gfen\_a).
J.A.M. acknowledges UK Science and Technology Facilities Council (STFC) support from grant ST/T000384/1. P.A. acknowledges funding from his STFC Ernest Rutherford Fellowship (No. ST/R004285/2). 
DJP was supported by the European Research Council (ERC) under the European Union's Horizon 2020 research and innovation programme (grant agreement No 724326) and the C1 grant TRACEspace of Internal Funds KU Leuven.
DYK thanks STFC for the grant ST/T000252/1, and the budgetary funding of Basic Research program II.16.
SKP is grateful to FWO Vlaanderen for a senior postdoctoral fellowship (No. 12ZF420N).
GN acknowledges the Rita Levi Montalcini 2017 fellowship funded by the Italian Ministry of Education, University and Research.

\bibliographystyle{spbasic}
\bibliography{references}
\end{document}